\ifdefined\pdfminorversion
\fi
\documentclass[11pt,a4paper]{article}

\usepackage[utf8]{inputenc}
\usepackage[T1]{fontenc}
\usepackage{lmodern}
\usepackage{microtype}
\usepackage[margin=1in]{geometry}
\usepackage{amsmath,amssymb,amsthm,mathtools}
\usepackage{physics}
\usepackage{graphicx}
\usepackage{subcaption}
\usepackage{booktabs}
\usepackage{siunitx}
\usepackage{xcolor}
\usepackage{cite}
\usepackage{hyperref}
\usepackage{cleveref}
\usepackage{authblk}
\usepackage{enumitem}
\usepackage{orcidlink}
\hypersetup{
    colorlinks=true,
    linkcolor=blue!60!black,
    citecolor=blue!60!black,
    urlcolor=blue!60!black
}
 
\title{\hfill ~\\[-30mm]
\phantom{h} \hfill\mbox{\small IPPP/26/58}
\\[0.5cm]
\vspace{13mm}   \textbf{Robust Quantum Machine Learning for Collider Event Selection under Detector Variability}}

\author[1]{Christopher Brown\, \orcidlink{0000-0002-7766-6615} }
\author[2, 3]{Michael Spannowsky\, \orcidlink{0000-0002-8362-0576} }
\author[4]{Simon Williams\, \orcidlink{0000-0001-8540-0780} }
\affil[1]{\small \textit{CERN, Esplanade des Particules 1, 1211 Geneva 23, Switzerland}\vspace{0.2cm}}
\affil[2]{\small \textit{Institute for Theoretical Physics, Campus S\"ud, Karlsruhe Institute of Technology, D-76128 Karlsruhe, Germany}\vspace{0.2cm}}
\affil[3]{\small \textit{Institute for Quantum Materials and Technologies, Karlsruhe Institute of Technology, Karlsruhe 76131, Germany}\vspace{0.2cm}}
\affil[4]{\small \textit{Institute for Particle Physics Phenomenology, Durham University, Durham, DH1 3LE, UK}}

\date{}
 
\begin{document}

\maketitle
 
% =====================================================================
\begin{abstract}
\noindent
Robust machine-learning methods are becoming increasingly important for high-energy physics data analysis as experiments enter the era of higher luminosity and future higher-energy colliders. Detector degradation, changing running conditions and calibration drift can shift data distributions, causing models trained on clean reference samples to degrade after deployment. We investigate whether parameterised quantum models provide a useful inductive bias for robust collider-event selection in two complementary settings. In the unsupervised study, quantum autoencoders trained on background events are compared with classical and variational autoencoders for anomaly detection. In the supervised study, quantum classifiers with data reuploading are trained to distinguish a supersymmetric signal from background and are compared with linear and multilayer-perceptron classifiers. All models are trained under reference conditions and subsequently evaluated under controlled feature-level smearing while their parameters and preprocessing transformations are held fixed. On clean inputs, the quantum autoencoders achieve competitive anomaly-detection performance, including in the low-false-positive-rate regime relevant for triggering, while the deeper data-reuploading classifier attains discrimination comparable to the non-linear classical baseline. Under smearing, the quantum models generally exhibit smaller shifts in their output scores and retain their discrimination more effectively than the expressive classical baselines. These results suggest that parameterised quantum models can provide a useful robustness inductive bias for collider-event selection and motivate further studies with realistic detector systematics, finite-shot statistics and quantum-device noise.
\end{abstract}

\section{Introduction}

Machine-learning models are routinely trained and validated on data drawn from a fixed reference distribution, yet deployed in environments where that distribution shifts over time. Sensor degradation, changing operating conditions, calibration drift and previously unseen inputs can all move the data away from the training distribution, such that a model that performs well during validation may degrade sharply after deployment. \emph{Robustness}, understood here as the ability to retain stable predictions and discrimination performance under such distribution shifts, is therefore a property of a learning system as important as its reference accuracy. In high-energy physics, methods that explicitly reduce dependence on nuisance parameters or incorporate systematic uncertainties during training have consequently become an important part of robust machine learning~\cite{Louppe:2016ylz,Englert:2018cfo}.

Robustness is determined in large part by the hypothesis class associated with a model family. Highly flexible, over-parameterised models can fit the training distribution very accurately, but may also learn features that are unstable under perturbations of the input. More constrained model classes encode inductive biases that can, in favourable cases, lead to more stable behaviour under shift. This view reframes the choice of model family as a question about stability under shift, not only about expressive power, and it motivates asking the same question of model classes that are structurally distinct from the standard neural-network families. 

Parameterised quantum circuits are one such class~\cite{Benedetti_2019}. Through their data encoding, circuit geometry, entangling structure and measurement, they define constrained families of functions that can be interpreted as non-linear feature maps or kernel models in high-dimensional Hilbert spaces~\cite{Havlicek:2018nqz, Schuld:2021hzr}, and that are not straightforwardly reproduced by their classical counterparts. Most of the work on quantum machine learning has asked whether this structure yields an accuracy advantage on clean benchmark data~\cite{Blance:2020ktp, Blance:2020nhl, Guan_2021, delgado2022quantumcomputingdataanalysis, terashi2021event, Ngairangbam:2021yma, 1p1q, Maier:2025ppr, Gonski:2026jgu}. Robustness studies of quantum classifiers have largely considered adversarially chosen perturbations, establishing both vulnerabilities and robustness guarantees under specific threat models~\cite{Liu:2019drq,Dowling:2024eja}. The detector-induced distribution shifts studied here are instead stochastic, physically motivated perturbations of reconstructed observables. We ask whether the restrictions imposed by a quantum model class can provide a useful inductive bias against this distinct source of drift in the learned model response.

We investigate this question in two complementary learning settings. First, we consider quantum autoencoders (QAEs)~\cite{Romero_2017, Ngairangbam:2021yma, 1p1q} as an unsupervised architecture for anomaly detection. Classical inputs are embedded into quantum states and processed by a parameterised circuit trained to compress background events into a smaller latent quantum register. Successful compression is quantified through the fidelity of the discarded, or \emph{trash}, register with a fixed vacuum state. The resulting anomaly score is therefore defined by a learned quantum compression map rather than by a pointwise reconstruction error, and can be evaluated without constructing an explicit decoder. Second, we study supervised quantum classifiers constructed using data reuploading. In these models, alternating data-encoding~\cite{PerezSalinas2020datareuploading,Schuld:2020enb} and trainable circuit layers generate an increasingly expressive decision function as the number of reuploading layers is increased. This provides a controlled means of varying the expressivity of the quantum model whilst retaining the same basic circuit architecture. Comparing classifiers of increasing circuit depth with linear and multilayer-perceptron baselines allows us to examine the relationship between reference classification performance and sensitivity to input perturbations. Taken together, the unsupervised and supervised studies test whether robust behaviour is specific to quantum compression and fidelity-based anomaly scores, or whether it arises more generally across distinct parameterised quantum model classes.

Although the question addressed here is general, we investigate it in the setting of high-energy physics. Quantum machine learning forms part of a rapidly developing programme of quantum-computing research in high-energy physics. This broader programme encompasses applications to the simulation of quantum field theories~\cite{Jordan_2012, Klco:2018zqz, Preskill:2018fag, Banuls:2019bmf, PRXQuantum.4.027001, Ingoldby:2024fcy, Abel:2024kuv, PRXQuantum.5.037001, Pavesic:2026yiz}, including studies of real-time dynamics and scattering processes~\cite{Martinez:2016yna, jordan2019, Jha:2024jan, Abel:2025zxb, qr72-51v1, Ingoldby:2025bdb}. Quantum algorithms have also been developed for collider calculations and event simulation, including numerical integration~\cite{Ramirez-Uribe:2021ubp, deLejarza:2024pgk, Williams:2025hza}, parton showers and event generation~\cite{Bauer:2019qxa, Bepari:2020xqi, Bepari:2021kwv, Gustafson:2022dsq}, and perturbative QCD calculations~\cite{Chawdhry:2023jks, Chawdhry2026}. Within collider physics specifically, quantum machine-learning methods have been explored for event classification, anomaly detection, particle reconstruction and online data selection~\cite{terashi2021event, Ngairangbam:2021yma, Maier:2025ppr, 1p1q, mott2017solving, Blance:2020nhl, Guan_2021, delgado2022quantumcomputingdataanalysis}. In this work, we study the robustness of quantum machine-learning algorithms in the context of online event selection at collider experiments. Collider trigger systems provide a particularly stringent testbed for robust machine learning. Trigger decisions are made from compact, low-level event representations whose dimensionality is comparable to the number of inputs that near-term quantum devices can process. Compact learned or physics-motivated event representations provide a tractable setting in which parameterised quantum models can be studied without the extreme dimensional reductions required by many offline collider analyses. At the same time, trigger models must operate under strict latency and resource constraints, making compact models and shallow inference circuits operationally relevant rather than merely convenient. Real-time implementations of autoencoder-based anomaly detection have been demonstrated on field-programmable gate arrays~\cite{Govorkova:2021utb}, and hardware acceleration has recently been investigated for QAEs in the same collider-trigger setting~\cite{Ge:2026eya}.

Trigger systems are also unusually exposed to distribution shift. Changing pile-up conditions, evolving detector calibrations, radiation damage and hardware ageing can continuously distort the input distribution during data taking. A trigger cannot generally be retrained each time the detector response changes, and even a small change in the background acceptance can be operationally significant when decisions are made at very high event rates. The trigger environment therefore represents a deliberately demanding, near-worst-case benchmark for domain shift. Demonstrating robustness in this setting constitutes a stringent test of the inductive biases of the quantum models, whilst also addressing an application of independent interest to collider physics.

To reproduce this environment, we use the ADC2021~\cite{adc2021} and supersymmetry (SUSY)~\cite{susy_279} benchmarks. ADC2021 represents events using low-level reconstructed object-level observables, whereas the SUSY dataset combines low-level observables with derived kinematic quantities. Feature-level perturbations are applied before the embedding stage, so that the resulting distribution shift propagates through the complete model pipeline. In the unsupervised study, QAEs with local and non-local entangling structures are compared with classical and variational autoencoders trained on samples drawn from the same background distribution. In the supervised study, data-reuploading classifiers of increasing depth are compared with linear and multilayer-perceptron (MLP) classifiers trained on the same labelled samples. For both settings, we evaluate reference discrimination performance together with the displacement of the model outputs and the retention of global discrimination performance under smearing.

Across both learning settings, the quantum models achieve competitive reference performance whilst exhibiting enhanced stability under feature-level perturbations. The QAEs retain strong anomaly-detection performance on clean inputs and show smaller shifts in their anomaly-score distributions than the classical baselines under smearing. In the supervised setting, increasing the number of data-reuploading layers improves the discrimination power of the quantum classifier, whilst its output remains substantially less sensitive to input perturbations than that of the non-linear classical model. Taken together, these results place the quantum models in a favourable region of the robustness performance trade-off and suggest that the constrained structure of parameterised quantum circuits can provide a useful inductive bias for robustness across both unsupervised and supervised learning settings.

The paper is organised as follows. In Section~\ref{sec:unsupervised} we introduce the unsupervised anomaly-detection study, including the QAE architecture, its reference performance relative to classical baselines, and its robustness under feature-level perturbations. In Section~\ref{sec:supervised} we introduce the supervised data-reuploading classifier, study its dependence on circuit depth, and compare its performance and robustness with classical classifiers. We summarise our conclusions in Section~\ref{sec:conclusions}.

%%%%%%%%%%%%%%%%%%%%%%%%%%%%%%%%%%%%%%%%%%%%%%%%%%%%%%%%%%%%%
%%%%%%%%%%%%%%%%%%%%%%%%%%%%%%%%%%%%%%%%%%%%%%%%%%%%%%%%%%%%%
%%%%%%%%%%%%%%%%%%%%%%%%%%%%%%%%%%%%%%%%%%%%%%%%%%%%%%%%%%%%%

\section{Unsupervised Quantum Anomaly Detection}\label{sec:unsupervised}

We first investigate robustness in an unsupervised setting, in which models are trained exclusively on reference background events. We compare reconstruction-based classical autoencoders (AEs) with quantum autoencoders (QAEs) whose anomaly scores are obtained from the fidelity of a learned quantum compression. We first introduce the quantum model and its training objective, before comparing its reference anomaly-detection performance and response to feature-level perturbations with the classical baselines.

%%%%%%%%%%%%%%%%%%%%%%%%%%%%%%%%%%%%%%%%%%%%%%%%%%%%%%%%%%%%%
%%%%%%%%%%%%%%%%%%%%%%%%%%%%%%%%%%%%%%%%%%%%%%%%%%%%%%%%%%%%%

\subsection{Quantum autoencoder}\label{sec:method_qae}

\begin{figure*}[t]
    \centering

    \captionsetup{
        font=small,
        skip=5pt
    }
    \captionsetup[subfigure]{
        font=small,
        labelfont=bf,
        justification=centering,
        singlelinecheck=true,
        skip=3pt
    }

    % Main architecture
    \begin{subfigure}[t]{0.95\textwidth}
        \centering
        \includegraphics[
            width=\linewidth
        ]{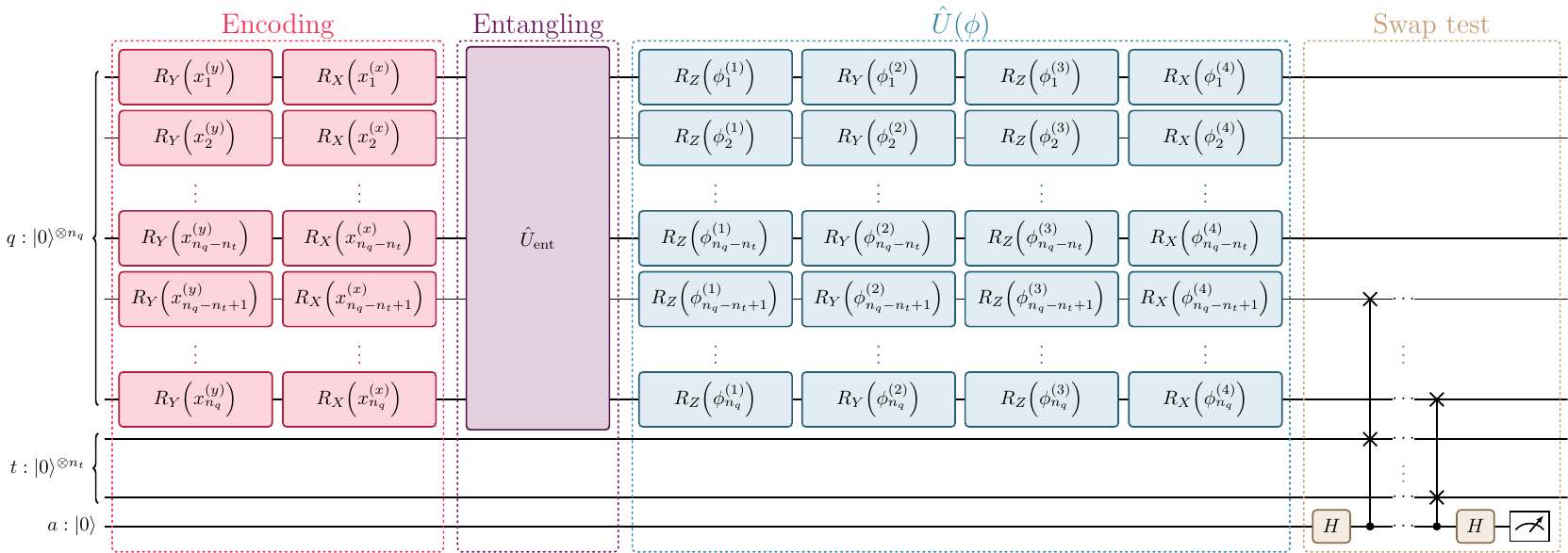}
        \label{fig:qae-circuit}
    \end{subfigure}

    \vspace{0.2em}

    \begin{subfigure}[b]{\textwidth}
        \centering
        \includegraphics[
            width=\linewidth
        ]{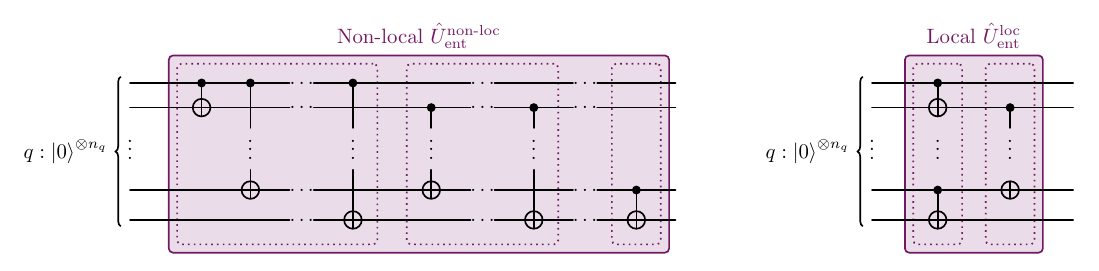}
        \label{fig:uent-brickwork}
    \end{subfigure}

    \caption{
        Quantum autoencoder architecture and entangling ans\"atze. The circuit (top) consists of angle encoding of the input features using \(R_Y\) and \(R_X\) rotations, an entangling block \(U_{\mathrm{ent}}\), and a parameterised unitary \(\hat{U}(\phi)\) with trainable parameters \(\phi\). A \textsc{swap} test evaluates the fidelity between the state of the \(n_t\) trash qubits and the reference vacuum state \(\lvert 0\rangle^{\otimes n_t}\). The non-local and local implementations of \(U_{\mathrm{ent}}\), employing long- and short-range \textsc{cnot} connectivity, respectively, are shown below.
    }
    \label{fig:qae-architecture}
\end{figure*}

A QAE~\cite{Romero_2017} learns a parameterised unitary that compresses a family of input states into a smaller latent quantum register. In the anomaly-detection setting considered here, the model is trained exclusively on background events, such that inputs that are poorly represented by the learned background subspace are compressed less efficiently and assigned larger anomaly scores. Classical feature vectors are first encoded into quantum states, after which the compression map is implemented by a parameterised quantum circuit. The resulting fidelity-based anomaly score has a different functional form from the reconstruction losses used by the AE baselines and therefore induces a distinct inductive bias. We first introduce the construction for a general input and then specify the circuit architectures and training configuration used in this work.

A variational quantum circuit consists of three stages: state-preparation, parameterised unitary evolution and measurement~\cite{Benedetti_2019}. For a classical input containing $2n_q$ features, we group the features into pairs $\mathbf{x}_i=(x_i^{(x)},x_i^{(y)})$, with one pair assigned to each of the $n_q$ data qubits. Each feature is rescaled using constants determined from the background training sample and held fixed for all subsequent data. The rescaled input, $\widetilde{\mathbf{x}}_i$, is then loaded into the quantum register using angle encoding,
\begin{equation}\label{eq:qae_encoding}
    \ket{x} = \bigotimes_{i=1}^{n_q} R_x\!\left(\widetilde{x}_i^{(x)}\right) R_y\!\left(\widetilde{x}_i^{(y)}\right) \ket{0}_i~,
\end{equation}
where $\ket{x}$ denotes the encoded input state, and $\ket{0}$ is the vacuum state, represented by the zero state of the single-qubit computational basis $\{\ket{0}, \ket{1}\}$. The ordering in Equation~\eqref{eq:qae_encoding} corresponds to first applying the $R_y(\widetilde{x}_i^{(y)})$ rotation and then $R_x(\widetilde{x}_i^{(x)})$, as shown in Figure~\ref{fig:qae-architecture}. This encoding maps two classical features to each qubit, such that an input of dimension $2n_q$ is represented on the device using $n_q$ qubits. 

Angle encoding requires only single-qubit rotations for state preparation and therefore avoids the additional circuit depth associated with amplitude encoding. The encoded state is, however, initially separable across the qubit register. Correlations between the input features must consequently be learned by the subsequent variational circuit. The features are rescaled to a bounded interval before encoding, limiting the angular range supplied to the rotation gates. Trainable embedding parameters additionally allow the data map to adapt during optimisation. Alternative encodings have also been shown to give competitive anomaly-detection performance for collider data~\cite{1p1q,duffy2024}.

The encoded state is evolved by a parameterised unitary $\hat{U}_{\rm QAE}(\boldsymbol{\phi})$ acting on the full register of data qubits,
\begin{equation}\label{eq:qae_evolve}
    \ket{\psi(x;\boldsymbol{\phi})} = \hat{U}_{\rm QAE}(\boldsymbol{\phi})\ket{x}~.
\end{equation}
As the state-preparation circuit consists of independent single-qubit rotations, the variational circuit must include entangling operations in order to capture correlations between features encoded on different qubits. As shown in Figure~\ref{fig:qae-architecture}, we decompose the QAE unitary as
\begin{equation}\label{eq:qae_parameterised_unitary}
    \hat{U}_{\rm QAE}(\boldsymbol{\phi}) = \hat{U}(\boldsymbol{\phi})\hat{U}_{\rm ent}~,
\end{equation}
where $\hat{U}_{\rm ent}$ is an entangling layer constructed from a sequence of \textsc{cnot} gates, and $\hat{U}(\boldsymbol{\phi})$ is a trainable layer of single-qubit rotations applied across the data register, such that
\begin{equation}
\hat{U}(\boldsymbol{\phi}) = \bigotimes_{i=1}^{n_q} R_X\!\left(\phi_i^{(4)}\right) R_Z\!\left(\phi_i^{(3)}\right) R_Y\!\left(\phi_i^{(2)}\right) R_Z\!\left(\phi_i^{(1)}\right)~.
\end{equation}
Here, $\phi_i^{(n)}$ denotes the trainable parameter associated with the $n$-th rotation acting on qubit $i$. With the operator ordering in Equation~\eqref{eq:qae_parameterised_unitary}, the entangling layer is applied before the trainable rotation layer.

The connectivity of $\hat{U}_{\rm ent}$ determines both the correlations that can be represented efficiently and the circuit resources required for its implementation. We consider two choices of entangling layer. The first is the non-local ans\"atz shown in the lower-left panel of Figure~\ref{fig:qae-architecture}. This architecture applies a \textsc{cnot} gate between every pair of data qubits,
\begin{equation}\label{eq:nonlocal}
\hat{U}_{\rm ent}^{\rm non-loc} = \prod^{n_q-1}_{i=1} \prod^{n_q}_{j=1+1} {\rm CNOT}_{i\rightarrow j}~.
\end{equation} 
The non-local ans\"atz therefore provides all-to-all entangling connectivity, however it requires ($n_q(n_q-1)/2$ \textsc{cnot} gates and all-to-all connectivity between non-neighbouring qubits. 

We consequently consider a second, more hardware-efficient architecture, referred to as the local ans\"atz. As shown in the lower-right panel of Figure~\ref{fig:qae-architecture}, this architecture restricts the entangling operations to nearest-neighbour qubits and arranges them in a ``brick-wall'' pattern
\begin{equation}
    \hat{U}_{\rm ent}^{\rm loc} = \left[ \prod_{i=1}^{\lfloor (n_q-1)/2\rfloor} \operatorname{CNOT}_{2i\rightarrow(2i+1)} \right] \left[ \prod_{i=1}^{\lfloor n_q/2\rfloor} \operatorname{CNOT}_{(2i-1)\rightarrow 2i} \right].
\end{equation}
The first sublayer acts on the disjoint pairs $((1,2),(3,4),\ldots)$, while the second acts on $((2,3),\allowbreak (4,5),\allowbreak \ldots)$. Gates within each sublayer can be applied in parallel, allowing all nearest-neighbour connections to be implemented using $(n_q-1)$ \textsc{cnot} gates across two entangling sublayers. Since both entangling layers consist entirely of fixed \textsc{cnot} gates, neither introduces additional trainable parameters. The two architectures therefore contain the same $4n_q$ variational parameters but differ in their connectivity, entangling-gate count and circuit depth. We treat them as distinct QAE architectures, rather than as ideal and noisy implementations of a common circuit. Their comparison isolates the effect of entangling connectivity and circuit-resource requirements on reference anomaly-detection performance and robustness.

A unitary acting on the full data register preserves the dimension of the Hilbert space and therefore does not by itself constitute the compression required for an autoencoder. The information bottleneck is introduced by partitioning the $n_q$ output qubits of the $q$ register in Figure~\ref{fig:qae-architecture} into a latent subsystem $q_{\rm L}$, containing $n_q-n_t$ qubits, and a trash subsystem $q_{\rm T}$, containing $n_t$ qubits. The QAE is trained to map the family of background states approximately onto states of the form
\begin{equation}\label{eq:qae_compression}
    \hat{U}_{\rm QAE}(\boldsymbol{\phi})\ket{x}_{q} \simeq \ket{z(x)}_{q_{\rm L}} \otimes \ket{0}^{\otimes n_t}_{q_{\rm T}}~,
\end{equation}
where $\ket{z(x)}_{q_{\rm L}}$ contains the compressed representation of the input. The target state of the trash subsystem is therefore the computational vacuum
\begin{equation}\label{eq:qae_vacuum}
    \ket{0}_{q_{\rm T}} \equiv \ket{0}^{\otimes n_t}_{q_{\rm T}}.
\end{equation}
Successful compression requires the trash subsystem to be driven towards this reference state and disentangled from the latent subsystem. In the circuit of Figure~\ref{fig:qae-architecture}, its overlap with the vacuum is evaluated using an additional $n_t$-qubit reference register $t$, initialised in $\ket{0}^{\otimes n_t}_{t}$, together with the ancilla qubit $a$ used in the swap test.

For a given input $x$, the reduced state of the trash subsystem is obtained by tracing over the latent subsystem,
\begin{equation}\label{eq:qae_trash}
    \rho_{q_{\rm T}}(x;\boldsymbol{\phi}) = \mathrm{Tr}_{q_{\rm L}} \!\left[ \hat{U}_{\rm QAE}(\boldsymbol{\phi}) \ket{x}_{q}\!\bra{x} \hat{U}_{\rm QAE}^{\dagger}(\boldsymbol{\phi}) \right]~.
\end{equation}
The quality of the compression is quantified by the fidelity of this state with the vacuum in the computational basis,
\begin{equation}\label{eq:qae_fidelity}
    f_{\rm vac}(x;\boldsymbol{\phi}) = {}_{q_{\rm T}}\!\bra{0} \rho_{q_{\rm T}}(x;\boldsymbol{\phi}) \ket{0}_{q_{\rm T}}~.
\end{equation}
Since the reference state is pure, this fidelity is equal to unity when the trash subsystem is exactly in the vacuum state of the computational basis and decreases when information remains encoded outside of this state. Residual entanglement with the latent subsystem also produces a mixed reduced state and therefore lowers the vacuum fidelity. Training the autoencoder consequently amounts to driving $f_{\rm vac}(x;\boldsymbol{\phi})$ towards unity for the background training sample.

A practical advantage of this construction is that the compression quality depends only on the encoder output and the trash subsystem. Unlike an encode-decode autoencoder, no explicit decoder needs to be constructed or executed in order to evaluate the anomaly score~\cite{Ngairangbam:2021yma}. This avoids the additional circuit depth associated with a decoding unitary and removes the decoder as a further source of implementation error. In the circuit shown in Figure~\ref{fig:qae-architecture}, the fidelity in Equation~\eqref{eq:qae_fidelity} is evaluated by comparing $q_{\rm T}$ with the vacuum reference register $t$ using the swap test.
 
The QAE is trained in an unsupervised manner using background events only. Given a background training sample $D_{\rm bkg}$, the circuit parameters are chosen to maximise the mean vacuum fidelity of the trash subsystem, or equivalently to minimise the loss,
\begin{equation}\label{eq:qae_loss}
    \mathcal{L}_{\rm QAE}(\boldsymbol{\phi}) = 1 - \frac{1}{|{D}_{\rm bkg}|} \sum_{x\in{D}_{\rm bkg}} f_{\rm vac}(x;\boldsymbol{\phi})~.
\end{equation}
Denoting the optimised parameters by $\boldsymbol{\phi}^{\star}$, the per-event anomaly score is defined as
\begin{equation}\label{eq:qae_score}
    s_{\rm QAE}(x) = 1-f_{\rm vac}(x;\boldsymbol{\phi}^{\star})~.
\end{equation}
A circuit trained in this way learns a compression adapted to the background distribution. Background-like events that are well represented by the learned latent subspace leave the trash subsystem close to the vacuum state in the computational basis and therefore receive small anomaly scores. Events that are poorly represented by this subspace retain encoded information outside the vacuum state, or residual correlations between $q_{\rm L}$ and $q_{\rm T}$, and consequently receive larger scores. Since the QAE score is defined through the fidelity of a learned quantum compression rather than through a reconstruction error in the original feature space, its response to feature-level perturbations need not coincide with that of an AE trained on the same inputs. This comparison is presented in Section~\ref{sec:unsupervised_robustness}.
 
In the implementation used here, the $2n_q$ components of the shared learned embedding are rescaled to $[0,\pi]$ using constants determined from the background training sample. The data are then encoded using the successive $R_y$ and $R_x$ rotations of Equation~\eqref{eq:qae_encoding}.

%%%%%%%%%%%%%%%%%%%%%%%%%%%%%%%%%%%%%%%%%%%%%%%%%%%%%%%%%%%%%
%%%%%%%%%%%%%%%%%%%%%%%%%%%%%%%%%%%%%%%%%%%%%%%%%%%%%%%%%%%%%

\subsection{Model setup and anomaly-detection performance}\label{sec:qae:performance}

We evaluate the anomaly-detection models using the ADC2021 dataset~\cite{adc2021}. Each event is stored using a fixed set of 19 object slots: one missing-transverse-energy (MET) entry, the four highest-$p_T$ electrons, the four highest-$p_T$ muons and the ten highest-$p_T$ jets. Candidates within each particle category are ordered by decreasing $p_T$; when an event contains fewer electrons, muons or jets than the corresponding maximum, the unused slots are zero-padded. We use the three kinematic fields $(p_T,\eta,\phi)$ stored for each slot, giving an input dimension of $d_{\rm in}=19\times3=57$. Since optimisation of the QAEs requires repeated classical simulation of the underlying circuits, the original event representation $\mathbf{x}_{\rm raw}\in\mathbb{R}^{d_{\rm in}}$ is first mapped to an eight-dimensional learned feature space using an MLP,
\begin{equation}\label{eq:qae_mlp_embedding}
    \mathbf{x} = g_{\boldsymbol{\eta}}(\mathbf{x}_{\rm raw})~, \qquad \mathbf{x}\in\mathbb{R}^{8}~.
\end{equation}
For the $nq=4$ system studied here, each QAE has 16 trainable parameters. The embedding network is trained with an unsupervised contrastive loss on background events and its parameters $\boldsymbol{\eta}$ are subsequently frozen during autoencoder training. The embedding model is a lightweight MLP with  2,520 total parameters that condenses the 57-dimension physical input space down to 8 embedding dimensions. The training was performed over 10 epochs, batch size of 1024, and an AdamW optimiser with a  learning rate of 0.01. A cosine decay was used to schedule the learning rate over the training.  This method of first embedding the physical input features to a lower dimensional space and performing the anomaly detection in the embedding space reduces the complexity of the task for the downstream autoencoder and is used to reduce the overall model size of anomaly detection algorithms \cite{anomalypreservingneuralembeddings, CMS-DP-2025-061}. The resulting representation is shared by the quantum and classical autoencoders, ensuring that all models are compared using the same fixed input features.

The ADC2021 dataset consists of five different samples; a background sample made up of SM events, a neutral scalar boson A decaying to four leptons $A\to4\ell$, a neutral scalar boson decaying to two tau leptons $h^{0}\to\tau\tau $, a charged scalar boson decaying to a tau and a neutrino $h^{\pm}\to\tau\nu $, and a leptoquark decaying to a $b$ quark and $\tau$ lepton. The additional blackbox sample in the ADC2021 dataset was not used. The dataset was split randomly into a training and two testing samples with a 60, 20, 20 split. One test sample was used to evaluate model performance under reference conditions, while the other was reserved for the smearing study. The total number of samples in each of these splits is detailed in Table~\ref{tbl:samplesizes}. For training the embedding and anomaly detection algorithms only the background dataset was used with the same training events for both the embedding training and anomaly detection algorithm training.

\begin{table}[t!]
\centering
    \begin{tabular}{|l|l|l|l|l|}
\hline
Sample name & Total Number of samples & Training Set & Test Set & Smearing Test Set \\
\hline
Background & 4,000,000 & 2,400,000 & 800,000 & 800,000 \\
$A \to 4\ell$ & 55,969 & 33,581 & 11,193 & 11,193 \\
$h^0 \to \tau\tau$ & 691,283 & 414,769 & 138,256 & 138,256 \\
$h^\pm \to \tau\nu$ & 760,272 & 456,163 & 152,054 & 152,054 \\
$LQ \to b\tau$ & 340,544 & 204,326 & 68,108 & 68,108 \\
\hline
\end{tabular}
\caption{Numbers of events in the full, training, test and smearing-test samples for the ADC2021~\cite{adc2021} background and four benchmark signals.}
\label{tbl:samplesizes}
\end{table}

%%%  vs QAE output scores
\begin{figure}[!t]
    \centering
    \captionsetup[subfigure]{skip=1pt}
    \captionsetup{skip=3pt}

    \begin{subfigure}[t]{0.48\textwidth}
        \centering
        \includegraphics[
            width=\linewidth,
            height=0.28\textheight,
            keepaspectratio
        ]{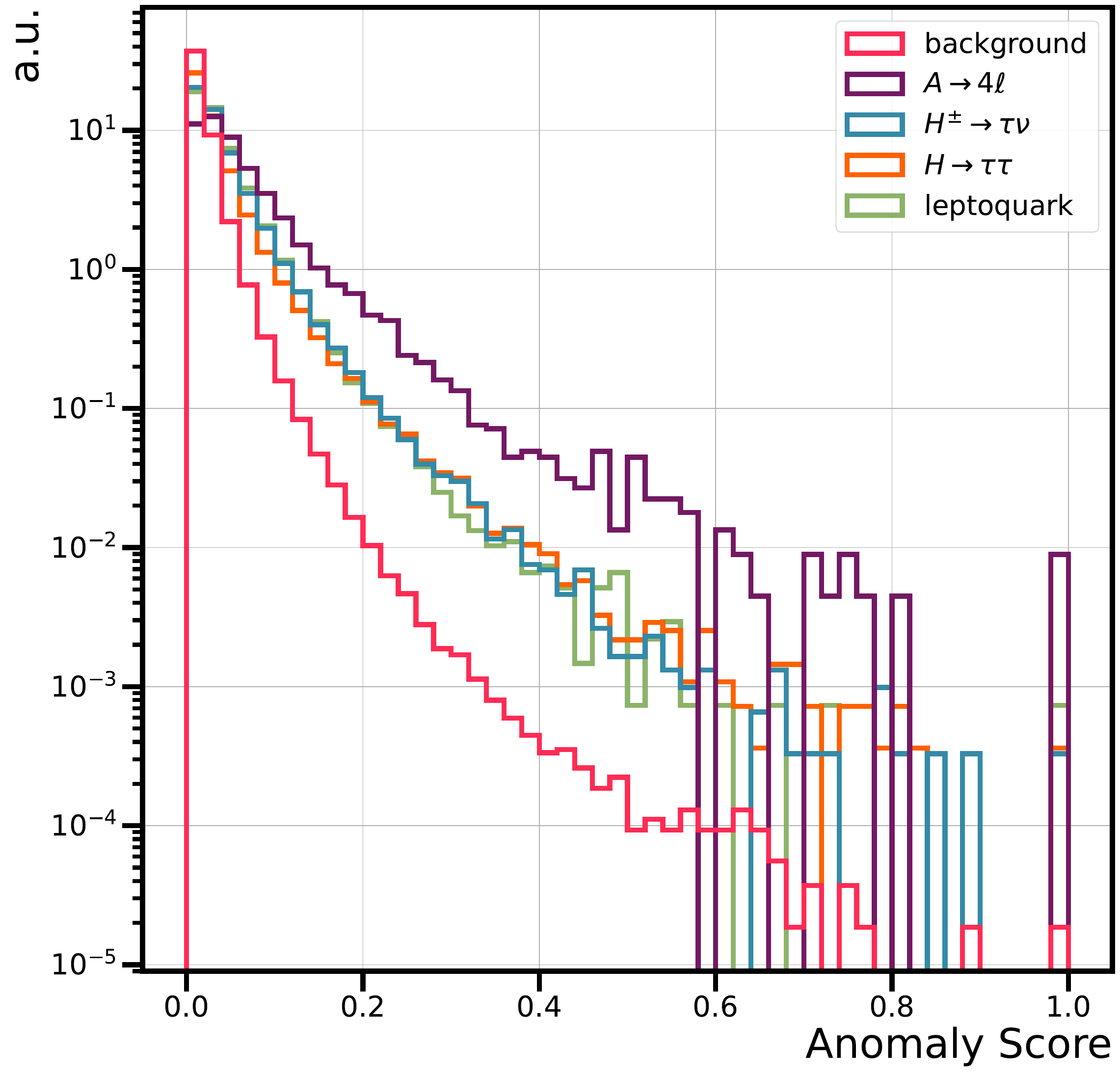}
        \caption{}
        \label{fig:CAE_outputscores}
    \end{subfigure}
    \hfill
    \begin{subfigure}[t]{0.48\textwidth}
        \centering
        \includegraphics[
            width=\linewidth,
            height=0.28\textheight,
            keepaspectratio
        ]{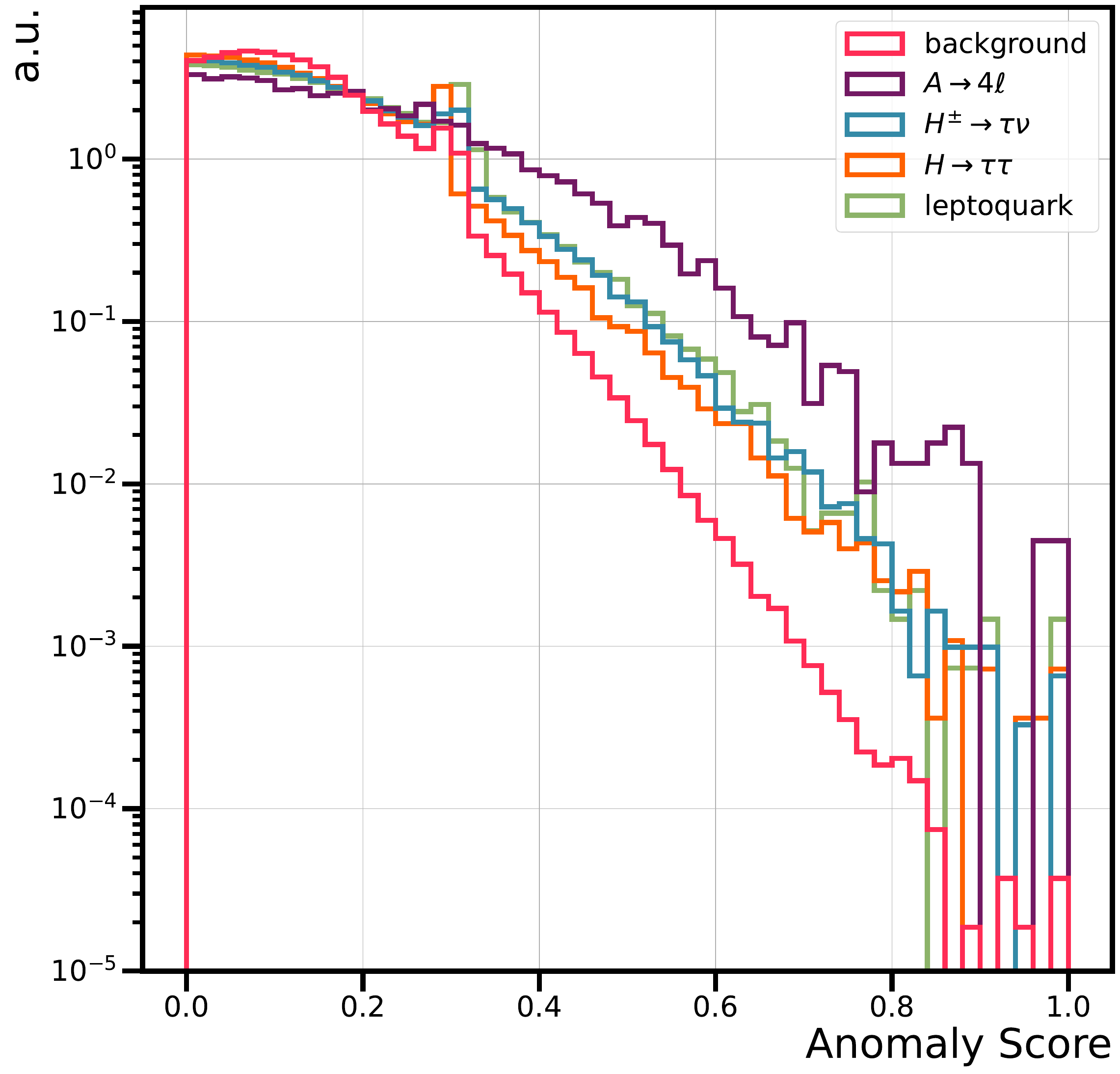}
        \caption{}
        \label{fig:VAE_outputscores}
    \end{subfigure}

    \begin{subfigure}[t]{0.48\textwidth}
        \centering
        \includegraphics[
            width=\linewidth,
            height=0.28\textheight,
            keepaspectratio
        ]{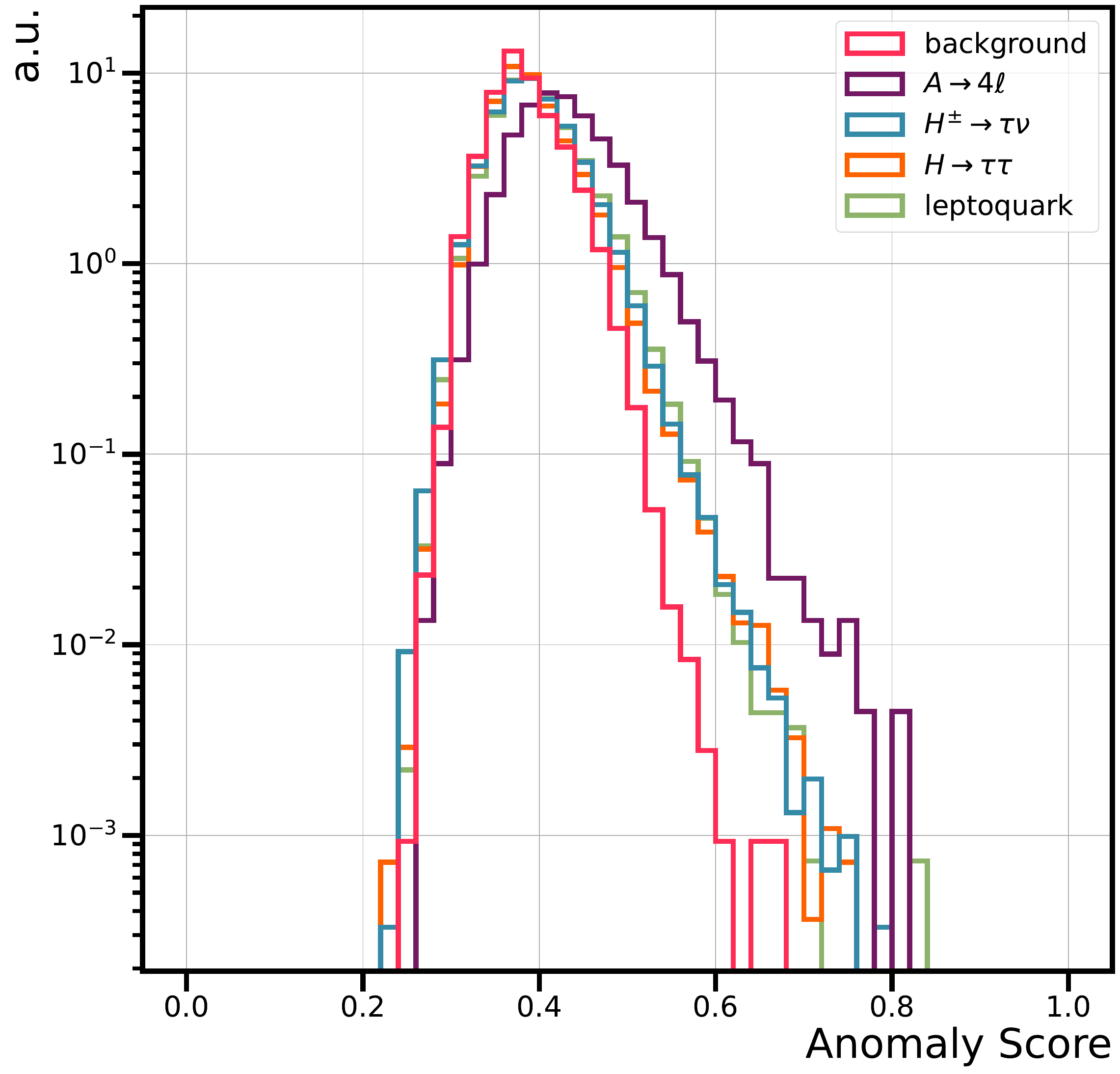}
        \caption{}
        \label{fig:QAE_outputscores}
    \end{subfigure}
    \hfill
    \begin{subfigure}[t]{0.48\textwidth}
        \centering
        \includegraphics[
            width=\linewidth,
            height=0.28\textheight,
            keepaspectratio
        ]{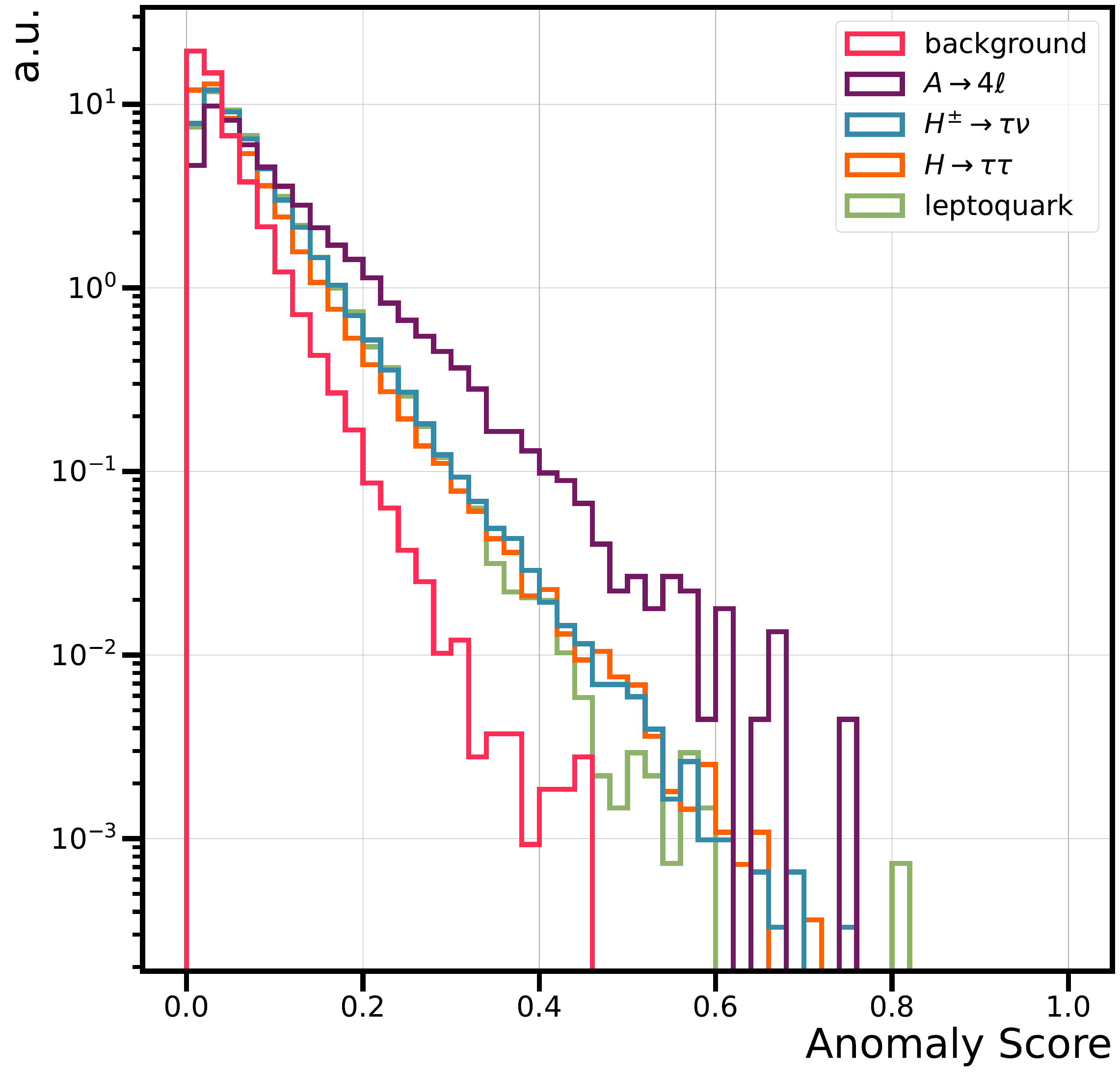}
        \caption{}
        \label{fig:HW_QAE_outputscores}
    \end{subfigure}

    \caption{Anomaly-score distributions obtained under reference
    detector conditions using the shared MLP embedding. Panels
    (a)--(d) show the results for the autoencoder,
    variational autoencoder, non-local quantum autoencoder and local
    quantum autoencoder, respectively. In each panel, the background
    distribution is compared with those of the four benchmark signals.}
    \label{fig:unsupervised_output_scores}
\end{figure}

For the QAEs, we set $n_q=4$, such that the eight components of $\mathbf{x}$ are encoded pairwise onto the four data qubits using Equation~\eqref{eq:qae_encoding}. We choose $n_t=3$ trash qubits, leaving a one-qubit latent subsystem. The circuit is therefore trained to compress the information required to represent the background states into a single latent qubit, whilst returning the three trash qubits to the computational vacuum $\ket{000}_{q_{\rm T}}$. For this configuration, the vacuum fidelity is
\begin{equation}
    f_{\rm vac}(x;\boldsymbol{\phi}) = {}_{q_{\rm T}}\!\bra{000} \rho_{q_{\rm T}}(x;\boldsymbol{\phi}) \ket{000}_{q_{\rm T}}~.
\end{equation}
The swap-test readout additionally requires a three-qubit reference register prepared in $\ket{000}_{t}$ and a single ancillary qubit. The complete simulated circuit therefore contains four data qubits, three reference qubits and one swap-test ancilla.

We compare the local and non-local entangling architectures introduced in Figure~\ref{fig:qae-architecture}. Both models use the same learned input representation, one-qubit latent subsystem and trainable single-qubit rotation layer, and differ only in the connectivity of $\hat{U}_{\rm ent}$. Each model contains $4n_q$ variational parameters. The circuit parameters are optimised using an Adam Optimiser with a learning rate of 0.15 on 500 embedded background events for 60 training epochs with no batching of the data in each epoch. The circuits are simulated using \textsc{PennyLane}~\cite{pennylane}, extracting exact expectation values from the simulation without finite-shot sampling or device noise. The learning rate was scheduled over training using a cosine annealing down to a minimum learning rate of $10^{-5}$. Circuit weights were initialised as a random array between 0 and 0.01, and a validation set of 500 samples was used to monitor the validation loss during training.  

As classical benchmarks, we use a standard autoencoder (AE) and a variational autoencoder (VAE), following their established use for model-independent new-physics searches at the LHC~\cite{Cerri:2018anq,Govorkova:2021utb}. Both models receive the same frozen eight-dimensional representation as the QAEs and are trained on events drawn from the same background distribution. The AE and VAE are trained using $10^6$ embedded background events. The AE is trained by minimising the mean-squared reconstruction error using an Adam optimiser with learning rate of $5\times10^{-6}$ over 70 epochs and a batch size of 1024. It utilised a learning rate reduction on plateau schedule and an early stopping both based on the validation loss which used a 10\% split of the training data. The VAE combines a reconstruction term with a Kullback-Leibler divergence that regularises the latent distribution. It was also trained using an Adam optimiser with a learning rate of $5\times10^{-6}$ and batch size of 1024. The scheduling and early stopping were the same as the AE model. The corresponding per-event anomaly scores at inference for the AE were defined as the mean-squared reconstruction error of the embedded events before and after the anomaly detection layers. The VAE uses the $l^2$ norm of the mean vector in the model's latent space ($\mu^2$) as the anomaly score for each embedded event.

%%% QAE and HW_QAE figure
\begin{figure}[t!]
    \centering
    \captionsetup[subfigure]{font=small,skip=2pt}
    \captionsetup{skip=4pt}
    \begin{subfigure}[t]{0.48\textwidth}
        \centering
        \includegraphics[
            width=\linewidth,
            height=0.28\textheight,
            keepaspectratio
        ]{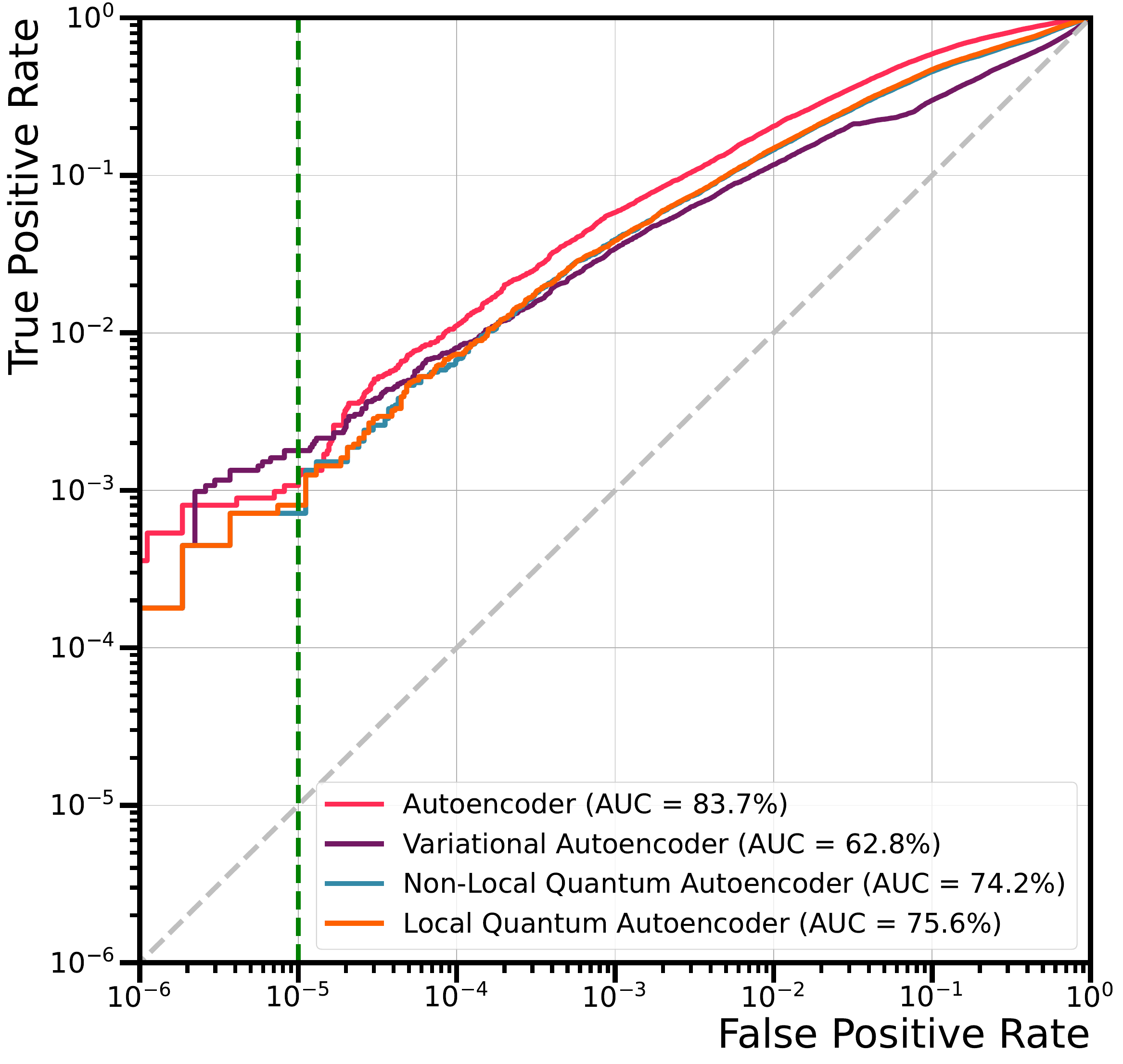}
        \caption{}
        \label{fig:CAE_ROC}
    \end{subfigure}
    \hfill
    \begin{subfigure}[t]{0.48\textwidth}
        \centering
        \includegraphics[
            width=\linewidth,
            height=0.28\textheight,
            keepaspectratio
        ]{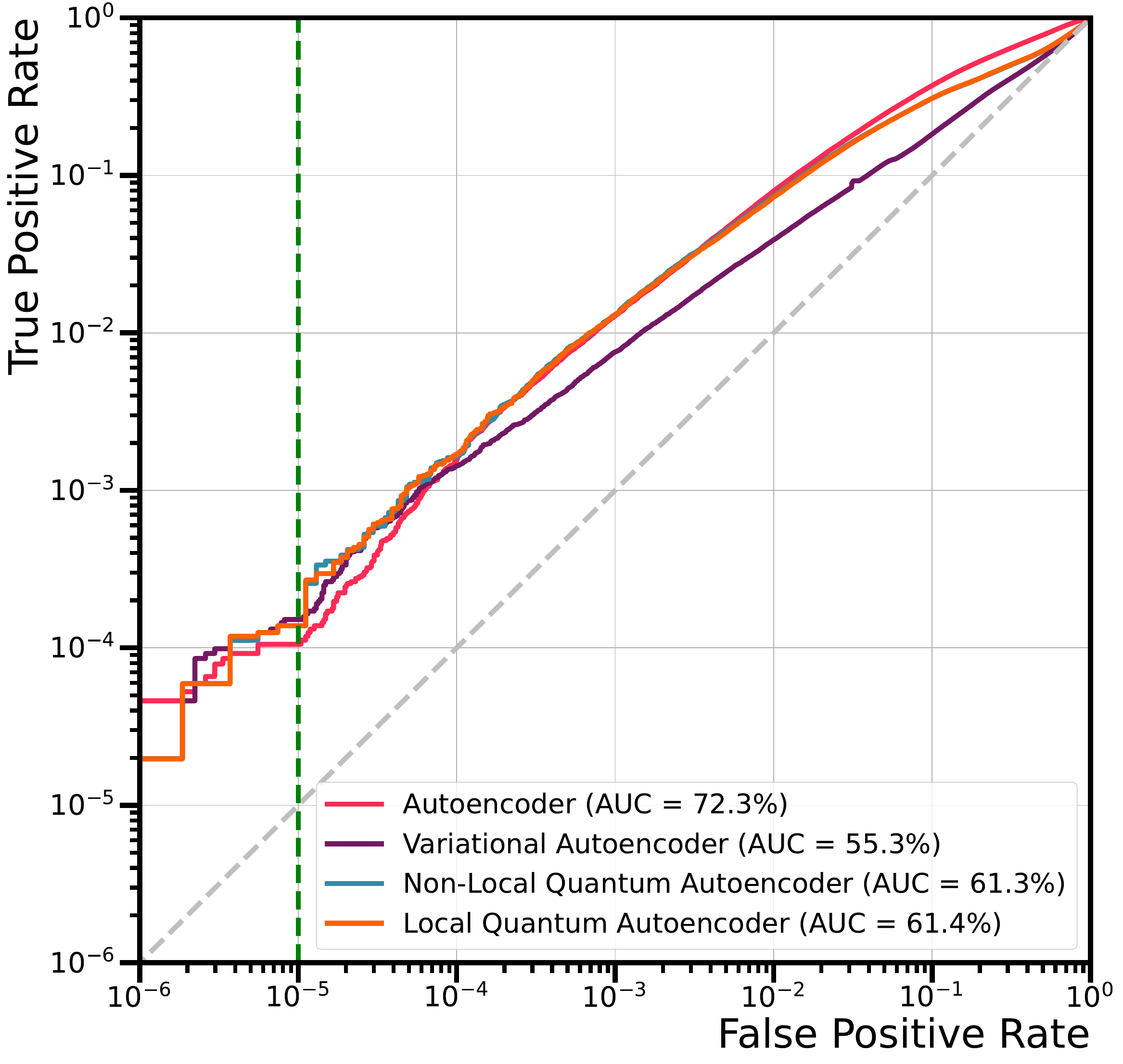}

        \caption{}
        \label{fig:VAE_ROC}
    \end{subfigure}

    \begin{subfigure}[t]{0.48\textwidth}
        \centering
        \includegraphics[
            width=\linewidth,
            height=0.28\textheight,
            keepaspectratio
        ]{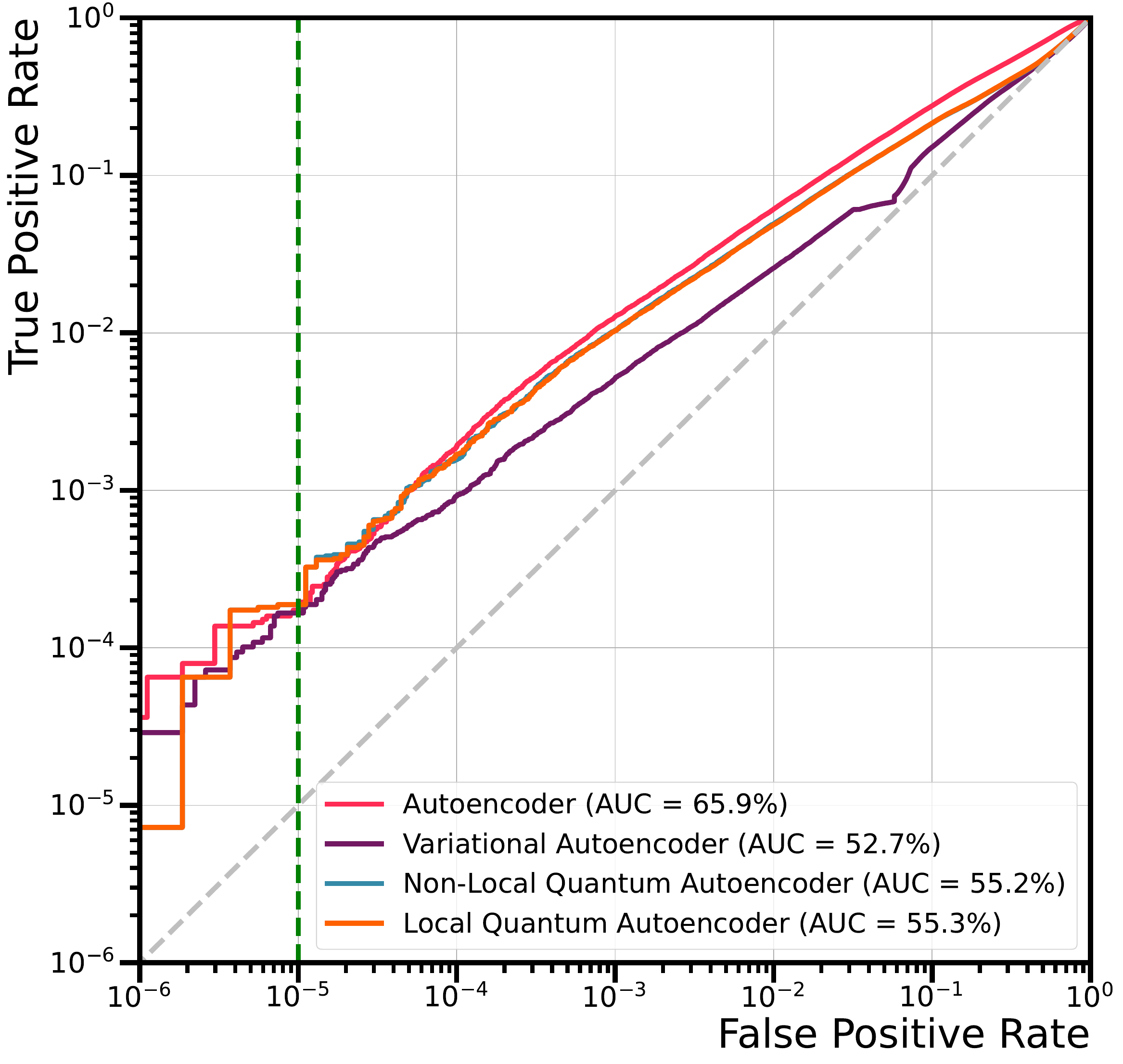}
        \caption{}
        \label{fig:QAE_ROC}
    \end{subfigure}
    \hfill
    \begin{subfigure}[t]{0.48\textwidth}
        \centering
        \includegraphics[
            width=\linewidth,
            height=0.28\textheight,
            keepaspectratio
        ]{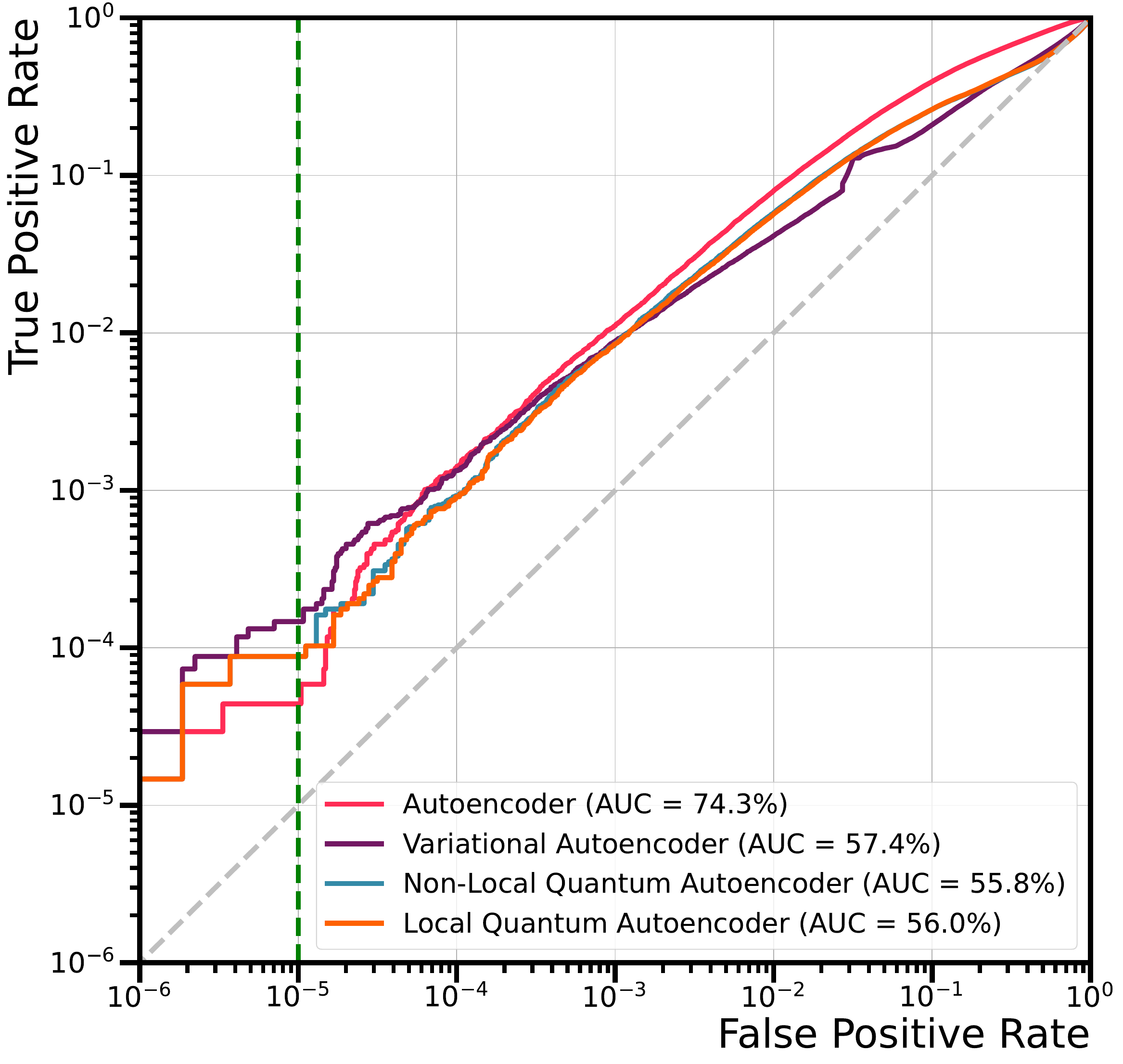}
        \caption{}
        \label{fig:HW_QAE_ROC}
    \end{subfigure}

    \caption{
        Anomaly-detection performance in the absence of detector-induced distribution shift using the shared MLP embedding. Panels (a)--(d) show the ROC curves for the $A\to4\ell$, $H^\pm\to\tau\nu$, $H\to\tau\tau$, and leptoquark benchmark signals, respectively. Each panel compares the classical autoencoder, variational autoencoder, non-local quantum autoencoder, and local quantum autoencoder, with the corresponding AUC values quoted in the legends. The vertical dashed line indicates a false-positive rate of $10^{-5}$, representative of the low-background operating regime relevant for collider triggers.
        }
    \label{fig:unsupervised_ROC}
\end{figure}

We first compare the anomaly-detection performance of the models in the absence of detector-induced distribution shift. The anomaly-score distributions for the four models are shown in Figure~\ref{fig:unsupervised_output_scores}, with the corresponding ROC curves presented in Figure~\ref{fig:unsupervised_ROC}. The discrimination performance depends strongly on the benchmark signal. The $A\to4\ell$ signal is the most readily identified by all four models, whereas $H\to\tau\tau$ generally exhibits the greatest overlap with the background. All four models nevertheless learn non-trivial discrimination. In particular, each QAE contains only $16$ trainable circuit parameters, compared with $312$ trainable parameters in the AE, and is optimised using only $500$ embedded background events, rather than the $10^6$ events used to train the classical models. Under these reference conditions, the AE provides the strongest overall discrimination, with AUC values of $83.7\%$, $72.3\%$, $65.9\%$, and $74.3\%$ for $A\to4\ell$, $H^\pm\to\tau\nu$, $H\to\tau\tau$, and the leptoquark signal, respectively, but at the cost of a substantially larger model and training sample. These results establish the clean-data performance against which the response to detector-induced distribution shift is assessed below.

The QAEs give their strongest discrimination for $A\to4\ell$, for which the non-local and local architectures achieve AUC values of $74.2\%$ and $75.6\%$, respectively. For $H^\pm\to\tau\nu$, the corresponding values are $61.3\%$ and $61.4\%$, while for $H\to\tau\tau$ they are $55.2\%$ and $55.3\%$. The two architectures obtain AUC values of $55.8\%$ and $56.0\%$ for the leptoquark benchmark. The VAE gives AUC values between $52.7\%$ and $62.8\%$ and is the weakest-performing model for three of the four signals, although it marginally exceeds the two QAEs for the leptoquark benchmark. These trends are also visible in the anomaly-score distributions, with $A\to4\ell$ exhibiting the clearest displacement towards larger scores, whilst the remaining signals show greater overlap with the background.

The two quantum architectures exhibit closely comparable performance. Despite its more restricted entangling connectivity, the local QAE achieves a marginally larger AUC for each benchmark, outperforming the non-local QAE by $1.4$ percentage points for $A\to4\ell$ and by no more than $0.2$ percentage points for the remaining signals. Their ROC curves also remain close across most of the false-positive-rate range. These results indicate that the relevant correlations can be captured using only local entangling operations within the present setup, without requiring direct long-range connections. The local QAE therefore retains the discrimination performance of the more highly connected architecture while offering a circuit structure better suited to near-term quantum hardware which has limited connectivity between qubits.

The vertical dashed line in Figure~\ref{fig:unsupervised_ROC} indicates a background efficiency of $\epsilon_B=10^{-5}$, corresponding to the low-background regime relevant for collider triggers. Under the associated bandwidth constraint, performance is characterised by the signal efficiency at fixed background efficiency. At this operating point, the models achieve comparable signal efficiencies across the benchmark signals, despite the larger differences observed in their integrated AUC values. The corresponding ratio, $\epsilon_S/\epsilon_B\sim10^2$, indicates that the selection enhances the signal-to-background ratio by approximately two orders of magnitude. The larger AUC of the AE therefore does not yield a systematic improvement in the trigger-relevant high-score region.

It is worth emphasising that each QAE contains only $16$ trainable circuit parameters and is optimised using $500$ embedded background events, whereas the AE contains $312$ trainable parameters and the classical  autoencoders are trained using $10^6$ events. The unequal training-set sizes mean that these results should not be interpreted as a controlled comparison of sample efficiency; this dependence is examined separately in Appendix~\ref{app:training-size}. Nevertheless, within this compact and data-limited setting, the quantum circuits learn effective anomaly scores and outperform the VAE for three of the four benchmark signals, demonstrating that useful discrimination can be achieved with substantially fewer trainable parameters and training examples.

%%%%%%%%%%%%%%%%%%%%%%%%%%%%%%%%%%%%%%%%%%%%%%%%%%%%%%%%%%%%%
%%%%%%%%%%%%%%%%%%%%%%%%%%%%%%%%%%%%%%%%%%%%%%%%%%%%%%%%%%%%%

\subsection{Robustness under domain shift}\label{sec:unsupervised_robustness}

We next investigate the response of the anomaly-detection models to a controlled shift in the input distribution. In a collider experiment, changes in detector resolution, calibration and operating conditions can alter the reconstructed values of the observables supplied to a trigger model. We emulate such changes through a controlled feature-level smearing of the reconstructed collider $p_T$ observables. This does not attempt to reproduce a specific detector effect in full detail, but provides a systematic means of varying the discrepancy between the training and deployment distributions. For an event with reference representation $\mathbf{x}_{\rm raw}$, we denote the corresponding representation at smearing strength $\lambda$ by
\begin{equation}\label{eq:smearing_map}
    \mathbf{x}_{\rm raw}^{(\lambda)} = \mathcal{S}_{\lambda} \!\left(\mathbf{x}_{\rm raw}\right)~,
\end{equation}
where $\mathcal{S}_{\lambda}$ denotes the smearing prescription and $\lambda=0$ corresponds to the reference detector conditions. Specifically, each unmasked input feature $x_{{\rm raw},i}$ is transformed according to
\begin{equation}\label{eq:smearing_2}
x_{{\rm raw},i}^{(\lambda)} = x_{{\rm raw},i}\left(1+\lambda\epsilon_i\right)~, 
\qquad
\epsilon_i\sim\mathcal{N}(0,1)~,
\qquad
\epsilon_i\in[-1,1]~,
\end{equation}
where the Gaussian perturbations are sampled independently for each $p_T$ feature of every event in the test sample. Smeared values are clipped at zero to prevent unphysical negative values. Entries corresponding to absent reconstructed objects, identified by $p_T=0$ in the reference event, are masked before smearing and remain unchanged. This prevents the smearing procedure from introducing artificial objects into the event.

The smeared observables are subsequently passed through the same frozen embedding network $g_{\boldsymbol{\eta}}$ from Equation~\eqref{eq:qae_mlp_embedding}, using the feature-scaling constants determined from the reference training sample. The embedding network and autoencoder parameters are held fixed throughout. The distribution shift therefore propagates through the complete inference pipeline without retraining or recalibration.

%%% Smear AEvs QAE vs HW_QAE
\begin{figure}[t]
    \centering

    \begin{subfigure}[t]{0.48\textwidth}
        \centering
        \includegraphics[height=0.23\textheight]
        {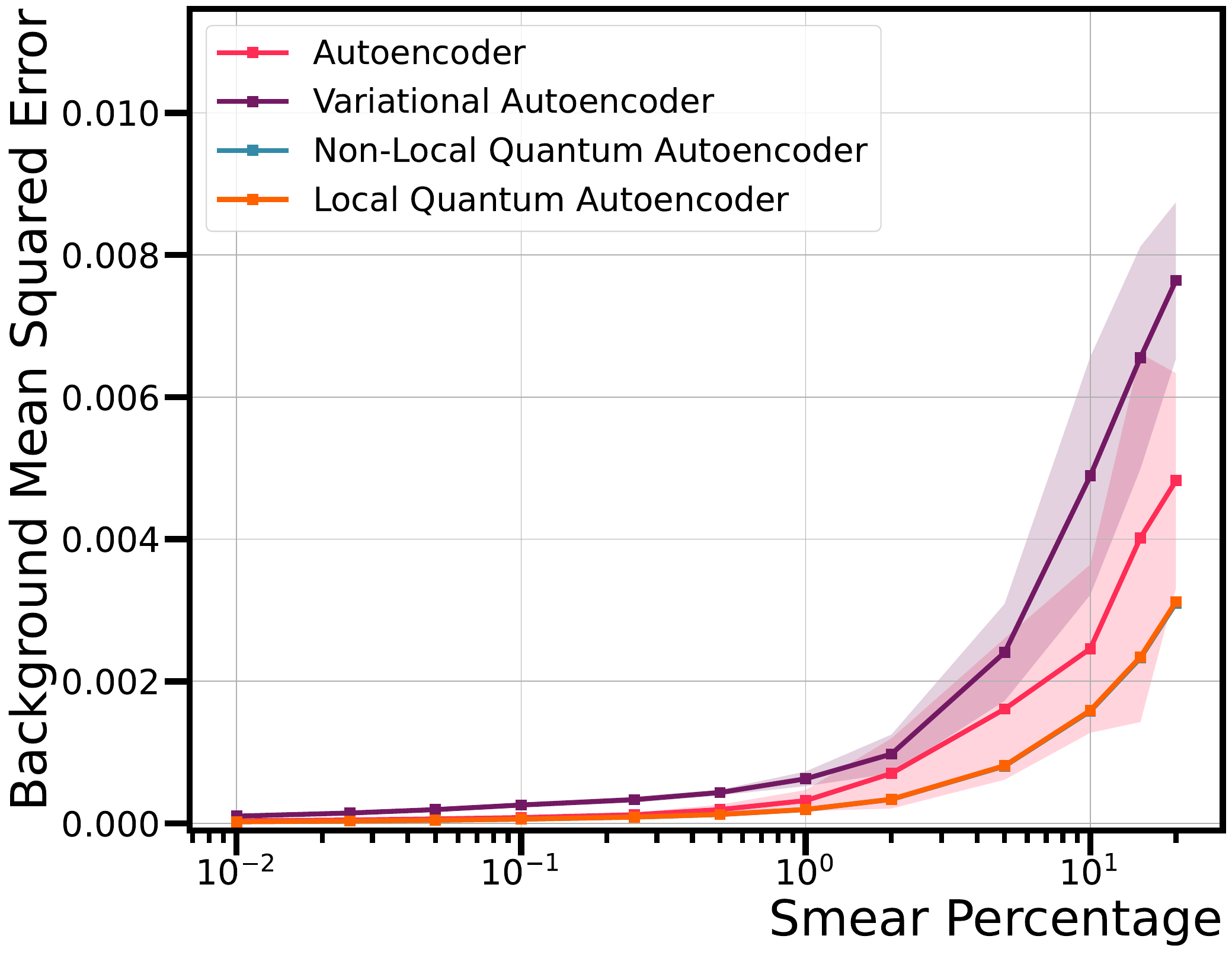}
        \label{fig:background_wd_smear}
    \end{subfigure}
    \hfill
    \begin{subfigure}[t]{0.48\textwidth}
        \centering
        \includegraphics[height=0.23\textheight]
        {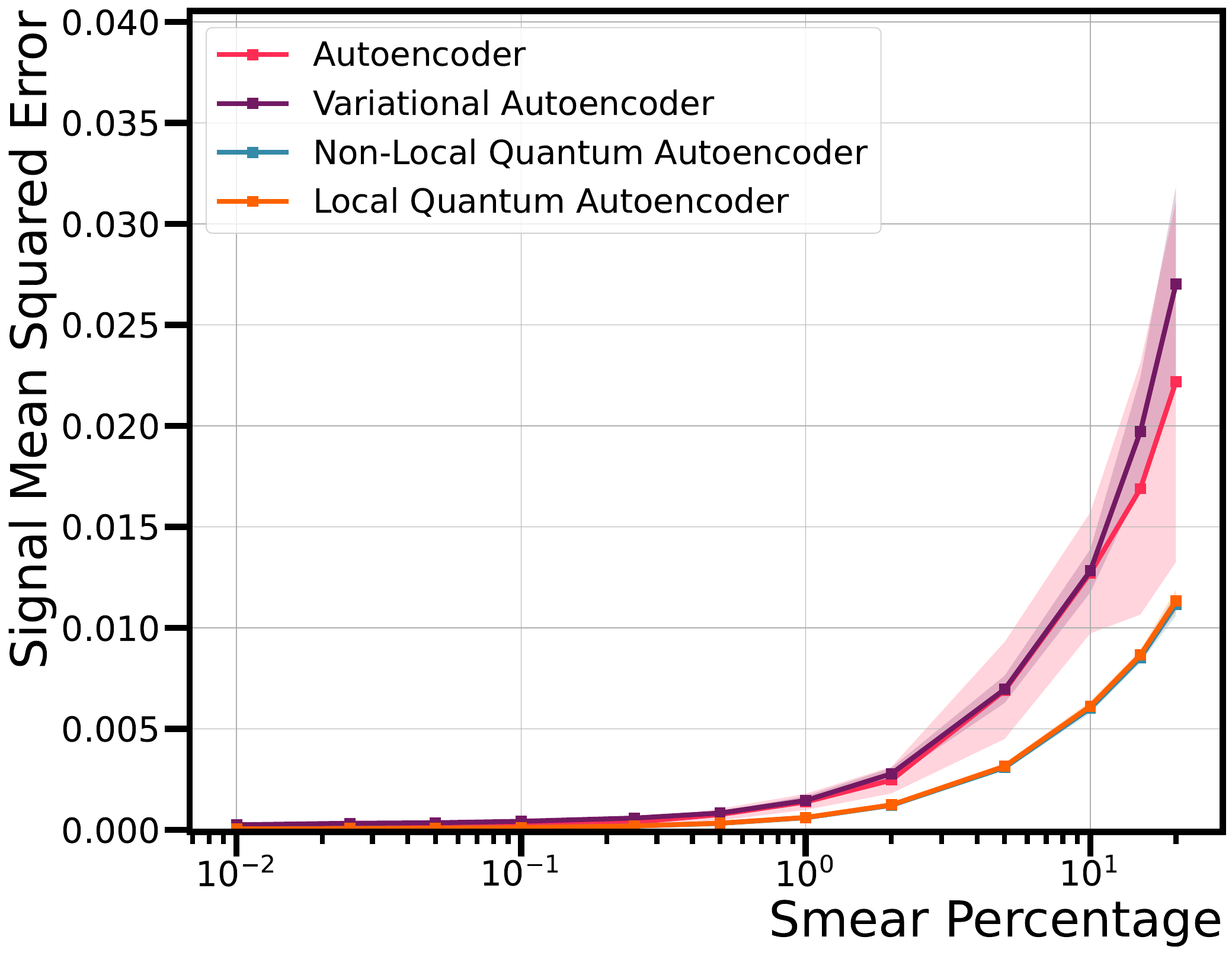}
        \label{fig:signal_wd_smear}
    \end{subfigure}

    \caption{Mean-squared deviation of the anomaly score from its reference value as a function of the detector-smearing strength. Results are shown for background events in the left panel and $A\rightarrow4\ell$ events in the right panel, comparing the classical autoencoder, variational autoencoder, non-local quantum autoencoder and local quantum autoencoder. The model parameters, learned embedding and feature-scaling constants are held fixed throughout. The shaded bands indicate the standard deviation across ten independent slices of the test set with input smearing.}
    \label{fig:mse_smearing}
\end{figure}

For each model $m$, we quantify the change in its response by comparing the anomaly score assigned to the smeared event with that assigned to the corresponding reference event. For a sample ${D}_{c}$ belonging to class $c\in\{\mathrm{bkg},\mathrm{sig}\}$, we define the mean-squared anomaly-score deviation as
\begin{equation}\label{eq:score_mse}
    \Delta_{m,c}^{2}(\lambda) = \frac{1}{|{D}_{c}|} \sum_{x\in\mathcal{D}_{c}} \left[ s_m\!\left(x^{(\lambda)}\right) - s_m(x) \right]^2~,
\end{equation}
where $s_m$ denotes the anomaly score of model $m$. Smaller values of $\Delta_{m,c}^{2}$ indicate that the model output is less sensitive to the applied perturbation. The smearing is repeated ten times, and the curves and shaded bands show the mean and standard deviation across these independent realisations.

Figure~\ref{fig:mse_smearing} shows the resulting anomaly-score deviation for the background and signal samples. For all four models, the deviation increases with the smearing strength, reflecting the growing displacement of the inputs from the reference distribution. At small smearing strengths, the responses remain close to their reference values and the differences between the models are limited. The separation becomes more pronounced once the applied perturbation is increased.

For the background sample, the classical models exhibit the largest score displacements at strong smearing. The VAE gives the largest mean-squared deviation, while the AE also becomes increasingly sensitive at the highest smearing strengths. By contrast, the local and non-local QAEs remain close to one another and exhibit substantially smaller changes in their anomaly scores. For the signal sample, the local QAE gives the smallest deviation across the range shown, while the classical models again display the largest response shifts at strong smearing. These results indicate that the anomaly scores produced by the QAEs are less sensitive to the applied feature-level perturbations.

\begin{figure}[t]
    \centering
    \includegraphics[
        width=0.48\textwidth,
        keepaspectratio
    ]{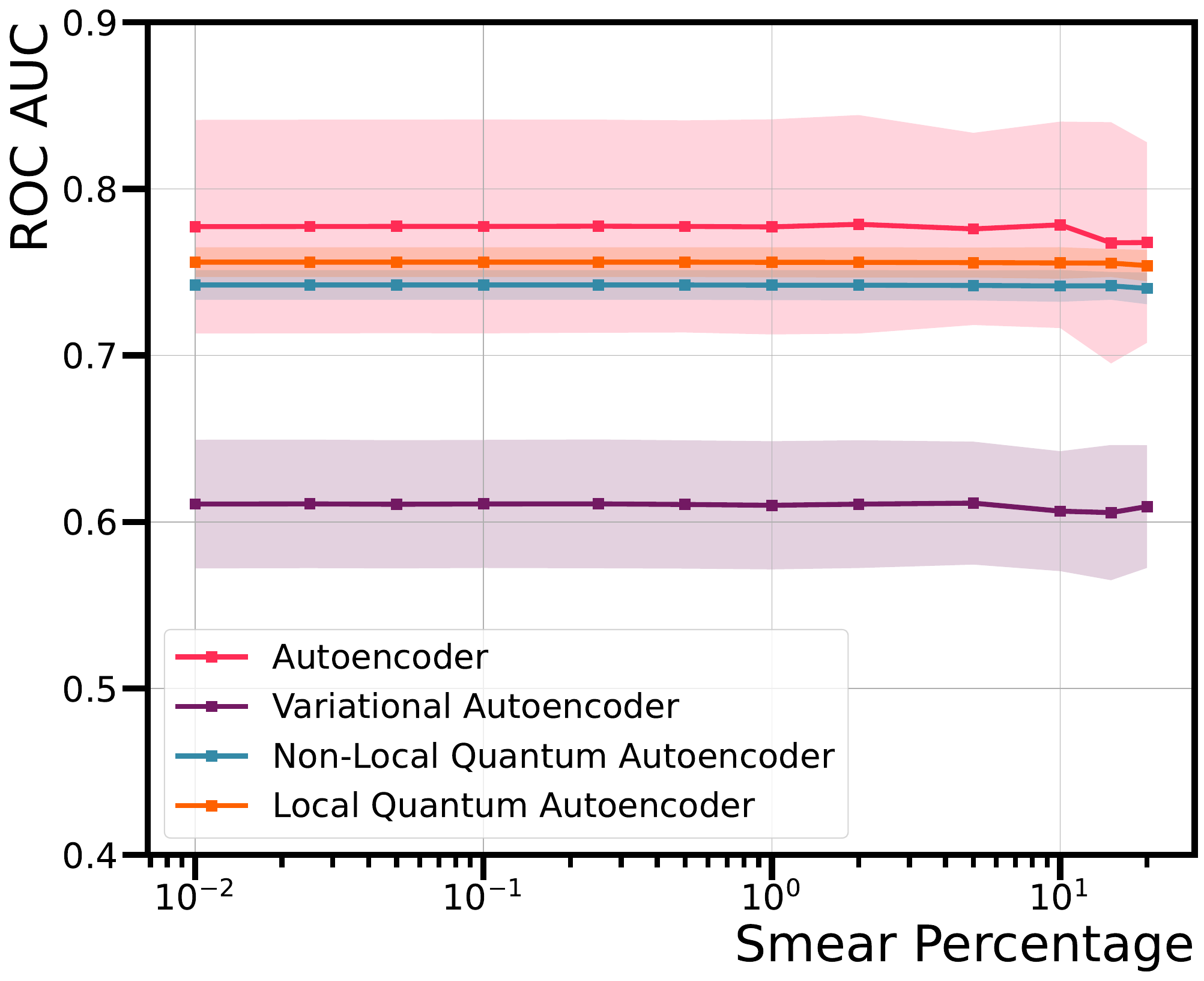}

    \caption{ROC AUC as a function of the detector-smearing strength for the $A\rightarrow4\ell$ sample. Results are shown for the classical autoencoder, variational autoencoder, non-local quantum autoencoder and local quantum autoencoder. The model parameters, learned embedding and feature-scaling constants are held fixed throughout. The shaded bands indicate the standard deviation across ten independent slices of the test set with input smearing.}
    \label{fig:auc_smearing}
\end{figure}

The mean-squared score deviation provides an event-level measure of the stability of the model response, but does not by itself establish that signal-background discrimination is retained. We therefore also evaluate the ROC AUC as a function of the smearing strength. As shown in Figure~\ref{fig:auc_smearing}, the ROC AUC remains approximately constant over the full range of smearing strengths considered. The AE retains the largest AUC, while the two QAEs remain close to one another and the VAE gives the weakest discrimination. No systematic degradation is observed for the AE or either QAE, while the reduction in the VAE AUC at the largest smearing strength remains comparable to the variation across smearing realisations.

The stability of the AUC indicates that the relative ordering of signal and background events is largely preserved under the applied perturbations. This does not contradict the increasing anomaly-score deviations shown in Figure~\ref{fig:mse_smearing}. The AUC is insensitive to common monotonic transformations of the model output and can therefore remain stable even when the numerical anomaly scores undergo a significant displacement. The two diagnostics consequently probe different aspects of robustness: the mean-squared deviation measures the stability of the event-level model response, whereas the AUC measures the retention of global signal-background discrimination.

Taken together, the results show that all four models retain broadly stable discrimination performance under the detector-induced distribution shift, while the quantum autoencoders exhibit a substantially more stable numerical response than the classical baselines.

The comparison between the two quantum architectures shows that the local QAE retains both the discrimination performance and robustness of the more highly connected model. Under reference detector conditions, the architectures achieve closely comparable discrimination, with the local QAE obtaining a marginally larger AUC for each benchmark. Under smearing, the local QAE exhibits a comparable anomaly-score deviation for the background sample and a smaller deviation for the signal sample. These results indicate that robust anomaly detection can be achieved using only local entangling operations within the present setup, without requiring direct long-range connections. The local QAE therefore provides a hardware-efficient architecture whose connectivity requirements are better matched to near-term quantum devices.

%%%%%%%%%%%%%%%%%%%%%%%%%%%%%%%%%%%%%%%%%%%%%%%%%%%%%%%%%%%%%
%%%%%%%%%%%%%%%%%%%%%%%%%%%%%%%%%%%%%%%%%%%%%%%%%%%%%%%%%%%%%
%%%%%%%%%%%%%%%%%%%%%%%%%%%%%%%%%%%%%%%%%%%%%%%%%%%%%%%%%%%%%

\section{Supervised Quantum Classifier}\label{sec:supervised}

We next investigate whether the robustness observed for QAEs extends to supervised learning. We consider a parameterised quantum classifier based on data reuploading, trained to distinguish labelled signal and background events. Repeated data-encoding and trainable circuit layers provide a controlled means of increasing the expressivity of the quantum model. We first introduce the classifier architecture and examine its dependence on circuit depth, before comparing its classification performance and sensitivity to detector-induced distribution shift with linear and non-linear classical baselines.

%%%%%%%%%%%%%%%%%%%%%%%%%%%%%%%%%%%%%%%%%%%%%%%%%%%%%%%%%%%%%
%%%%%%%%%%%%%%%%%%%%%%%%%%%%%%%%%%%%%%%%%%%%%%%%%%%%%%%%%%%%%

\subsection{Quantum classifier with data reuploading}

\begin{figure*}[t!]
    \centering

    \captionsetup{
        font=small,
        skip=5pt
    }
    \captionsetup[subfigure]{
        font=small,
        labelfont=bf,
        justification=centering,
        singlelinecheck=true,
        skip=3pt
    }

    % Main architecture
    \begin{subfigure}[t]{0.95\textwidth}
        \centering
        \includegraphics[
            width=\linewidth
        ]{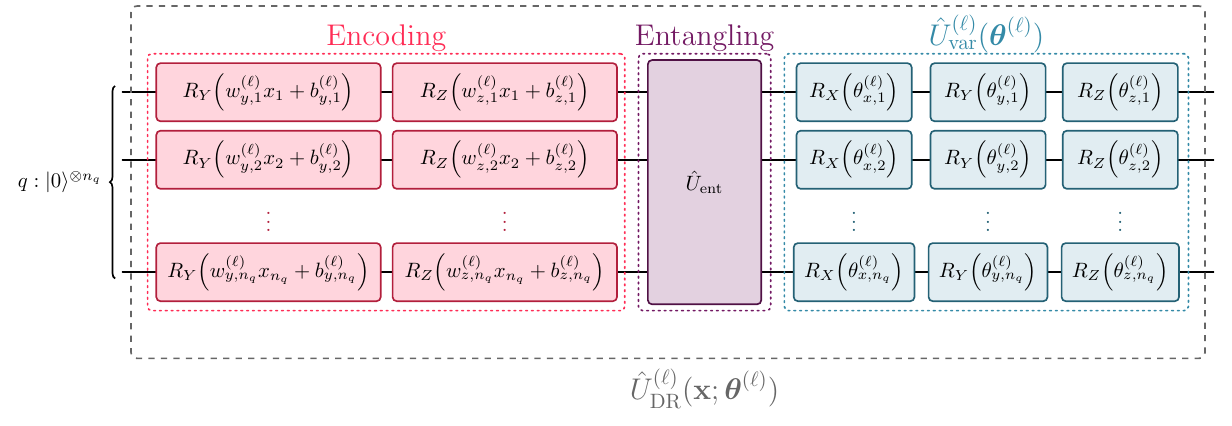}
        \label{fig:supervised-circuit}
    \end{subfigure}

    \vspace{-0.5cm}

    \begin{subfigure}[b]{0.95\textwidth}
        \centering
        \includegraphics[
            width=\linewidth
        ]{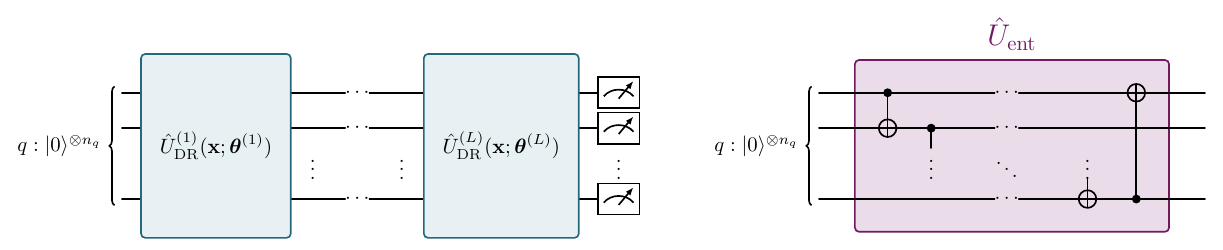}
        \label{fig:uent-supervised}
    \end{subfigure}

    \caption{
        Supervised quantum classifier architecture and entangling ans\"atz. The data-reuploading layer (top) consists of trainable angle encoding of the input features using weighted $R_Y$ and $R_Z$ rotations, an entangling block $\hat{U}_{\mathrm{ent}}$, and a parameterised unitary $\hat{U}_{\mathrm{var}}^{(\ell)}(\boldsymbol{\theta}^{(\ell)})$ comprising trainable $R_X$, $R_Y$, and $R_Z$ rotations. The complete classifier (bottom left) applies $L$ data-reuploading layers before measurement in the computational basis. The cyclic nearest-neighbour \textsc{cnot} implementation of $\hat{U}_{\mathrm{ent}}$ is shown at bottom right.
    }
    \label{fig:supervised-architecture}
\end{figure*}

A data-reuploading quantum classifier~\cite{PerezSalinas2020datareuploading,Schuld:2020enb} constructs a decision function by alternating data-dependent rotations with parameterised quantum operations. In contrast to a circuit in which the input is encoded only once at the beginning, data reuploading introduces the same classical features repeatedly throughout the circuit. Repeating the encoding can enlarge the accessible frequency spectrum of the quantum model and thereby generate increasingly rich non-linear dependence on the input without increasing the number of qubits~\cite{Schuld:2020enb}. In the supervised setting considered here, all circuit parameters are trained using labelled signal and background events.

Consider a real-valued classical feature vector of the form
\begin{equation}
    \mathbf{x} = \left(x_1,\ldots,x_{n_q}\right) \in\mathbb{R}^{n_q}~,
\end{equation}
where the feature $x_i$ is encoded on qubit $i$ of an $n_q$-qubit register, initialised in the state $\ket{0}^{\otimes n_q}$. As shown in Figure~\ref{fig:supervised-architecture}, the classifier is composed of $L$ data-reuploading layers, each containing a trainable data-encoding operation, a fixed entangling block and a set of parameterised single-qubit rotations. Within layer $\ell$, each feature determines two rotation angles through the affine maps
\begin{equation}\label{eq:supervised_angles}
    \alpha_{\mu,i}^{(\ell)}(x_i) = w_{\mu,i}^{(\ell)}x_i+b_{\mu,i}^{(\ell)}, \qquad \mu\in\{y,z\},
\end{equation}
where the weights $w_{\mu,i}^{(\ell)}$ control the scale of the feature-dependent contribution, while the biases $b_{\mu,i}^{(\ell)}$ provide a constant angular offset. The corresponding encoding operation is
\begin{equation}\label{eq:supervised_encoding}
    \hat E_{\ell}(\mathbf{x}) = \bigotimes_{i=1}^{n_q} R_Z\!\left[\alpha_{z,i}^{(\ell)}(x_i)\right] R_Y\!\left[\alpha_{y,i}^{(\ell)}(x_i)\right]~,
\end{equation}
where the ordering corresponds to first applying the $R_Y$ rotation and then the $R_Z$ rotation to embed the feature on each qubit. Since the weights and biases are independently trainable in each layer, successive reuploading layers can apply different encodings to the same classical input.

The encoding is followed by the cyclic nearest-neighbour entangling operation shown in Fig.~\ref{fig:supervised-architecture},
\begin{equation}\label{eq:supervised_entangler}
    \hat U_{\rm ent} = \operatorname{CNOT}_{n_q\rightarrow 1} \operatorname{CNOT}_{n_q-1\rightarrow n_q} \cdots \operatorname{CNOT}_{1\rightarrow 2}~.
\end{equation}
The rightmost gate acts first, such that adjacent qubits are coupled sequentially along the register before the final gate closes the chain by coupling the last qubit to the first. This construction introduces correlations between features assigned to different qubits using $n_q$ two-qubit gates. The same entangling operation is used in every data-reuploading layer.

A trainable rotation block is then applied independently to each qubit,
\begin{equation}\label{eq:supervised_variational}
    \hat U_{\rm var}^{(\ell)} (\boldsymbol{\theta}^{(l)}) = \bigotimes_{i=1}^{n_q} R_Z\!\left(\theta_{z,i}^{(\ell)}\right) R_Y\!\left(\theta_{y,i}^{(\ell)}\right) R_X\!\left(\theta_{x,i}^{(\ell)}\right)~.
\end{equation}
The complete operation implemented by layer $\ell$ is therefore
\begin{equation}\label{eq:data_reuploading_layer}
    \hat U_{\rm{DR}}^{(\ell)}(\mathbf{x}; \boldsymbol{\theta^{(\ell)}}) = \hat U_{\rm var}^{(\ell)}\, \hat U_{\rm ent}\, \hat E_{\ell}(\mathbf{x})~.
\end{equation}
We collectively denote all trainable encoding weights, biases and variational rotation angles by $\boldsymbol{\theta}$. It follows that the state prepared by an $L$-layer classifier is therefore
\begin{equation}\label{eq:data_reuploading_state}
    \ket{\psi_L(\mathbf{x};\boldsymbol{\theta})} = \hat U_{\rm{DR}}^{(L)}(\mathbf{x}; \boldsymbol{\theta}^{(L)}) \cdots \hat U_{\rm{DR}}^{(1)}(\mathbf{x}; \boldsymbol{\theta}^{(1)}) \ket{0}^{\otimes n_q}~.
\end{equation}
The same classical input is therefore embedded in every layer, while the intervening entangling and variational operations change the basis in which each subsequent encoding acts. Increasing $L$ modifies both the circuit depth and the functional dependence of its measured observables on the input.

The final state is measured in the computational basis. From the resulting bit strings, we construct a vector containing the local single-qubit expectation values and cyclic nearest-neighbour correlations,
\begin{equation}\label{eq:supervised_measurements}
    \mathbf{m}(\mathbf{x};\boldsymbol{\theta}) = \Big( \langle\hat Z_1\rangle,\ldots, \langle\hat Z_{n_q}\rangle,\langle\hat Z_1\hat Z_2\rangle,\ldots, \langle\hat Z_{n_q}\hat Z_1\rangle \Big) \in\mathbb{R}^{2n_q}~,
\end{equation}
Since these observables are diagonal in the computational basis, they can be estimated from the same set of measured bit strings. The single-qubit observables probe the response of individual qubits, while the two-qubit observables retain information about correlations generated by the entangling layers.

The measurement vector is mapped to a scalar classifier score using a single-neuron classical readout comprising an affine transformation followed by a sigmoid activation function,
\begin{equation}\label{eq:quantum_classifier_score}
    s_{\rm QC}(\mathbf{x};\boldsymbol{\theta},\boldsymbol{\gamma}) = \sigma\left( \mathbf{w}^{\mathsf{T}} \mathbf{m}(\mathbf{x};\boldsymbol{\theta})+b \right)~, \qquad \sigma(u) = \frac{1}{1+e^{-u}}~,
\end{equation}
where $\boldsymbol{\gamma}=\{\mathbf{w},b\}$ denotes the trainable readout parameters, with $\mathbf{w}\in\mathbb{R}^{2n_q}$ and $b\in\mathbb{R}$. Scores close to zero correspond to background-like events, while scores close to unity correspond to signal-like events. For the seven-qubit classifiers used here, the measurement vector contains 14 observables, such that the readout contains 14 weights and one bias. 

% \note{Define $h_{\boldsymbol{\gamma}}$ explicitly, including its activation function, and state its exact number of weights and biases. Include these readout parameters in the quoted total parameter count of every quantum classifier.}

For a labelled training sample ${D}_{\rm sup}=\{(\mathbf{x}_j,y_j)\}$, with $y_j=0$ for background and $y_j=1$ for signal, the circuit and readout parameters are trained by minimising the binary cross-entropy,
\begin{equation}\label{eq:quantum_classifier_loss}
    \mathcal{L}_{\rm QC} = -\frac{1}{|{D}_{\rm sup}|} \sum_{(\mathbf{x},y)\in\mathcal{D}_{\rm sup}} \left[ y\log s_{\rm QC}(\mathbf{x}) +(1-y)\log\!\left(1-s_{\rm QC}(\mathbf{x})\right) \right]~.
\end{equation}
The classifier scores used in the subsequent analysis are evaluated using the optimised circuit and readout parameters.

Each layer contains $4n_q$ trainable data-encoding parameters, two weights and two biases per qubit, together with $3n_q$ variational rotation angles. The circuit therefore contains $7Ln_q$ trainable parameters for an $L$-layer model, in addition to those entering the classical readout. The number of qubits remains fixed as $L$ is increased, while the circuit depth and parameter count grow linearly with the number of layers.

We compare classifiers containing between one and four layers. We refer to the complete family as data-reuploading classifiers, including the $L=1$ baseline in which the input is encoded only once. Increasing $L$ introduces additional independently trainable encoding and variational operations, thereby enriching the functional dependence of the classifier on the input while retaining the same input representation, qubit register, entangling connectivity and measurement prescription. This provides a controlled means of studying how circuit depth affects both classification performance under reference detector conditions and sensitivity to detector-induced distribution shift.

%%%%%%%%%%%%%%%%%%%%%%%%%%%%%%%%%%%%%%%%%%%%%%%%%%%%%%%%%%%%%
%%%%%%%%%%%%%%%%%%%%%%%%%%%%%%%%%%%%%%%%%%%%%%%%%%%%%%%%%%%%%

\subsection{Model setup and classification performance}\label{sec:supervised_performance}

We evaluate the supervised models using the SUSY dataset~\cite{susy_279}, in which signal and background events are described by reconstructed collider observables. We use $1,084,869$ background samples and $915,131$ signal samples for training and two sets of $813,652$ background samples and $686,348$ signal samples for testing with and without smearing these were split randomly using the \textsc{scikit-learn}~\cite{scikit-learn} train test split to form a 40, 30, 30 dataset split. For the present study, we follow the methodology in \cite{terashi2021event} and use a seven-dimensional subset of the full feature vector, comprising $p_T^{lep1}$, $\eta^{lep1}$, $p_T^{lep2}$, $\eta^{lep2}$, and the derived kinematic variables $E_T^{\rm miss}$, $M^T_R$, and $M^T_{\Delta}$. 

Although the dataset was not constructed specifically as a trigger benchmark, this compact feature representation is also relevant to real-time event selection. Modern first-level trigger systems have access to increasingly detailed information, including reconstructed tracks, calorimeter objects and global event observables~\cite{CMS-DP-2025-061}, while remaining subject to strict latency and hardware constraints. Classifiers based on a small number of informative event-level quantities therefore provide a natural setting for studying fast signal-background discrimination in the collider trigger environment.

We train quantum classifiers containing between one and four layers of data reuploading using an Adam optimiser with learning rate 0.15, scheduled with a cosine annealing over 120 epochs. We trained using a balanced sample of 1000 labelled signal and background events, 500 events of each type with no batching applied. All circuit and readout parameters are optimised using the binary cross-entropy loss function from Equation~\eqref{eq:quantum_classifier_loss}. The $L=1$ circuit provides a baseline in which the input is encoded only once, while the deeper models introduce additional independently trainable encoding, entangling and variational operations. All quantum calculations are performed using exact circuit simulation in \textsc{PennyLane}~\cite{pennylane}, without finite-shot sampling or device-noise effects. The initialisation of the circuit parameters was performed by creating a random array of values between 0 and 0.001.

For comparison, we train a linear classifier and a classical MLP using the same seven input features. The linear model consists of a single affine transformation and provides a baseline for discrimination using a linear decision boundary.  The MLP contains a 7, 6, 4, 1 node structure with 81 parameters. Both the linear and MLP models are trained with an Adam optimiser with an initial learning rate of 0.005 over 20 epochs and a batch size of 1024. They use a reduction of the learning rate on plateau scheduler and early stopping based on the validation loss that used a 10\% split of the training data. The classical models are trained on $10^6$ labelled events using a binary cross-entropy loss and are evaluated on the same test sample as the quantum classifiers.

\begin{figure}[t!]
    \centering
    \captionsetup[subfigure]{font=small,skip=2pt}
    \captionsetup{skip=4pt}
    \begin{subfigure}[t]{0.48\textwidth}
        \centering
        \includegraphics[
            width=\linewidth,
            height=0.28\textheight,
            keepaspectratio
        ]{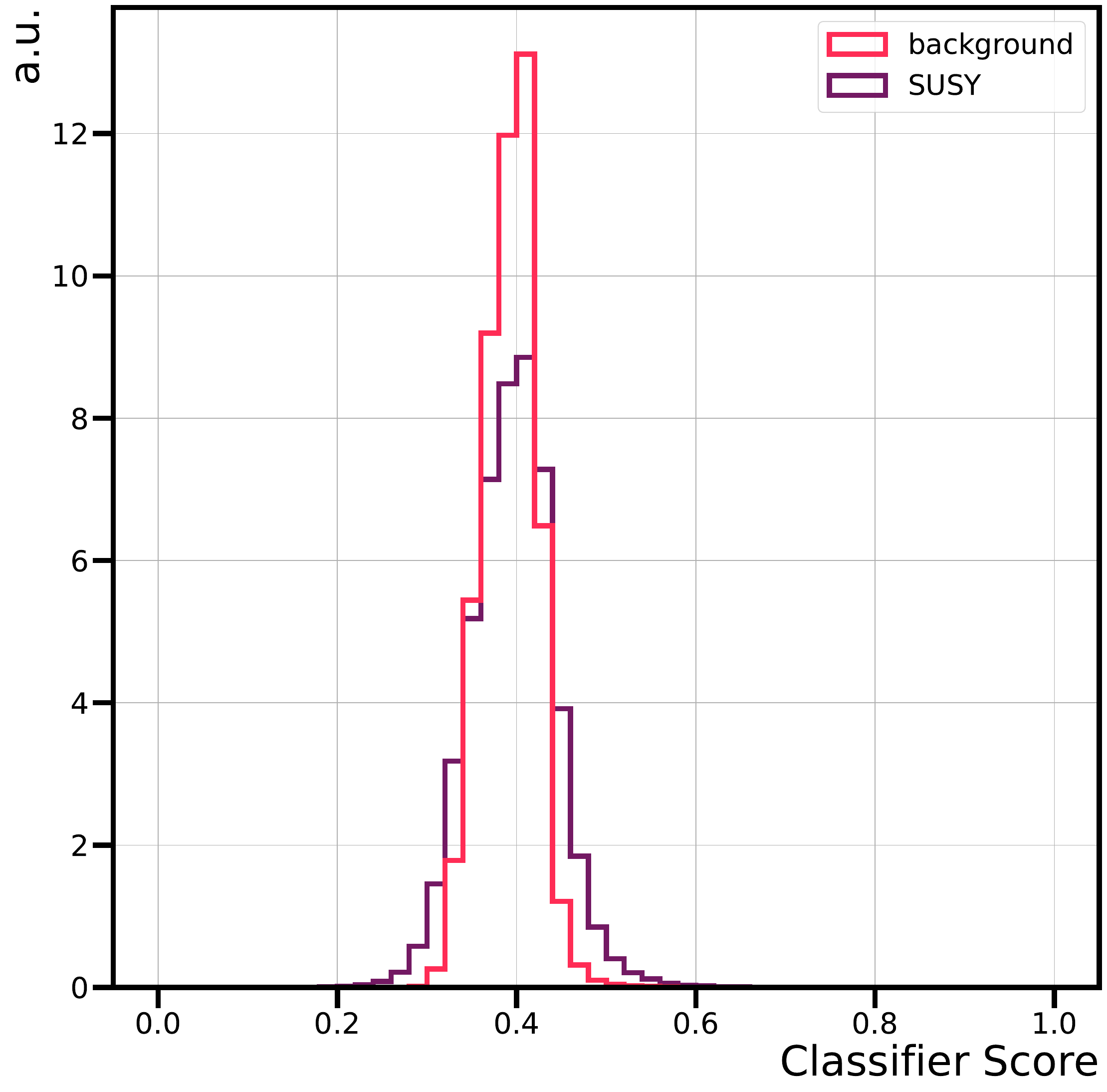}
        \caption{}
        \label{fig:linear_scores}
    \end{subfigure}
    \hfill
    \begin{subfigure}[t]{0.48\textwidth}
        \centering
        \includegraphics[
            width=\linewidth,
            height=0.28\textheight,
            keepaspectratio
        ]{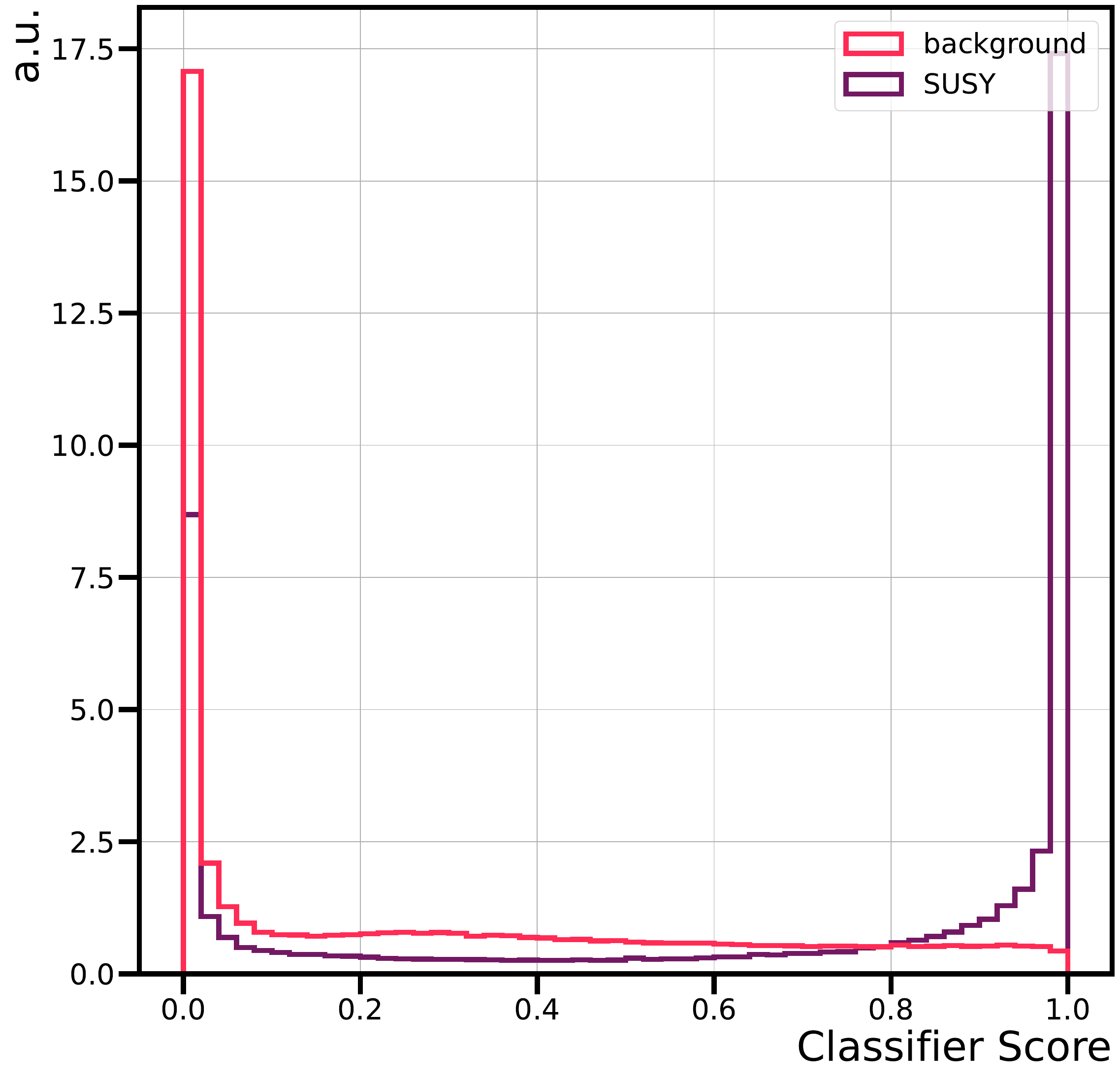}

        \caption{}
        \label{fig:MLP_scores}
    \end{subfigure}

    \begin{subfigure}[t]{0.48\textwidth}
        \centering
        \includegraphics[
            width=\linewidth,
            height=0.28\textheight,
            keepaspectratio
        ]{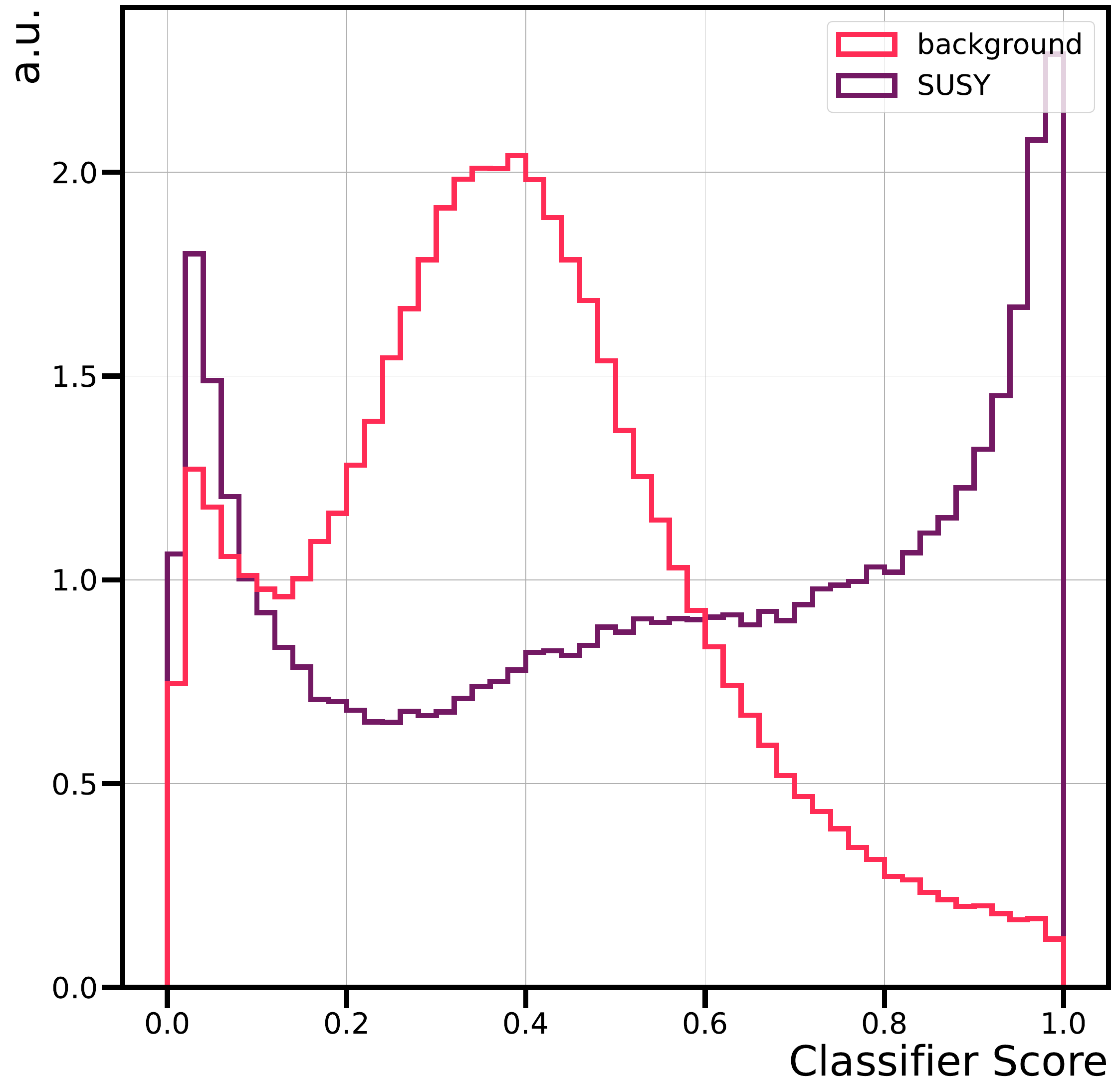}
        \caption{}
        \label{fig:1Layer_scores}
    \end{subfigure}
    \hfill
    \begin{subfigure}[t]{0.48\textwidth}
        \centering
        \includegraphics[
            width=\linewidth,
            height=0.28\textheight,
            keepaspectratio
        ]{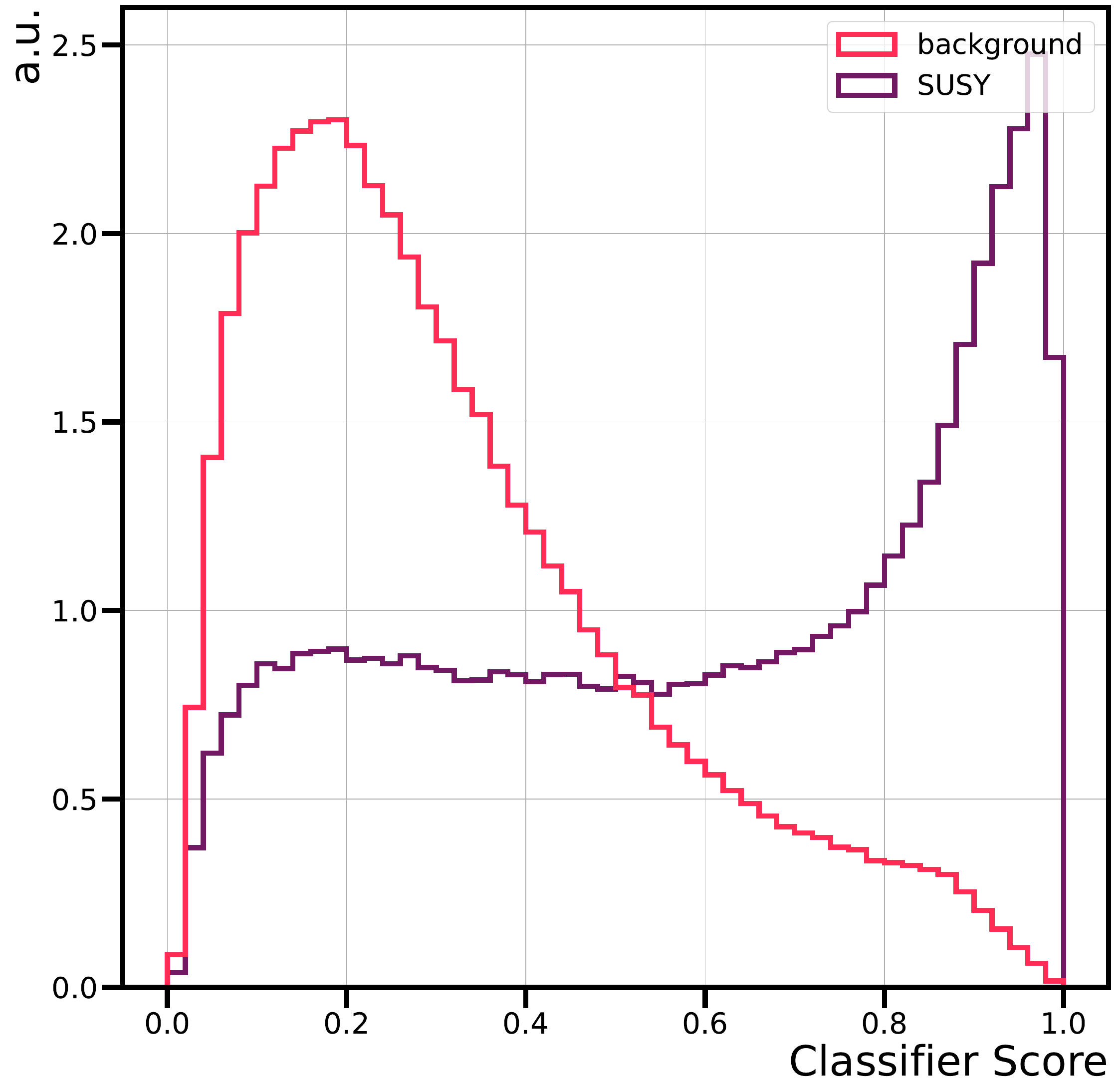}
        \caption{}
        \label{fig:4Layer_scores}
    \end{subfigure}

    \caption{Classifier-score distributions obtained under reference detector conditions for the supervised models. Panels (a)--(d) show the results for the linear classifier, multilayer perceptron, single-layer quantum classifier and four-layer quantum data-reuploading classifier, respectively. In each panel, the background distribution is compared with that of the SUSY signal sample.}
    \label{fig:supervised_scores}
\end{figure}

The classifier-score distributions obtained under reference detector conditions are shown in Figure~\ref{fig:supervised_scores}. The linear classifier exhibits substantial overlap between the background and signal distributions and does not establish a useful ordering of the two classes. The MLP produces a much clearer distinction, assigning a large fraction of the background and signal events scores close to zero and unity, respectively. The single-layer quantum classifier develops non-trivial separation, but considerable overlap remains. By contrast, the four-layer quantum model produces markedly more distinct score distributions, with background events preferentially assigned low scores and signal events accumulating towards unity. The comparison shows that the principal improvement in the quantum classifier arises from repeatedly introducing the input throughout the circuit. A single encoding layer captures some discriminating structure, but additional data-reuploading layers are required for the model to become competitive with the non-linear classical baseline.

%%% QAE and HW_QAE figure

\begin{figure}[t!]
    \centering
    \includegraphics[
        width=0.48\linewidth,
        keepaspectratio
    ]{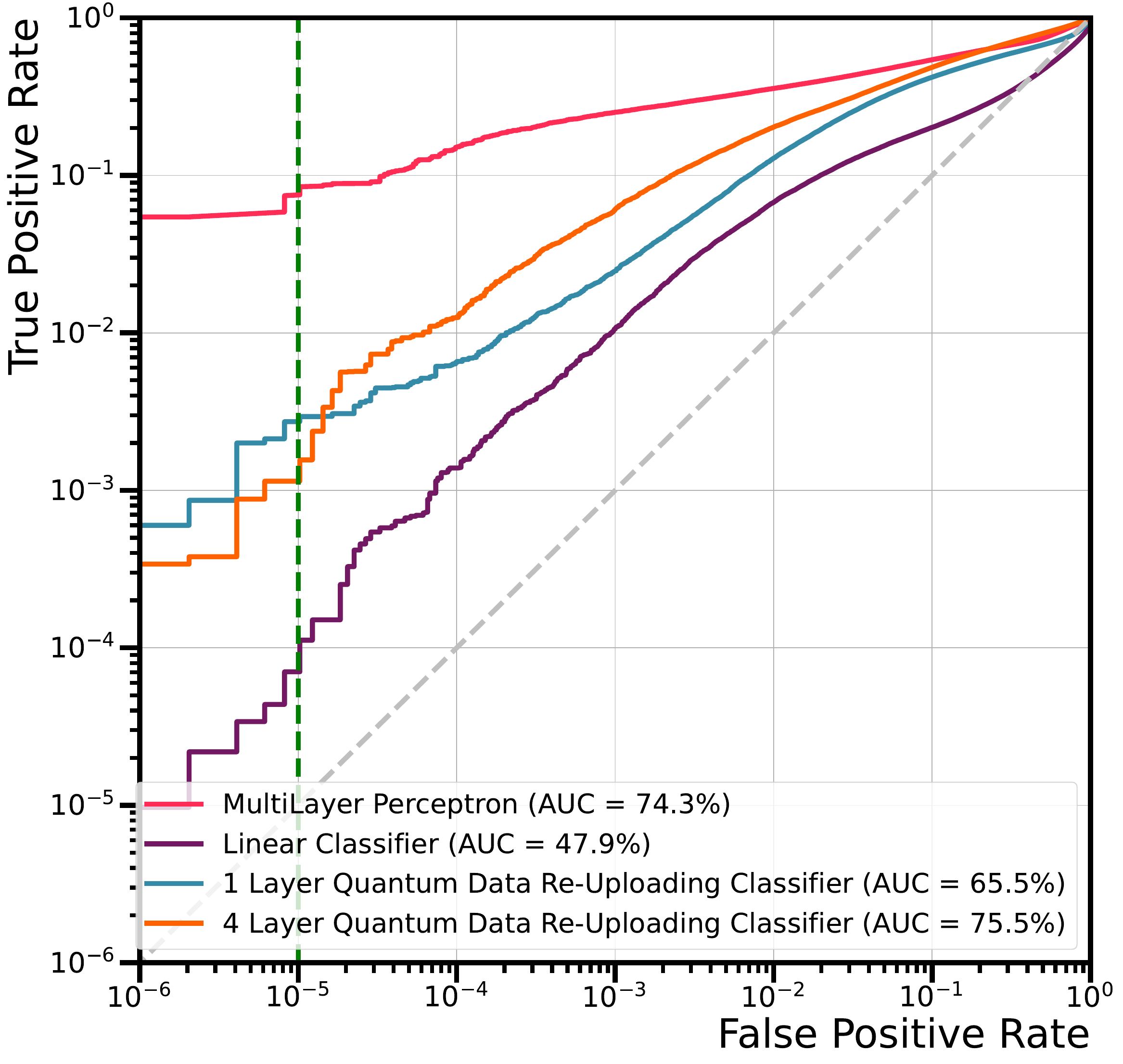}
    \caption{
    ROC curves obtained under reference detector conditions for the linear classifier, multilayer perceptron, single-layer quantum classifier and four-layer quantum data-reuploading classifier. The corresponding AUC values are reported in the legend. The grey dashed diagonal indicates random classification, while the vertical dashed line marks a false-positive rate of $10^{-5}$, representative of the low-background operating regime relevant to collider triggers.
    }
    \label{fig:supervised_roc}
\end{figure}

This behaviour is quantified by the ROC curves in Figure~\ref{fig:supervised_roc}. The linear classifier obtains an AUC of $47.9\%$, indicating that a linear decision boundary is unable to extract meaningful discrimination from the seven input features. The single-layer quantum classifier improves the AUC to $65.5\%$, but remains below the MLP, which reaches $74.3\%$. Increasing the quantum circuit to four data-reuploading layers raises the AUC to $75.5\%$, an improvement of $10.0$ percentage points over the single-layer circuit and marginally exceeding the non-linear classical baseline. The additional data-reuploading layers therefore close the performance gap observed for the shallow circuit. Within this setup, a fixed seven-qubit register can attain discrimination comparable to that of the approximately parameter-matched classical MLP, despite being trained using a substantially smaller labelled sample.

%%% SUSY ROC curves

The relative ordering changes in the low-background regime relevant to collider triggers. At a fixed false-positive rate of $10^{-5}$, marked by the vertical dashed lines in Figure~\ref{fig:supervised_roc}, the MLP retains the largest signal efficiency. The four-layer quantum classifier has a similar performance to the single-layer circuit and both substantially improve upon the linear baseline at this operating point. The four-layer model therefore gives the strongest integrated discrimination as measured by the AUC, whereas the MLP remains more effective in the extreme low-background tail. This distinction illustrates that the AUC and performance at a fixed trigger threshold probe complementary properties of the classifier response. These results establish the reference performance against which sensitivity to detector-induced distribution shift is assessed in the following subsection.

%%%%%%%%%%%%%%%%%%%%%%%%%%%%%%%%%%%%%%%%%%%%%%%%%%%%%%%%%%%%%
%%%%%%%%%%%%%%%%%%%%%%%%%%%%%%%%%%%%%%%%%%%%%%%%%%%%%%%%%%%%%

\subsection{Robustness under domain shift}\label{sec:supervised_robustness}

We next examine the robustness of the supervised classifiers under the same controlled feature-level smearing introduced in Section~\ref{sec:unsupervised_robustness}. Such perturbations emulate changes in reconstructed collider observables arising from finite detector resolution, calibration effects or variations in detector operating conditions. The smearing map $\mathcal{S}_{\lambda}$ defined in Equations~\eqref{eq:smearing_map} and \eqref{eq:smearing_2} is applied to the $p_T$ observables before evaluating the trained classifiers. No recalculation of the derived kinematic input variables $E_T^{miss}$, $M^T_R$, and $M^T_{\Delta}$ was performed.  The feature rescaling determined from the reference training sample and all model parameters are held fixed throughout. The resulting comparison therefore probes the response of each classifier to a shift in the input distribution, without retraining or adaptation to the smeared samples.

Following the event-level measure introduced for the anomaly-detection models in Equation~\eqref{eq:score_mse}, we quantify the change in the supervised classifier response through
\begin{equation}\label{eq:supervised_score_deviation}
    \Delta^{2}_{m,c}(\lambda) = \frac{1}{|{D}_{c}|} \sum_{\mathbf{x}\in\mathcal{D}_{c}} \left[ s_m\!\left(\mathbf{x}^{(\lambda)}\right) - s_m(\mathbf{x}) \right]^2,
\end{equation}
where $s_m$ denotes the output score of classifier $m$ and $c\in\{\mathrm{bkg},\mathrm{sig}\}$ labels the event class. Smaller values of $\Delta^{2}_{m,c}$ indicate that the numerical classifier response is less sensitive to the imposed input shift. As in the unsupervised study, the smearing is repeated ten times, with the curves and shaded bands showing the mean and standard deviation across the independent realisations.

\begin{figure}[tp]
    \centering
    \captionsetup[subfigure]{font=small,skip=2pt}
    \captionsetup{skip=4pt}

    \begin{subfigure}[t]{0.48\textwidth}
        \centering
        \includegraphics[height=0.28\textheight]{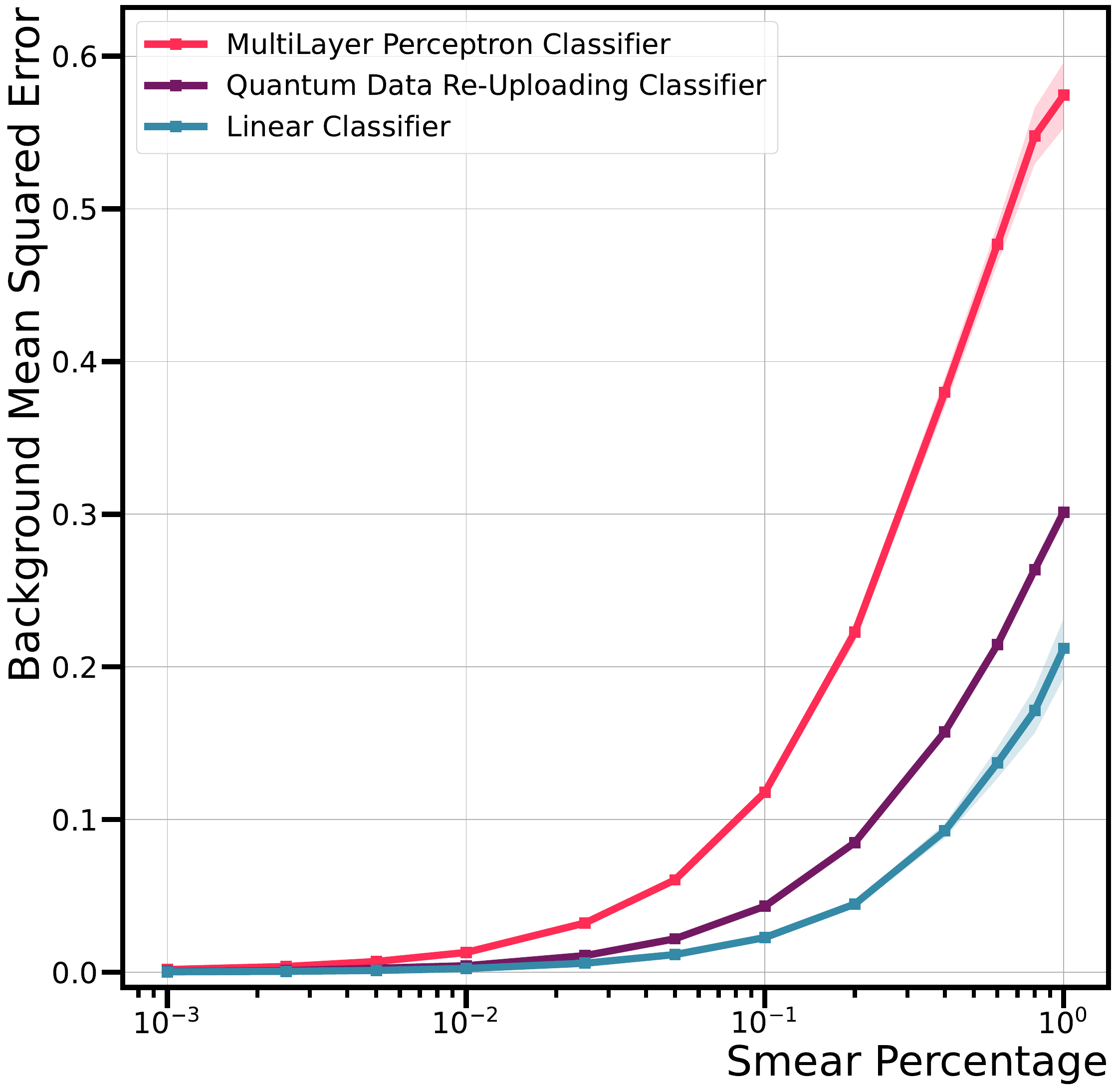}
        \caption{}
        \label{fig:1Layer_robustness_bkg}
    \end{subfigure}
    \hfill
    \begin{subfigure}[t]{0.48\textwidth}
        \centering
        \includegraphics[height=0.28\textheight]{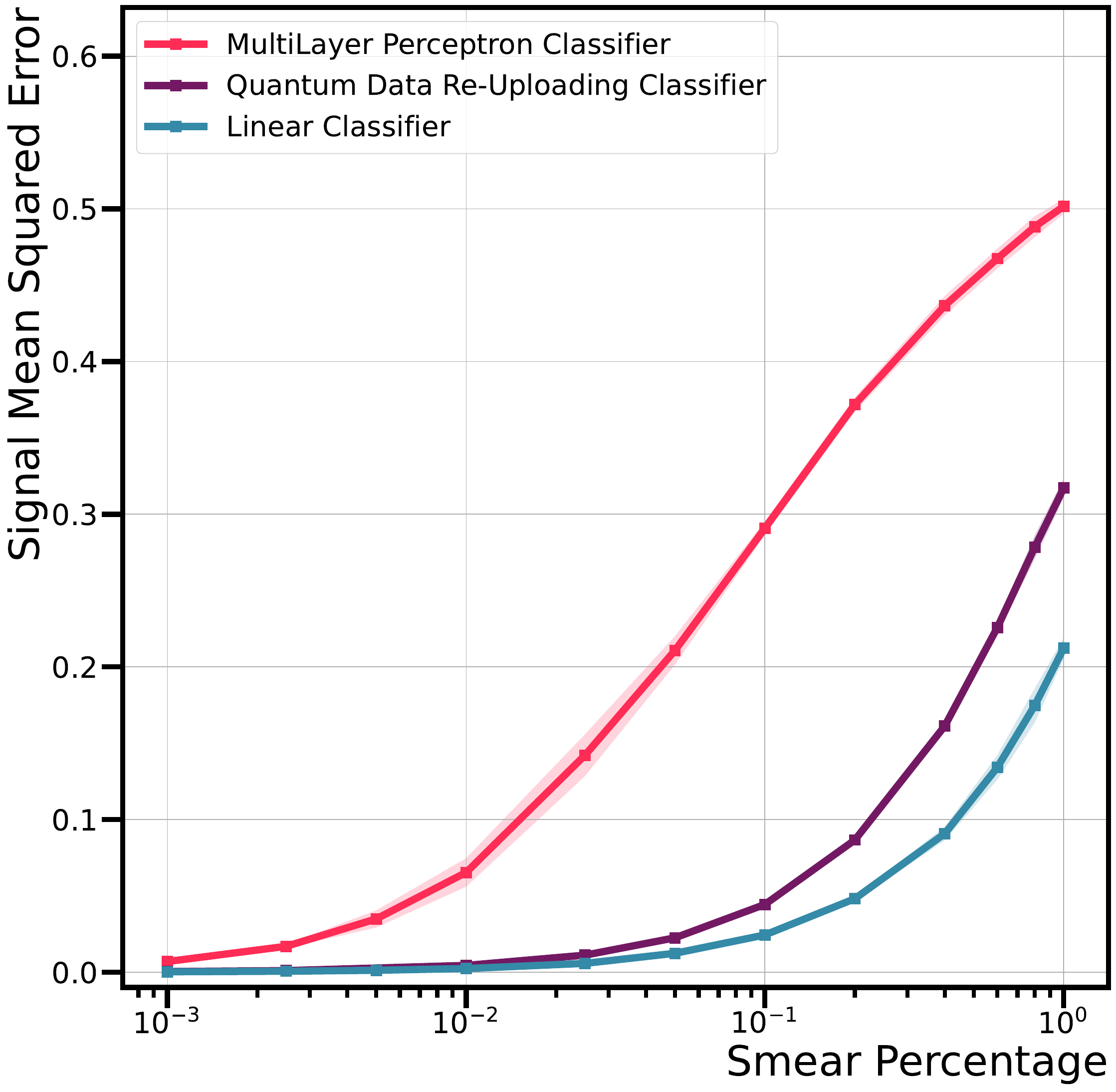}
        \caption{}
        \label{fig:2Layer_robustness_sig}
    \end{subfigure}

    \vspace{0.2em}

    \begin{subfigure}[t]{0.48\textwidth}
        \centering
        \includegraphics[height=0.28\textheight]{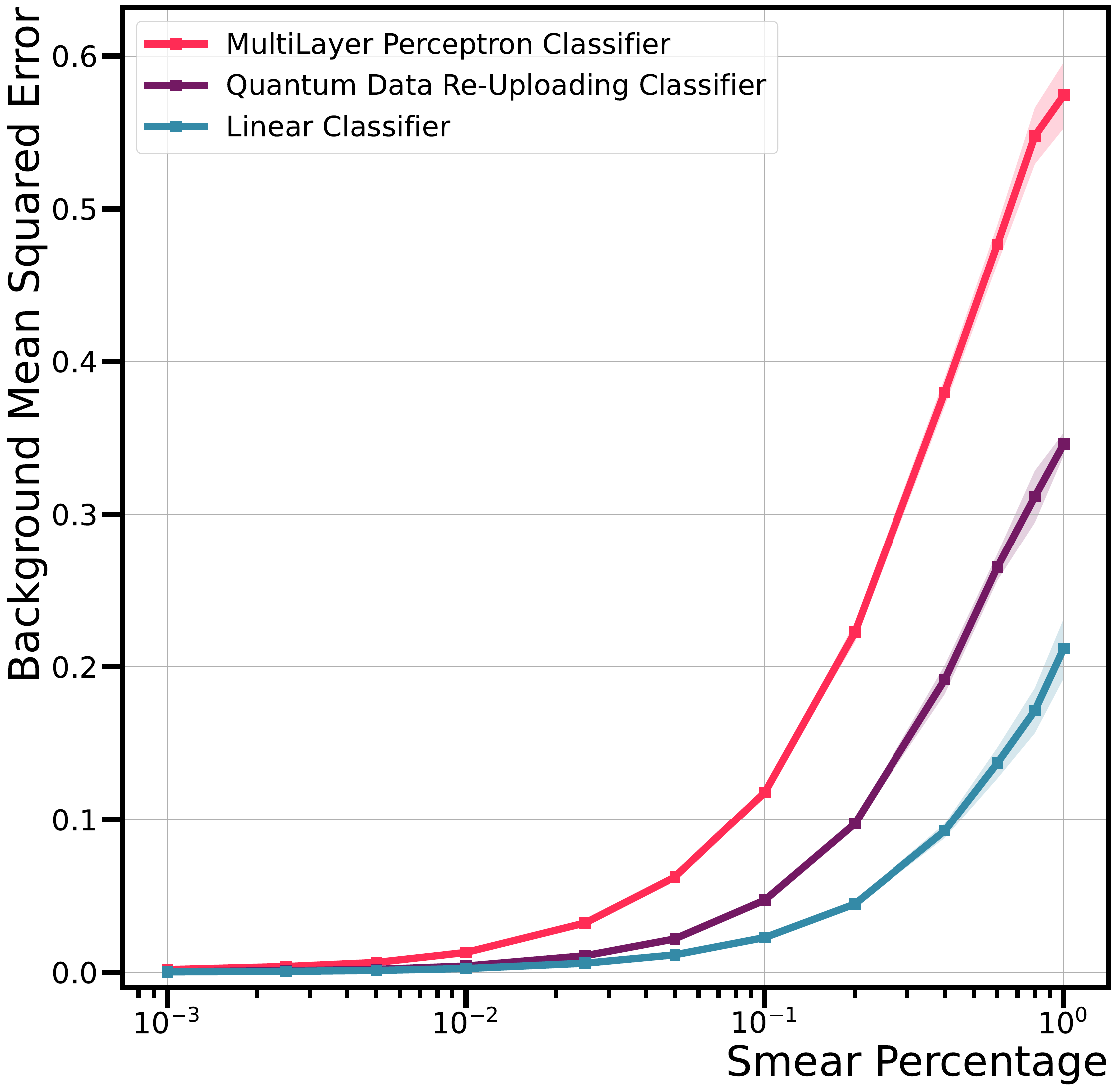}
        \caption{}
        \label{fig:4Layer_robustness_bkg}
    \end{subfigure}
    \hfill
    \begin{subfigure}[t]{0.48\textwidth}
        \centering
        \includegraphics[height=0.28\textheight]{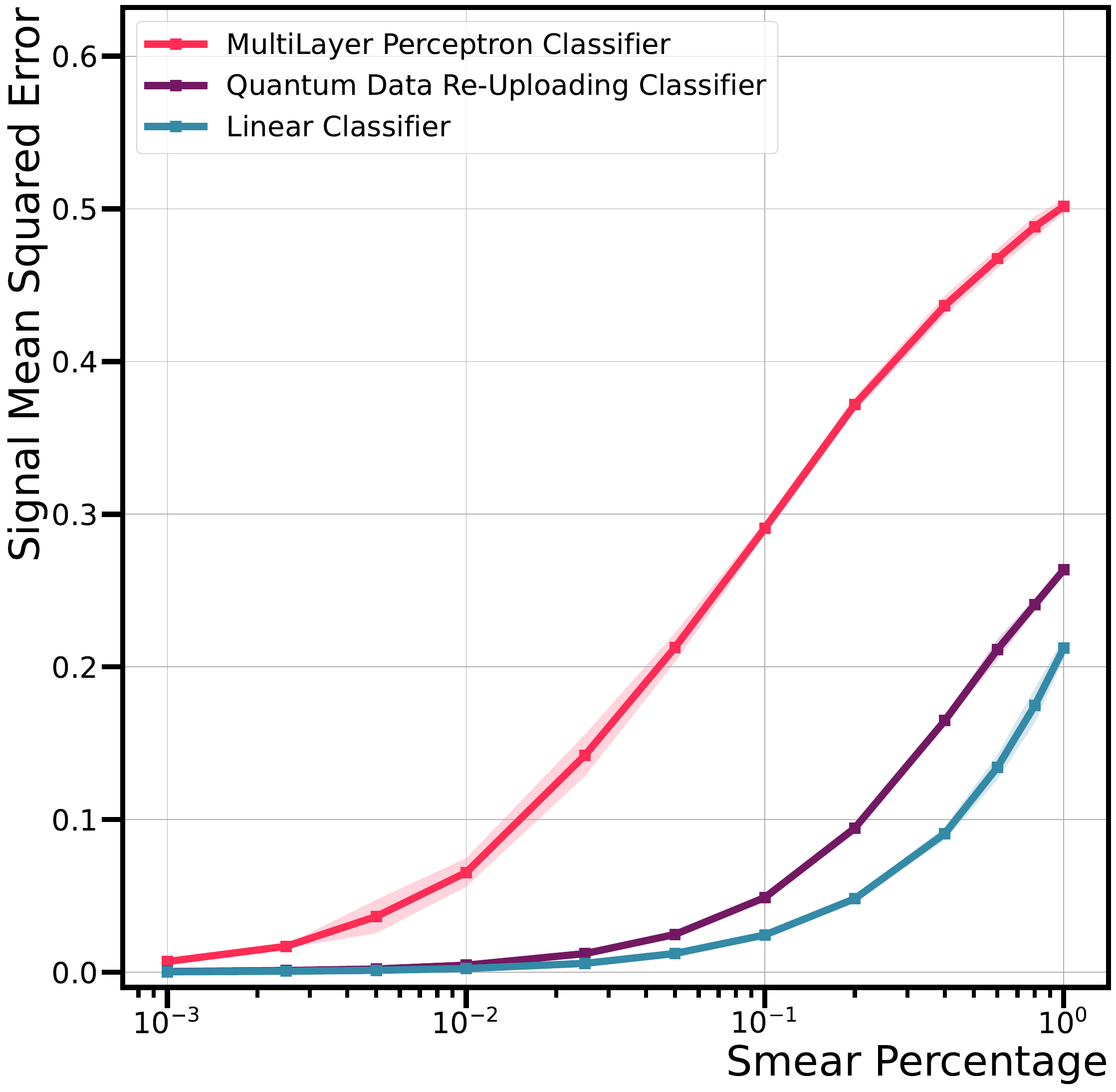}
        \caption{}
        \label{fig:4Layer_robustness_sig}
    \end{subfigure}
    \caption{
        Mean-squared deviation of the classifier score from its reference value as a function of the detector-smearing strength. Results are shown for background events in the left panels and SUSY signal events in the right panels. Panels (a) and (b) compare the single-layer quantum classifier with the linear classifier and multilayer perceptron, while panels (c) and (d) show the corresponding comparison for the four-layer quantum classifier. The model parameters and feature-scaling constants are held fixed throughout. The shaded bands indicate the standard deviation across ten independent smearing realisations.
        }
    \label{fig:supervised_score_deviation}
\end{figure}

Figure~\ref{fig:supervised_score_deviation} compares the resulting score deviations for the quantum, linear and MLP classifiers. For all models, the deviation increases with the smearing strength, reflecting the growing displacement of the inputs from the reference distribution. The differences between the models are small in the weakly perturbed regime but become increasingly pronounced as the smearing is increased.

%% Data reuploading performance

For both background and signal events, the MLP exhibits the largest change in its output score. Its deviation begins to increase at smaller smearing strengths and rises more rapidly than those of either the linear or quantum classifiers. The quantum data-reuploading classifier shows a substantially smaller response, remaining between the two classical baselines throughout the scan. The linear classifier gives the smallest score deviation and therefore does not strongly amplify the imposed perturbation.

The stability of the linear model must, however, be interpreted together with its reference classification performance. As shown in Figure~\ref{fig:supervised_roc}, the linear classifier provides little useful signal-background discrimination. Its small score displacement therefore reflects, at least in part, the limited dependence of its output on the input features. By contrast, the quantum classifier combines non-trivial discrimination with a considerably more stable numerical response than the non-linear MLP. It therefore occupies an intermediate and more useful region of the robustness-performance trade-off: more discriminating than the linear model, while less sensitive to detector-induced perturbations than the MLP.

\begin{figure}[t]
    \centering

    \begin{subfigure}[t]{0.48\textwidth}
        \centering
        \includegraphics[height=0.28\textheight]{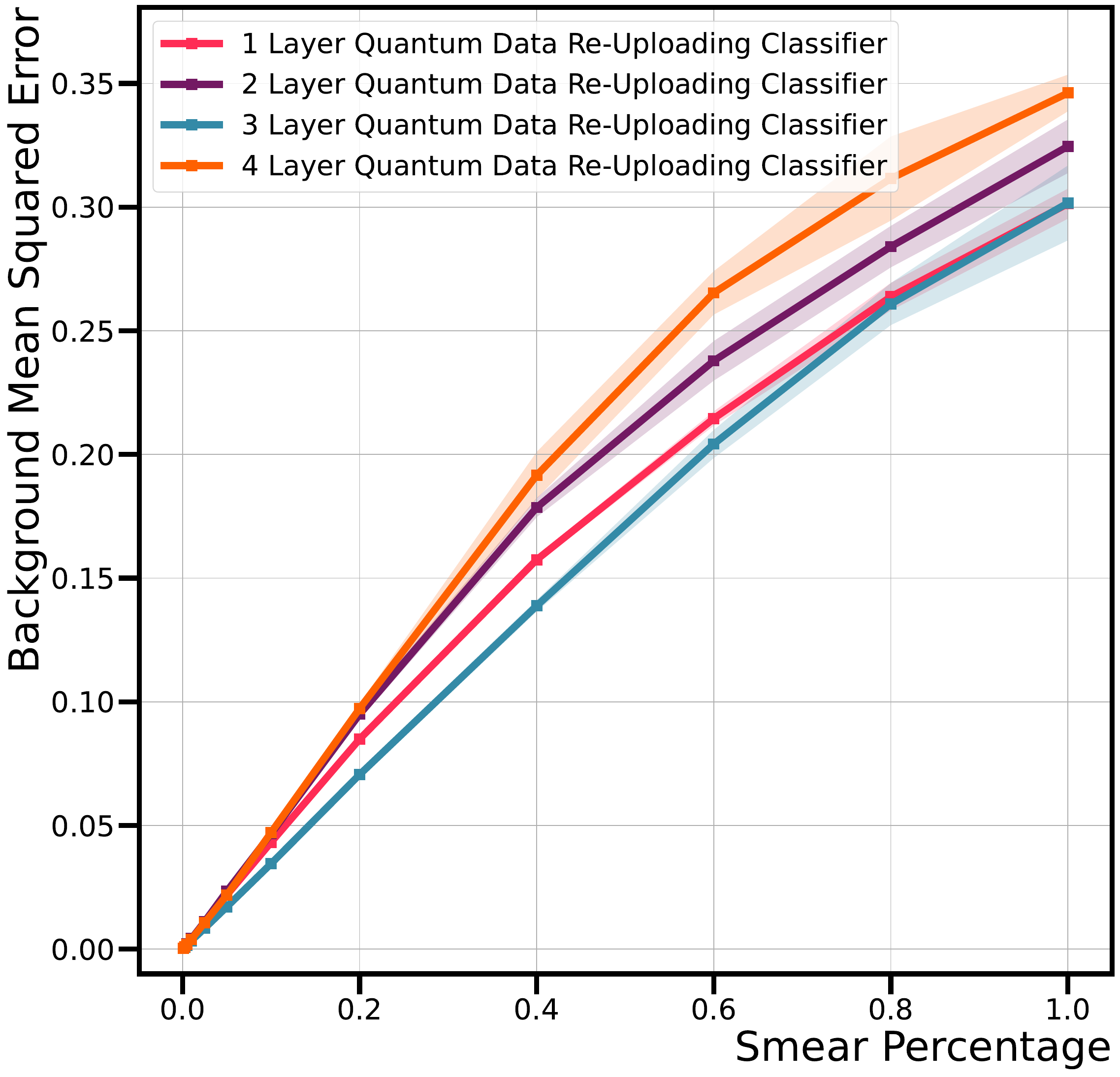}
        \caption{}
        \label{fig:dru_bkg}
    \end{subfigure}
    \hfill
    \begin{subfigure}[t]{0.48\textwidth}
        \centering
        \includegraphics[height=0.28\textheight]{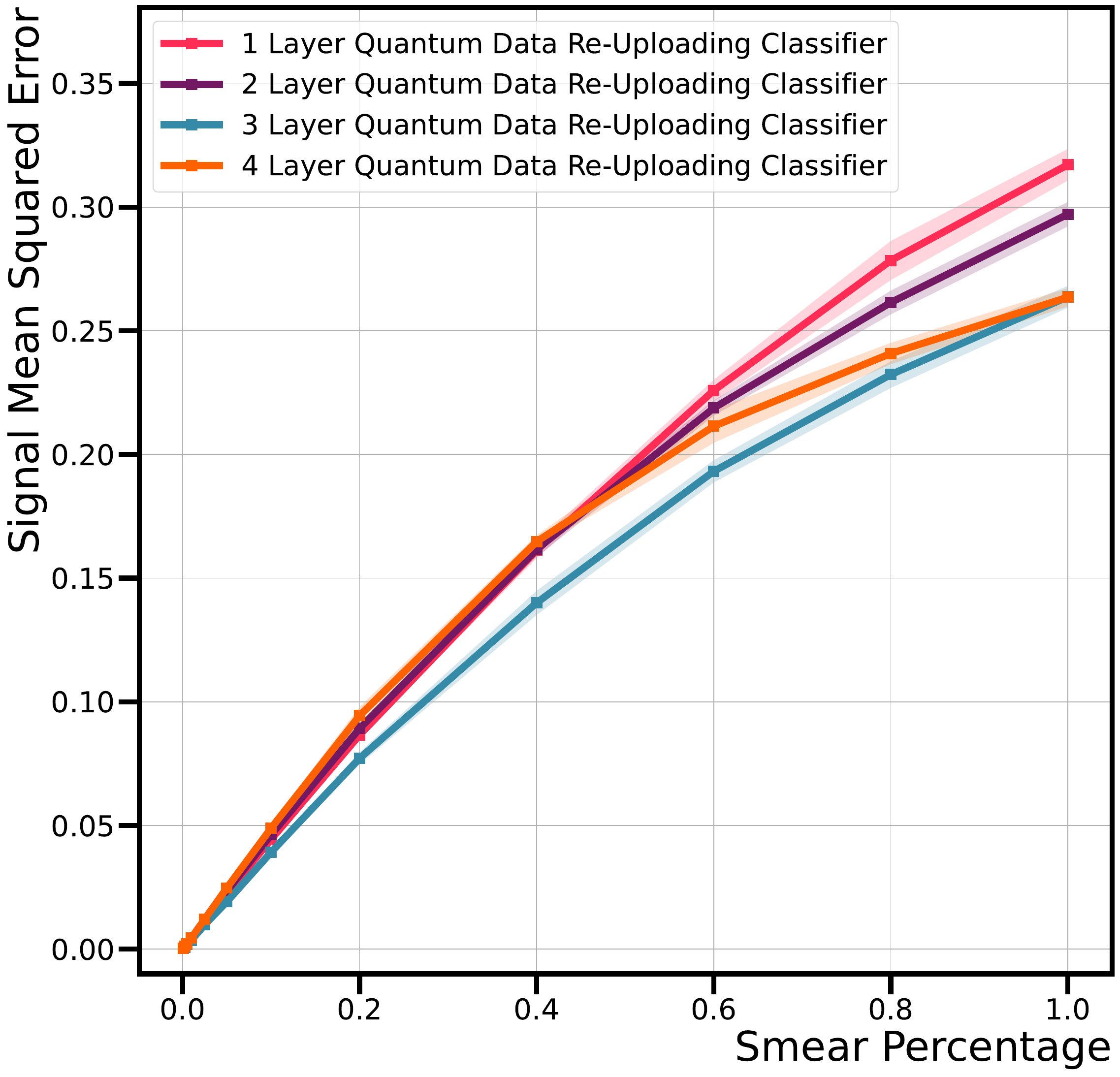}
        \caption{}
        \label{fig:dru_sig}
    \end{subfigure}

    \caption{
        Dependence of the quantum-classifier response on the data-reuploading depth. The mean-squared deviation of the classifier score from its reference value is shown as a function of the detector-smearing strength for background events in panel (a) and SUSY signal events in panel (b). Results are compared for classifiers containing between one and four data-reuploading layers. The shaded bands indicate the standard deviation across ten independent smearing realisations.
        }
    \label{fig:supervised_depth}
\end{figure}

To isolate the effect of circuit depth, Figure~\ref{fig:supervised_depth} compares the score deviations obtained using between one and four data-reuploading layers. The dependence on depth differs between the two event classes. For background events, the four-layer circuit develops the largest deviation at strong smearing, indicating that the deeper, more expressive classifier is more sensitive to perturbations in this region of feature space. The ordering is not strictly monotonic, however, with the three-layer circuit remaining among the least sensitive. For signal events, the single-layer model instead exhibits the largest deviation at strong smearing, while the three and four-layer circuits show the smallest response.

%%% Smear as a func of data-reuploading layers

Increasing the number of reuploading layers therefore changes how input perturbations propagate through the classifier, rather than producing a universal increase in the score displacement. In each layer, the same perturbed feature vector is supplied to a new set of trainable encoding rotations. A shift in one feature consequently modifies several rotation angles throughout a deeper circuit, with these changes propagated through the intervening entangling and variational operations. Repeated data encoding can therefore enhance the sensitivity of the learned decision function to particular input directions. The class-dependent and non-monotonic behaviour shows, however, that this response also depends on the trained circuit parameters and on the region of feature space occupied by the events.

\begin{figure}[t]
    \centering

    \begin{subfigure}[t]{0.48\textwidth}
        \centering
        \includegraphics[height=0.28\textheight]{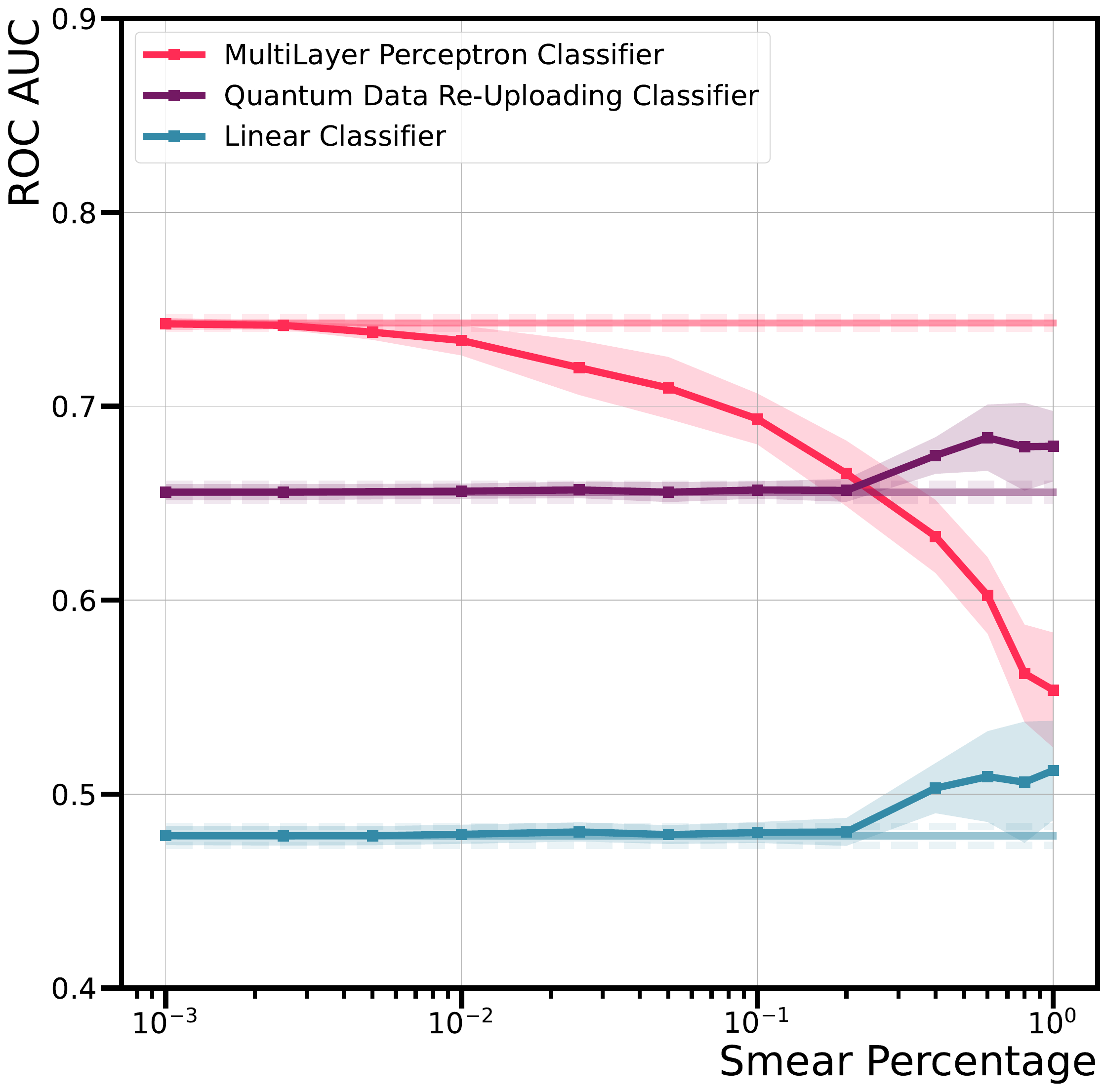}
        \caption{}
        \label{fig:supervised_roc_smear_l1}
    \end{subfigure}
    \hfill
    \begin{subfigure}[t]{0.48\textwidth}
        \centering
        \includegraphics[height=0.28\textheight]{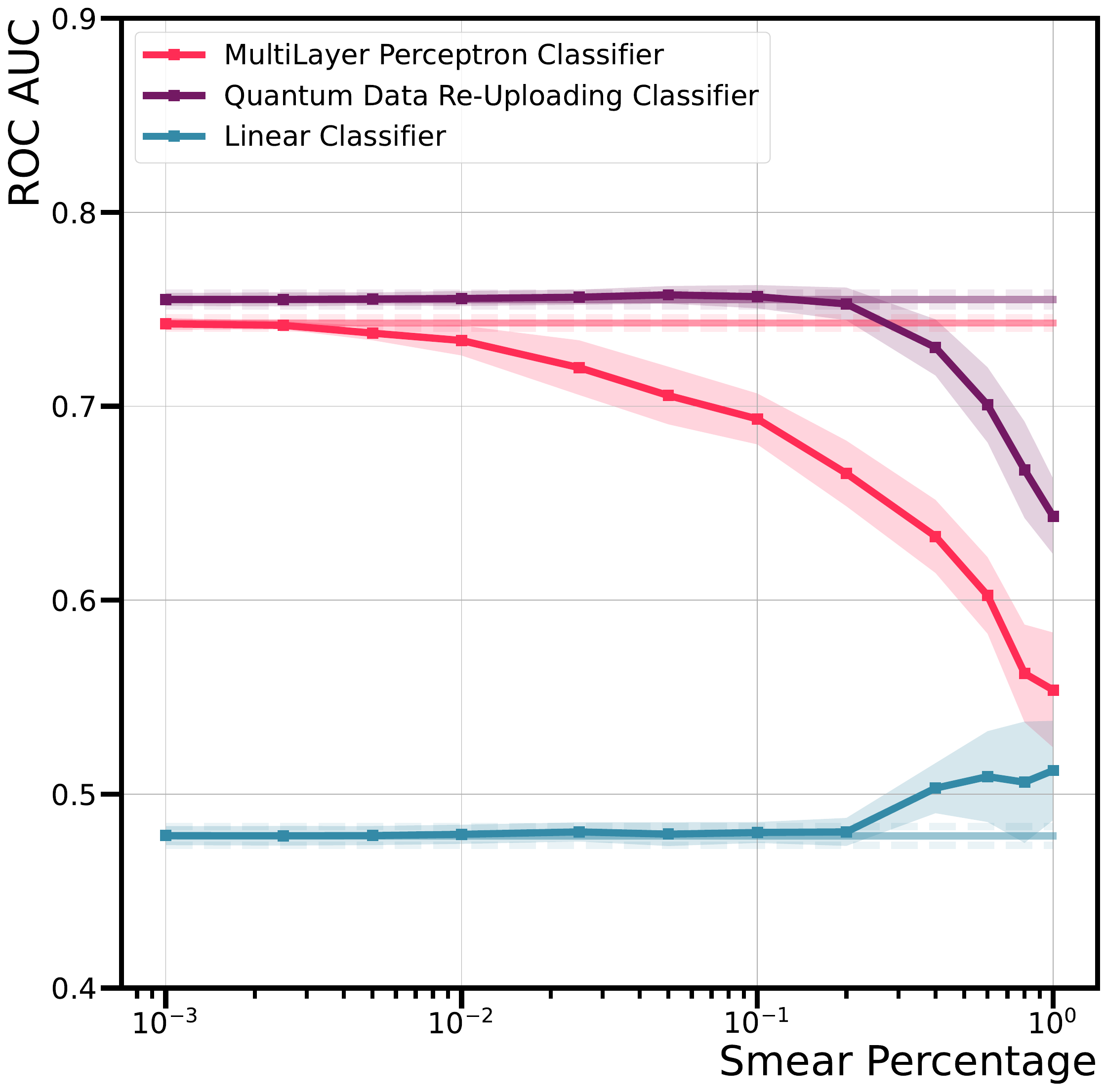}
        \caption{}
        \label{fig:supervised_roc_smear_l4}
    \end{subfigure}

    \caption{
        ROC AUC as a function of the detector-smearing strength for the SUSY signal sample. Panel (a) compares the linear classifier and multilayer perceptron with the single-layer quantum classifier, while panel (b) shows the corresponding comparison for the four-layer quantum data-reuploading classifier. Horizontal dashed lines indicate the reference AUC of each model. The shaded bands show the standard deviation across ten independent smearing realisations.
    }
    \label{fig:supervised_roc_vs_smear}
\end{figure}

The mean-squared score deviation measures the stability of the numerical classifier output for individual events, but does not by itself determine whether signal-background discrimination is retained. We therefore also evaluate the ROC AUC as a function of the smearing strength, as shown in Figure~\ref{fig:supervised_roc_vs_smear}.

%%% Smear as a func of data-reuploading layers

The MLP achieves strong discrimination under reference detector conditions, but its AUC decreases steadily as the smearing strength is increased. The degradation becomes particularly pronounced in the strongly shifted regime, where the model approaches substantially weaker discrimination. The linear classifier remains close to its reference AUC throughout, but this again reflects its limited initial performance rather than a useful retention of strong classification.

The single-layer quantum classifier also retains an approximately constant AUC across the scan. Its reference discrimination is weaker than that of the MLP, but its relative ordering of signal and background events is largely unaffected by the imposed perturbation. The four-layer quantum classifier begins with the strongest reference AUC and remains stable under weak and moderate smearing before degrading at the largest perturbations. Its loss of performance is nevertheless smaller than that of the MLP, and it retains the largest AUC in the strongly shifted regime.

The comparison reveals a trade-off between the reference performance and robustness of the quantum classifiers. Additional data-reuploading layers substantially improve the discrimination achieved under reference detector conditions, but also make the global classifier ordering more sensitive to sufficiently large shifts in the input distribution. The shallow quantum model is more stable but provides weaker discrimination, whereas the four-layer model gives stronger reference performance with a moderate increase in sensitivity.

Taken together, the event-level score deviations and AUC results show that the quantum data-reuploading classifiers respond less strongly to detector-induced input shifts than the non-linear classical baseline. In particular, the four-layer quantum model achieves discrimination comparable to the MLP under reference detector conditions, while exhibiting a smaller displacement of its output scores and retaining stronger global discrimination as the smearing is increased. The robustness observed for the quantum autoencoders therefore extends to a distinct supervised architecture and is not restricted to fidelity-based anomaly detection.

%%%%%%%%%%%%%%%%%%%%%%%%%%%%%%%%%%%%%%%%%%%%%%%%%%%%%%%%%%%%%
%%%%%%%%%%%%%%%%%%%%%%%%%%%%%%%%%%%%%%%%%%%%%%%%%%%%%%%%%%%%%
%%%%%%%%%%%%%%%%%%%%%%%%%%%%%%%%%%%%%%%%%%%%%%%%%%%%%%%%%%%%%

\section{Conclusions}\label{sec:conclusions}

We present a systematic study of the robustness of parameterised quantum models for collider-event selection under controlled shifts in the input distribution. Two complementary learning settings are considered: unsupervised anomaly detection using quantum autoencoders (QAEs), and supervised classification using quantum circuits with data reuploading. In both cases, the models are trained under a reference condition of clean detector data, and subsequently evaluated after applying a controlled feature-level smearing to the reconstructed input observables. The preprocessing transformations and trained model parameters are held fixed throughout, allowing the stability of the complete inference pipeline to be examined without retraining, adaptation, or explicit exposure to the smeared samples during training. In each setting, the quantum models are compared with optimised classical benchmarks.

In the unsupervised setting, the compact QAEs learned effective anomaly scores despite using substantially fewer training events and trainable parameters than the optimised classical autoencoder (AE), which was trained on the full background dataset. Although the AE obtained the largest integrated AUC across the four signal benchmarks, the difference was less pronounced in the low-background regime relevant to collider triggers, where the QAEs retained competitive signal efficiencies. The local and non-local quantum architectures produced closely comparable discrimination, with the local QAE obtaining a marginally larger AUC for each benchmark. Under feature-level smearing, the global discrimination of the classical and quantum autoencoders remained broadly stable, while the anomaly scores produced by the QAEs exhibited substantially smaller numerical shifts than those of the classical baselines. The local architecture retained this stability without requiring direct long-range entangling operations, making it well suited to the restricted qubit connectivity typical of near-term quantum hardware.

In the supervised setting, data reuploading substantially enhanced the discrimination achieved by the quantum classifier. Under reference conditions, the four-layer circuit obtained the largest integrated AUC of the models considered, marginally outperforming the non-linear classical baseline despite being trained on substantially fewer labelled events. The multilayer perceptron (MLP) retained a larger signal efficiency in the extreme low-background regime, but its output scores underwent the largest displacement under feature-level smearing and its discrimination degraded most strongly. By comparison, the quantum classifiers preserved their performance more effectively, providing the most favourable robustness-performance trade-off among the expressive models considered. The response to smearing varied non-monotonically with circuit depth and differed between the signal and background samples. The improvement in discrimination obtained through additional data-reuploading layers therefore did not require a corresponding loss of robustness.

Taken together, these results demonstrate that compact parameterised quantum models can achieve competitive performance in both unsupervised and supervised collider-event selection while exhibiting promising robustness under controlled shifts in the input distribution. Combined with their ability to learn effective decision functions from comparatively small training datasets, these findings identify collider-event selection as a promising setting for robust quantum machine learning. While the present study is based on exact circuit simulation and a simplified detector-inspired model of feature variability, it motivates further investigation using realistic detector systematics, finite-shot statistics and quantum-device noise. More broadly, our results establish robustness under distribution shift as an important measure of quantum-model performance alongside discrimination, data efficiency and resource requirements.

\appendix
\section{Anomaly-detection dependence on training-sample size}\label{app:training-size}

\begin{figure}[t!]
    \centering
    \captionsetup[subfigure]{font=small,skip=2pt}
    \captionsetup{skip=4pt}
    \begin{subfigure}[t]{0.48\textwidth}
        \centering
        \includegraphics[
            width=\linewidth,
            height=0.28\textheight,
            keepaspectratio
        ]{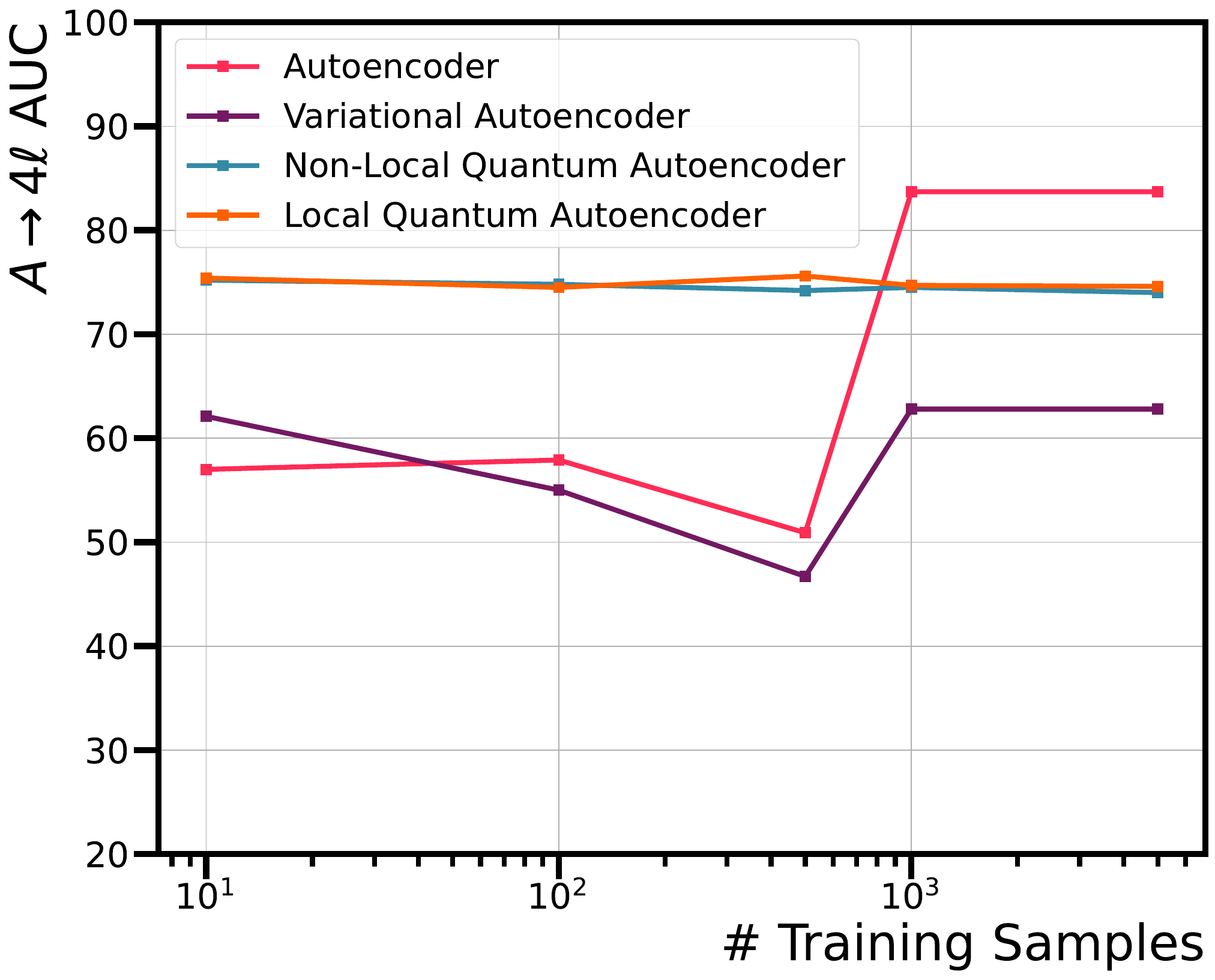}
        \caption{}
        \label{fig:training1}
    \end{subfigure}
    \hfill
    \begin{subfigure}[t]{0.48\textwidth}
        \centering
        \includegraphics[
            width=\linewidth,
            height=0.28\textheight,
            keepaspectratio
        ]{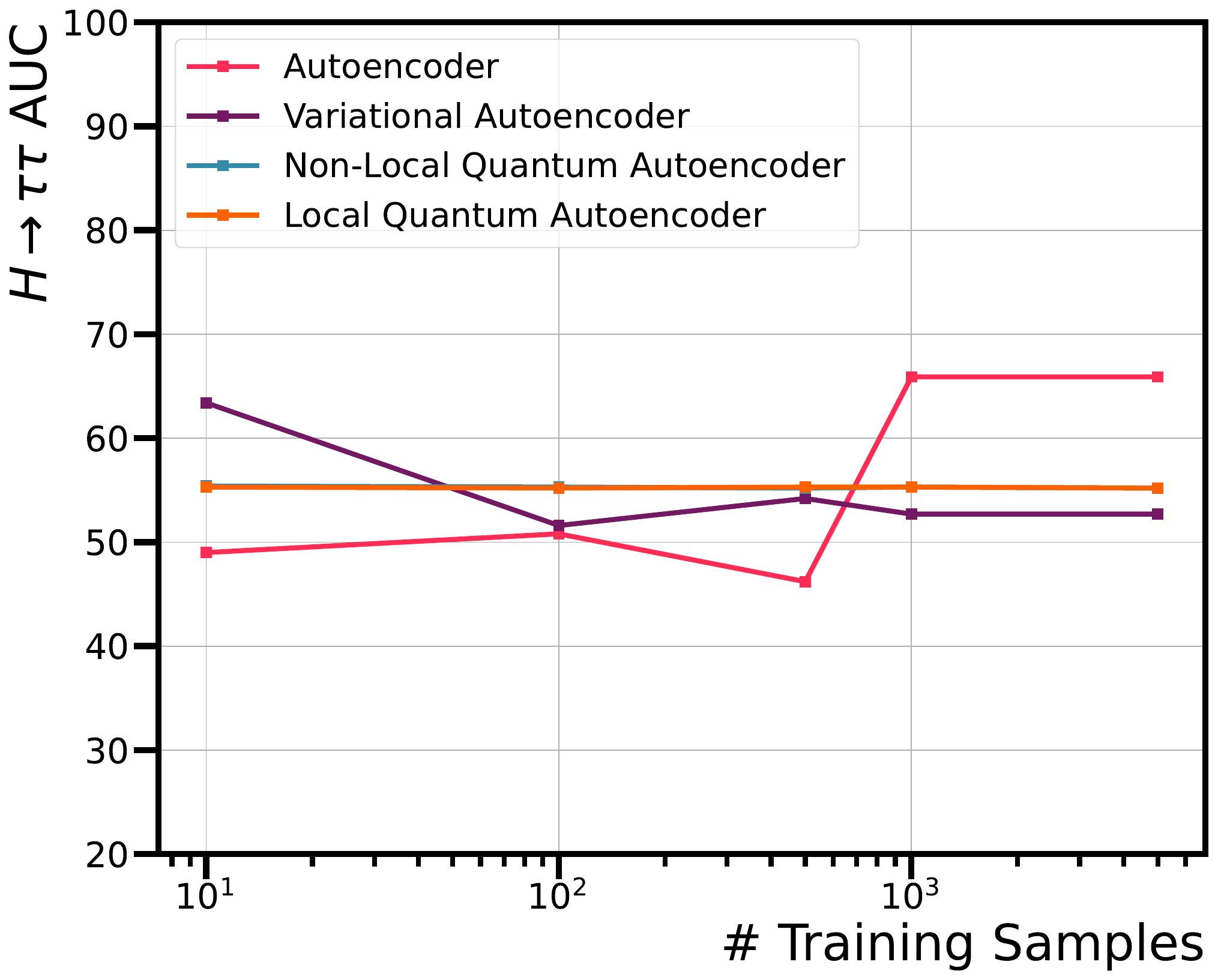}

        \caption{}
        \label{fig:training2}
    \end{subfigure}

    \begin{subfigure}[t]{0.48\textwidth}
        \centering
        \includegraphics[
            width=\linewidth,
            height=0.28\textheight,
            keepaspectratio
        ]{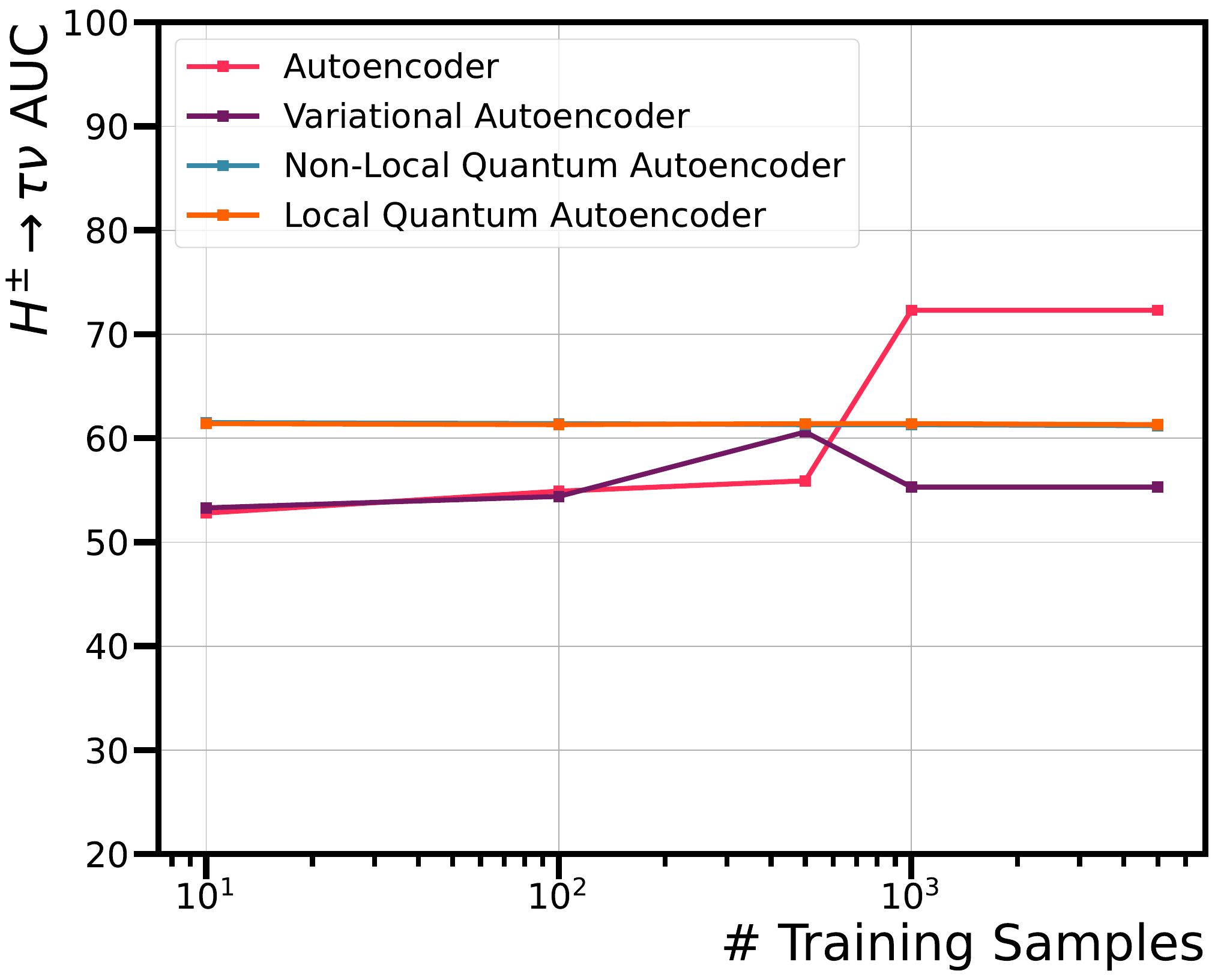}
        \caption{}
        \label{fig:training3}
    \end{subfigure}
    \hfill
    \begin{subfigure}[t]{0.48\textwidth}
        \centering
        \includegraphics[
            width=\linewidth,
            height=0.28\textheight,
            keepaspectratio
        ]{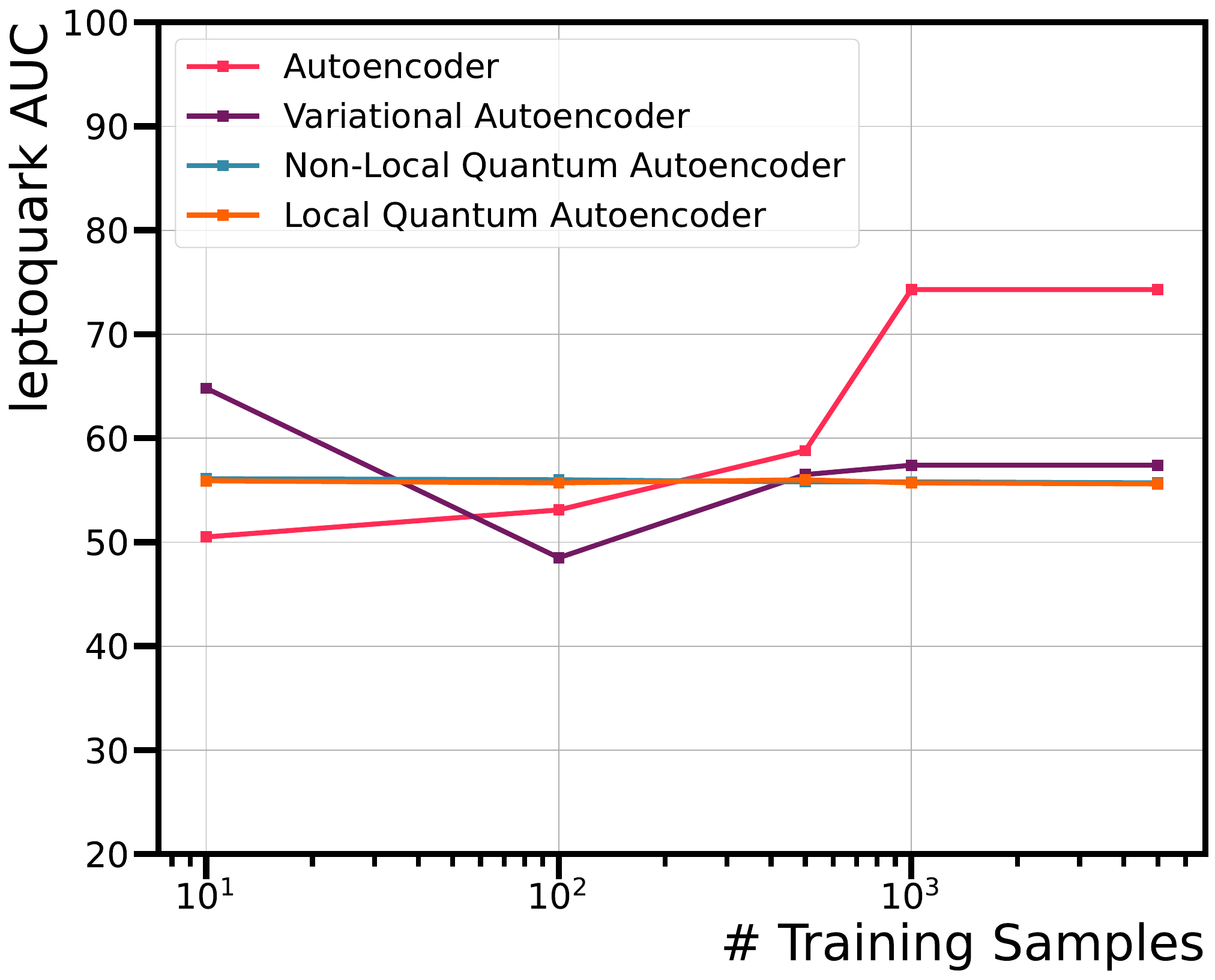}
        \caption{}
        \label{fig:training4}
    \end{subfigure}

    \caption{Dependence of the anomaly-detection performance on the number $N_{\mathrm{train}}$ of embedded background events used for training. The AUC, expressed as a percentage, is shown for the AE, VAE, non-local, and local QAE for (a) $A\rightarrow4\ell$, (b) $H\rightarrow\tau\tau$, (c) $H^{\pm}\rightarrow\tau\nu$, and (d) the leptoquark benchmark. For each model, the architecture and number of trainable parameters are held fixed across the scan. The QAE performance depends only weakly on the training-sample size, while the AE improves markedly between $N_{\mathrm{train}}=500$ and $N_{\mathrm{train}}=1000$.}
    \label{fig:training_size}
\end{figure}

The QAEs considered in the main analysis were trained using a comparatively small number of embedded background events. To examine the dependence of the anomaly-detection performance on this choice, we repeat the training of the AE, VAE, and local and non-local QAEs for different training-sample sizes. For each model, the architecture and number of trainable parameters are held fixed throughout the scan, such that the variation within each curve reflects the dependence on the amount of training data.

Figure~\ref{fig:training_size} shows that the performance of both QAEs depends only weakly on the training-sample size over the range considered. The local and non-local architectures remain closely comparable, and their AUCs exhibit only small variations as the number of training events is increased. In the low-data regime, $N_{\mathrm{train}}\leq 500$, the QAEs are therefore generally competitive with, and in most cases outperform, the classical baselines.

The AE displays a different dependence on the training-sample size. Its performance improves markedly between $N_{\mathrm{train}}=500$ and $N_{\mathrm{train}}=1000$, after which it remains approximately constant and exceeds the QAE performance for each benchmark. The VAE exhibits a non-monotonic dependence on $N_{\mathrm{train}}$, with no consistent improvement across the four signals.

%%% Smear with 500 training samples
\begin{figure}[t!]
    \centering
    \captionsetup[subfigure]{font=small,skip=2pt}
    \captionsetup{skip=4pt}
    \begin{subfigure}[t]{0.48\textwidth}
        \centering
        \includegraphics[
            width=\linewidth,
            height=0.28\textheight,
            keepaspectratio
        ]{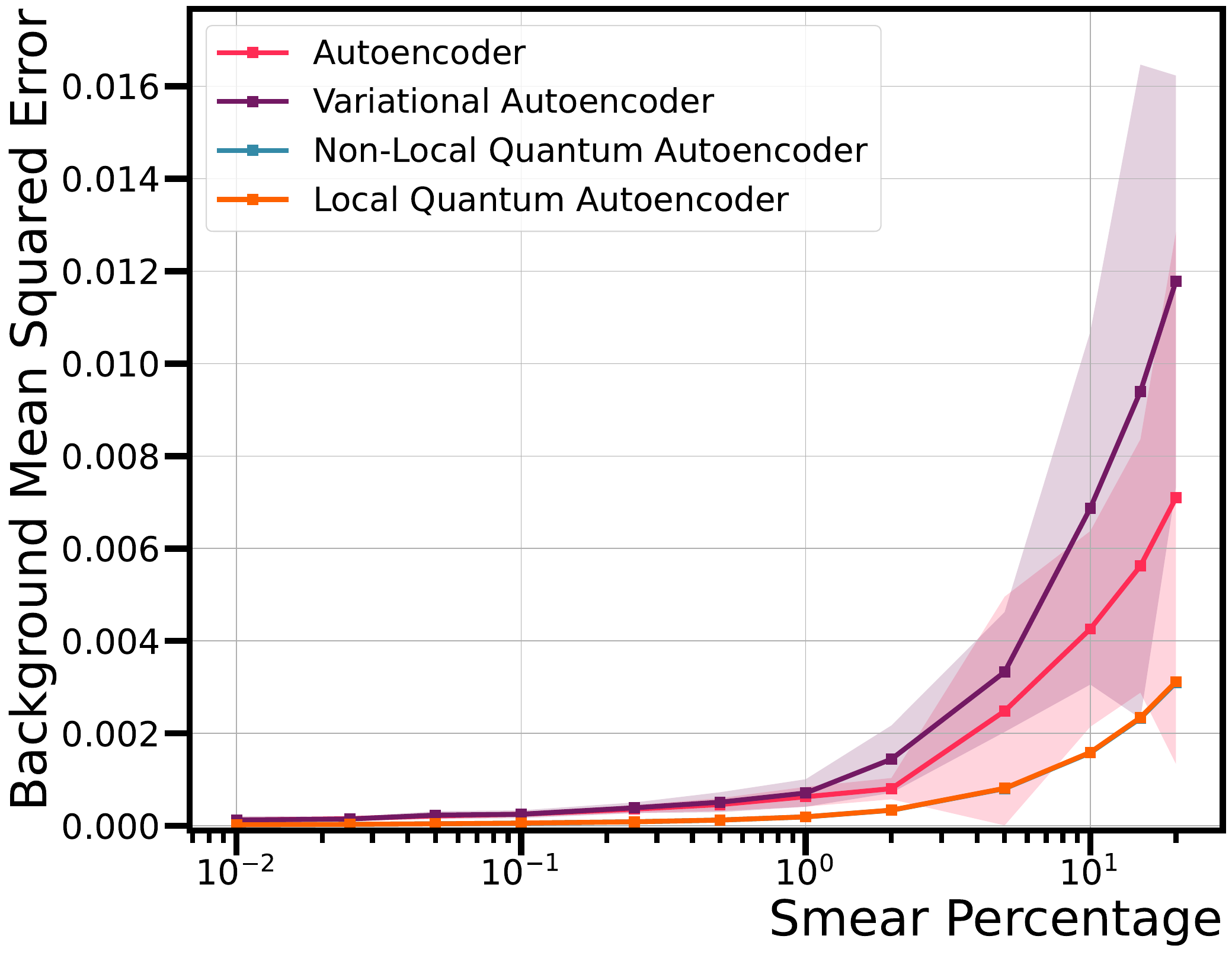}
        \caption{}
        \label{fig:500_bkg}
    \end{subfigure}
    \begin{subfigure}[t]{0.48\textwidth}
        \centering
        \includegraphics[
            width=\linewidth,
            height=0.28\textheight,
            keepaspectratio
        ]{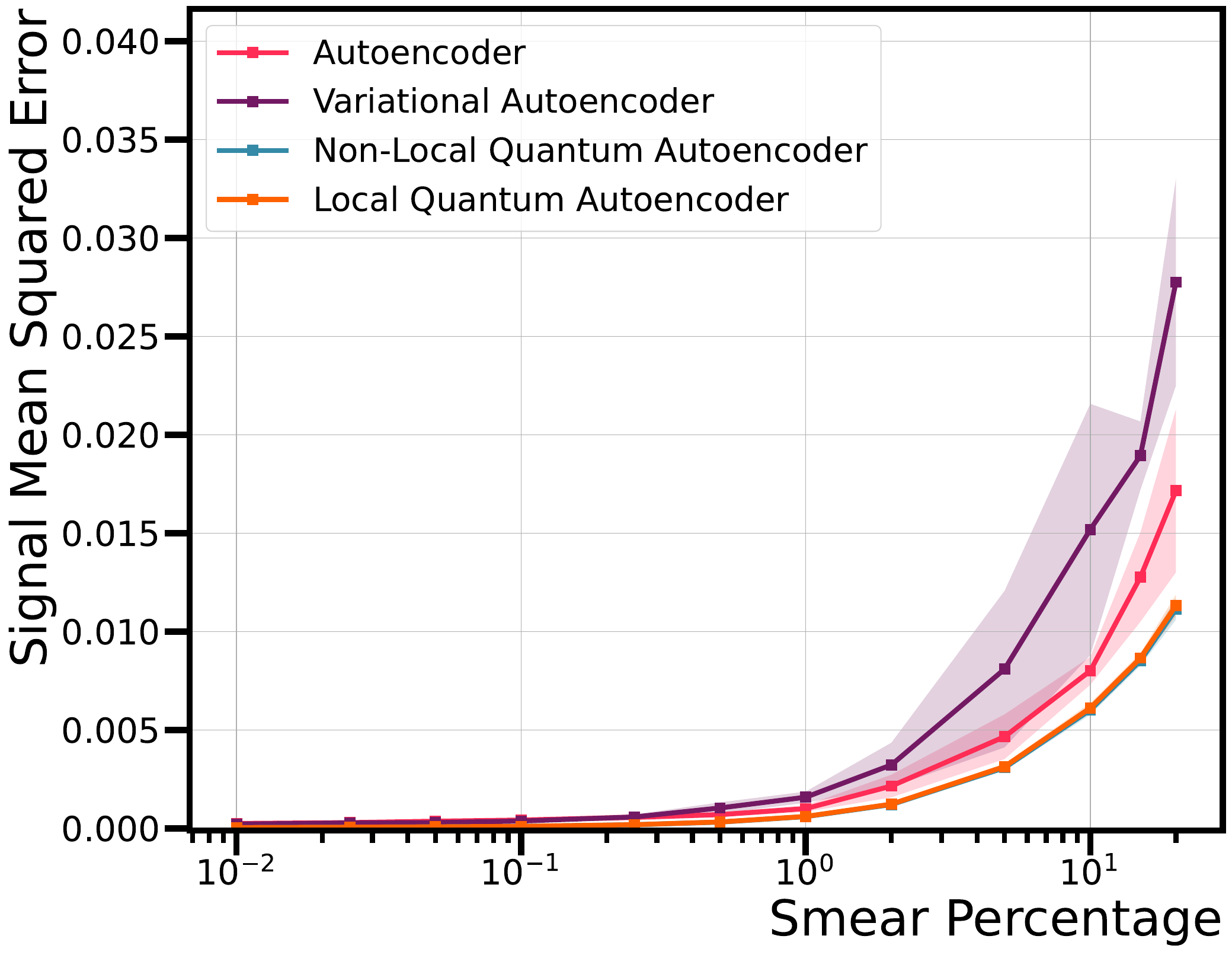}
        \caption{}
        \label{fig:500_sig}
    \end{subfigure}

    \caption{
        Mean-squared deviation of the anomaly score from its reference value as a function of the detector-smearing strength. Results are shown for background events in the left panel and $A\rightarrow4\ell$ events in the right panel, comparing the classical autoencoder, variational autoencoder, non-local quantum autoencoder and local quantum autoencoder. The model parameters, learned embedding and feature-scaling constants are held fixed throughout. The shaded bands indicate the standard deviation across ten independent slices of the test set with input smearing. All models were trained on 500 samples of the background dataset.
        }
    \label{fig:500samples}
\end{figure}

Figure~\ref{fig:500samples} shows the response to feature-level smearing when the classical and quantum models are each trained on 500 events. The classical model continues to exhibit larger deviations in its output as the smearing strength increases, indicating that the relative stability of the QAEs is not simply a consequence of differences in the number of training samples.

These results demonstrate that the discrimination performance of the QAEs is retained when their circuit parameters are optimised using very small samples of embedded background events. This should not be interpreted as establishing a general quantum advantage in sample efficiency, since the quantum and classical models differ in both architecture and parameter count. Within the models and training procedure considered here, however, the QAEs remain competitive in the low-data regime, whereas the AE achieves greater discrimination when trained on larger samples.

\clearpage
\bibliographystyle{inspire10}
\bibliography{refs}

@misc{pennylane,
      title={PennyLane: Automatic differentiation of hybrid quantum-classical computations}, 
      author={Ville Bergholm and Josh Izaac and Maria Schuld and Christian Gogolin and Shahnawaz Ahmed and Vishnu Ajith and M. Sohaib Alam and Guillermo Alonso-Linaje and B. AkashNarayanan and Ali Asadi and Juan Miguel Arrazola and Utkarsh Azad and Sam Banning and Carsten Blank and Thomas R Bromley and Benjamin A. Cordier and Jack Ceroni and Alain Delgado and Olivia Di Matteo and Amintor Dusko and Tanya Garg and Diego Guala and Anthony Hayes and Ryan Hill and Aroosa Ijaz and Theodor Isacsson and David Ittah and Soran Jahangiri and Prateek Jain and Edward Jiang and Ankit Khandelwal and Korbinian Kottmann and Robert A. Lang and Christina Lee and Thomas Loke and Angus Lowe and Keri McKiernan and Johannes Jakob Meyer and J. A. Montañez-Barrera and Romain Moyard and Zeyue Niu and Lee James O'Riordan and Steven Oud and Ashish Panigrahi and Chae-Yeun Park and Daniel Polatajko and Nicolás Quesada and Chase Roberts and Nahum Sá and Isidor Schoch and Borun Shi and Shuli Shu and Sukin Sim and Arshpreet Singh and Ingrid Strandberg and Jay Soni and Antal Száva and Slimane Thabet and Rodrigo A. Vargas-Hernández and Trevor Vincent and Nicola Vitucci and Maurice Weber and David Wierichs and Roeland Wiersema and Moritz Willmann and Vincent Wong and Shaoming Zhang and Nathan Killoran},
      year={2022},
      eprint={1811.04968},
      archivePrefix={arXiv},
      primaryClass={quant-ph},
      url={https://arxiv.org/abs/1811.04968}
}

@article{scikit-learn,
  title={Scikit-learn: Machine Learning in {P}ython},
  author={Pedregosa, F. and Varoquaux, G. and Gramfort, A. and Michel, V.
          and Thirion, B. and Grisel, O. and Blondel, M. and Prettenhofer, P.
          and Weiss, R. and Dubourg, V. and Vanderplas, J. and Passos, A. and
          Cournapeau, D. and Brucher, M. and Perrot, M. and Duchesnay, E.},
  journal={Journal of Machine Learning Research},
  volume={12},
  pages={2825--2830},
  year={2011}
}

@article{Guan_2021,
    doi = {10.1088/2632-2153/abc17d},
    url = {https://doi.org/10.1088/2632-2153/abc17d},
    year = {2021},
    month = {mar},
    publisher = {IOP Publishing},
    volume = {2},
    number = {1},
    pages = {011003},
    author = {Guan, Wen and Perdue, Gabriel and Pesah, Arthur and Schuld, Maria and Terashi, Koji and Vallecorsa, Sofia and Vlimant, Jean-Roch},
    title = {Quantum machine learning in high energy physics},
    journal = {Machine Learning: Science and Technology},
}

@misc{delgado2022quantumcomputingdataanalysis,
    title={Quantum computing for data analysis in high energy physics}, 
    author={Andrea Delgado and Kathleen E. Hamilton and Prasanna Date and Jean-Roch Vlimant and Duarte Magano and Yasser Omar and Pedrame Bargassa and Anthony Francis and Alessio Gianelle and Lorenzo Sestini and Donatella Lucchesi and Davide Zuliani and Davide Nicotra and Jacco de Vries and Dominica Dibenedetto and Miriam Lucio Martinez and Eduardo Rodrigues and Carlos Vazquez Sierra and Sofia Vallecorsa and Jesse Thaler and Carlos Bravo-Prieto and su Yeon Chang and Jeffrey Lazar and Carlos A. Argüelles and Jorge J. Martinez de Lejarza},
    year={2022},
    eprint={2203.08805},
    archivePrefix={arXiv},
    primaryClass={physics.data-an},
    url={https://arxiv.org/abs/2203.08805}, 
}

@article{terashi2021event,
    title={Event classification with quantum machine learning in high-energy physics},
    author={Terashi, Koji and Kaneda, Michiru and Kishimoto, Tomoe and Saito, Masahiko and Sawada, Ryu and Tanaka, Junichi},
    journal={Computing and Software for Big Science},
    volume={5},
    number={1},
    pages={2},
    year={2021},
    publisher={Springer}
}

@misc{Gonski:2026jgu,
    author = "Gonski, Julia and others",
    title = "{Machine Learning on Heterogeneous, Edge, and Quantum Hardware for Particle Physics (ML-HEQUPP)}",
    eprint = "2602.22248",
    archivePrefix = "arXiv",
    primaryClass = "physics.ins-det",
    reportNumber = "FERMILAB-PUB-26-0135-CSAID-ETD-PPD",
    month = "2",
    year = "2026"
}

@article{Blance:2020ktp,
    author = "Blance, Andrew and Spannowsky, Michael",
    title = "{Unsupervised event classification with graphs on classical and photonic quantum computers}",
    eprint = "2103.03897",
    archivePrefix = "arXiv",
    primaryClass = "hep-ph",
    doi = "10.1007/JHEP08(2021)170",
    journal = "JHEP",
    volume = "21",
    pages = "170",
    year = "2020"
}

@article{Blance:2020nhl,
    author = "Blance, Andrew and Spannowsky, Michael",
    title = "{Quantum Machine Learning for Particle Physics using a Variational Quantum Classifier}",
    eprint = "2010.07335",
    archivePrefix = "arXiv",
    primaryClass = "hep-ph",
    reportNumber = "IPPP/20/48",
    doi = "10.1007/JHEP02(2021)212",
    journal = "JHEP",
    volume = "02",
    pages = "212",
    year = "2021"
}

@article{Ngairangbam:2021yma,
    author = "Ngairangbam, Vishal S. and Spannowsky, Michael and Takeuchi, Michihisa",
    title = "{Anomaly detection in high-energy physics using a quantum autoencoder}",
    eprint = "2112.04958",
    archivePrefix = "arXiv",
    primaryClass = "hep-ph",
    reportNumber = "OU-HET-1125, IPPP/21/54",
    doi = "10.1103/PhysRevD.105.095004",
    journal = "Phys. Rev. D",
    volume = "105",
    number = "9",
    pages = "095004",
    year = "2022"
}

@article{1p1q,
    title = {One particle - one qubit: Particle physics data encoding for quantum machine learning},
    author = {Bal, Aritra and Klute, Markus and Maier, Benedikt and Oughton, Melik and Pezone, Eric and Spannowsky, Michael},
    journal = {Phys. Rev. D},
    volume = {112},
    issue = {7},
    pages = {076004},
    numpages = {9},
    year = {2025},
    month = {Oct},
    publisher = {American Physical Society},
    doi = {10.1103/l8y2-87vq},
    url = {https://link.aps.org/doi/10.1103/l8y2-87vq}
}

@article{Maier:2025ppr,
    author = "Maier, Benedikt and Spannowsky, Michael and Williams, Simon",
    title = "{Continuous-variable photonic quantum extreme learning machines for fast collider-data selection}",
    eprint = "2510.13994",
    archivePrefix = "arXiv",
    primaryClass = "quant-ph",
    reportNumber = "IPPP/25/63",
    doi = "10.1140/epjqt/s40507-026-00507-w",
    journal = "EPJ Quant. Technol.",
    volume = "13",
    number = "1",
    pages = "70",
    year = "2026"
}

@misc{duffy2024,
      title={Unsupervised Beyond-Standard-Model Event Discovery at the LHC with a Novel Quantum Autoencoder}, 
      author={Callum Duffy and Mohammad Hassanshah and Marcin Jastrzebski and Sarah Malik},
      year={2024},
      eprint={2407.07961},
      archivePrefix={arXiv},
      primaryClass={quant-ph},
      url={https://arxiv.org/abs/2407.07961}, 
}

@article{PerezSalinas2020datareuploading,
  doi = {10.22331/q-2020-02-06-226},
  url = {https://doi.org/10.22331/q-2020-02-06-226},
  title = {Data re-uploading for a universal quantum classifier},
  author = {P{\'{e}}rez-Salinas, Adri{\'{a}}n and Cervera-Lierta, Alba and Gil-Fuster, Elies and Latorre, Jos{\'{e}} I.},
  journal = {{Quantum}},
  issn = {2521-327X},
  publisher = {{Verein zur F{\"{o}}rderung des Open Access Publizierens in den Quantenwissenschaften}},
  volume = {4},
  pages = {226},
  month = feb,
  year = {2020}
}

@article{Havlicek:2018nqz,
    author = "Havlicek, Vojtech and C{\'o}rcoles, Antonio D. and Temme, Kristan and Harrow, Aram W. and Kandala, Abhinav and Chow, Jerry M. and Gambetta, Jay M.",
    title = "{Supervised learning with quantum-enhanced feature spaces}",
    eprint = "1804.11326",
    archivePrefix = "arXiv",
    primaryClass = "quant-ph",
    doi = "10.1038/s41586-019-0980-2",
    journal = "Nature",
    volume = "567",
    pages = "209--212",
    year = "2019"
}

@misc{Schuld:2021hzr,
    author = "Schuld, Maria",
    title = "{Supervised quantum machine learning models are kernel methods}",
    eprint = "2101.11020",
    archivePrefix = "arXiv",
    primaryClass = "quant-ph",
    month = "1",
    year = "2021"
}

@article{Romero_2017,
   title={Quantum autoencoders for efficient compression of quantum data},
   volume={2},
   ISSN={2058-9565},
   url={http://dx.doi.org/10.1088/2058-9565/aa8072},
   DOI={10.1088/2058-9565/aa8072},
   number={4},
   journal={Quantum Science and Technology},
   publisher={IOP Publishing},
   author={Romero, Jonathan and Olson, Jonathan P and Aspuru-Guzik, Alan},
   year={2017},
   month=Aug, pages={045001} 
}

@article{Benedetti_2019,
   title={Parameterized quantum circuits as machine learning models},
   volume={4},
   ISSN={2058-9565},
   url={http://dx.doi.org/10.1088/2058-9565/ab4eb5},
   DOI={10.1088/2058-9565/ab4eb5},
   number={4},
   journal={Quantum Science and Technology},
   publisher={IOP Publishing},
   author={Benedetti, Marcello and Lloyd, Erika and Sack, Stefan and Fiorentini, Mattia},
   year={2019},
   month=Nov, pages={043001} 
}

@article{Ingoldby:2025bdb,
    author = "Ingoldby, James and Spannowsky, Michael and Sypchenko, Timur and Williams, Simon and Wingate, Matthew",
    title = "{Real-Time Scattering on Quantum Computers via Hamiltonian Truncation}",
    eprint = "2505.03878",
    archivePrefix = "arXiv",
    primaryClass = "quant-ph",
    reportNumber = "IPPP/25/24",
    month = "5",
    year = "2025"
}

@article{Abel:2025zxb,
    author = "Abel, Steven and Spannowsky, Michael and Williams, Simon",
    title = "{Real-time scattering processes with continuous-variable quantum computers}",
    eprint = "2502.01767",
    archivePrefix = "arXiv",
    primaryClass = "quant-ph",
    reportNumber = "IPPP/24/82",
    doi = "10.1103/q36d-w649",
    journal = "Phys. Rev. A",
    volume = "112",
    number = "1",
    pages = "012614",
    year = "2025"
}

@article{Ingoldby:2024fcy,
    author = "Ingoldby, James and Spannowsky, Michael and Sypchenko, Timur and Williams, Simon",
    title = "{Enhancing quantum field theory simulations on NISQ devices with Hamiltonian truncation}",
    eprint = "2407.19022",
    archivePrefix = "arXiv",
    primaryClass = "quant-ph",
    reportNumber = "IPPP/24/37",
    doi = "10.1103/PhysRevD.110.096016",
    journal = "Phys. Rev. D",
    volume = "110",
    number = "9",
    pages = "096016",
    year = "2024"
}

@article{Abel:2024kuv,
    author = "Abel, Steven and Spannowsky, Michael and Williams, Simon",
    title = "{Simulating quantum field theories on continuous-variable quantum computers}",
    eprint = "2403.10619",
    archivePrefix = "arXiv",
    primaryClass = "quant-ph",
    reportNumber = "IPPP/24/10",
    doi = "10.1103/PhysRevA.110.012607",
    journal = "Phys. Rev. A",
    volume = "110",
    number = "1",
    pages = "012607",
    year = "2024"
}

@misc{jordan2019,
      title={Quantum Computation of Scattering in Scalar Quantum Field Theories}, 
      author={Stephen P. Jordan and Keith S. M. Lee and John Preskill},
      year={2019},
      eprint={1112.4833},
      archivePrefix={arXiv},
      primaryClass={hep-th},
      url={https://arxiv.org/abs/1112.4833}, 
}

@article{Jordan_2012,
   title={Quantum Algorithms for Quantum Field Theories},
   volume={336},
   ISSN={1095-9203},
   url={http://dx.doi.org/10.1126/science.1217069},
   DOI={10.1126/science.1217069},
   number={6085},
   journal={Science},
   publisher={American Association for the Advancement of Science (AAAS)},
   author={Jordan, Stephen P. and Lee, Keith S. M. and Preskill, John},
   year={2012},
   month=June, pages={1130–1133} 
}

@article{PRXQuantum.4.027001,
  title = {Quantum Simulation for High-Energy Physics},
  author = {Bauer, Christian W. and Davoudi, Zohreh and Balantekin, A. Baha and Bhattacharya, Tanmoy and Carena, Marcela and de Jong, Wibe A. and Draper, Patrick and El-Khadra, Aida and Gemelke, Nate and Hanada, Masanori and Kharzeev, Dmitri and Lamm, Henry and Li, Ying-Ying and Liu, Junyu and Lukin, Mikhail and Meurice, Yannick and Monroe, Christopher and Nachman, Benjamin and Pagano, Guido and Preskill, John and Rinaldi, Enrico and Roggero, Alessandro and Santiago, David I. and Savage, Martin J. and Siddiqi, Irfan and Siopsis, George and Van Zanten, David and Wiebe, Nathan and Yamauchi, Yukari and Yeter-Aydeniz, K\"ubra and Zorzetti, Silvia},
  journal = {PRX Quantum},
  volume = {4},
  issue = {2},
  pages = {027001},
  numpages = {70},
  year = {2023},
  month = {May},
  publisher = {American Physical Society},
  doi = {10.1103/PRXQuantum.4.027001},
  url = {https://link.aps.org/doi/10.1103/PRXQuantum.4.027001}
}

@article{Banuls:2019bmf,
    author = "Ba{\~n}uls, M. C. and others",
    title = "{Simulating Lattice Gauge Theories within Quantum Technologies}",
    eprint = "1911.00003",
    archivePrefix = "arXiv",
    primaryClass = "quant-ph",
    doi = "10.1140/epjd/e2020-100571-8",
    journal = "Eur. Phys. J. D",
    volume = "74",
    number = "8",
    pages = "165",
    year = "2020"
}

@article{Klco:2018zqz,
    author = "Klco, Natalie and Savage, Martin J.",
    title = "{Digitization of scalar fields for quantum computing}",
    eprint = "1808.10378",
    archivePrefix = "arXiv",
    primaryClass = "quant-ph",
    reportNumber = "INT-PUB-18-044",
    doi = "10.1103/PhysRevA.99.052335",
    journal = "Phys. Rev. A",
    volume = "99",
    number = "5",
    pages = "052335",
    year = "2019"
}

@article{qr72-51v1,
  title = {Scalable quantum simulations of scattering in scalar field theory on 120 qubits},
  author = {Zemlevskiy, Nikita A.},
  journal = {Phys. Rev. D},
  volume = {112},
  issue = {3},
  pages = {034502},
  numpages = {49},
  year = {2025},
  month = {Aug},
  publisher = {American Physical Society},
  doi = {10.1103/qr72-51v1},
  url = {https://link.aps.org/doi/10.1103/qr72-51v1}
}

@article{Martinez:2016yna,
    author = "Martinez, E. A. and others",
    title = "{Real-time dynamics of lattice gauge theories with a few-qubit quantum computer}",
    eprint = "1605.04570",
    archivePrefix = "arXiv",
    primaryClass = "quant-ph",
    doi = "10.1038/nature18318",
    journal = "Nature",
    volume = "534",
    pages = "516--519",
    year = "2016"
}

@article{Jha:2024jan,
    author = "Jha, Raghav G. and Milsted, Ashley and Neuenfeld, Dominik and Preskill, John and Vieira, Pedro",
    title = "{Real-time scattering in Ising field theory using matrix product states}",
    eprint = "2411.13645",
    archivePrefix = "arXiv",
    primaryClass = "hep-th",
    doi = "10.1103/9dxz-k5wb",
    journal = "Phys. Rev. Res.",
    volume = "7",
    number = "2",
    pages = "023266",
    year = "2025"
}

@article{Preskill:2018fag,
    author = "Preskill, John",
    title = "{Simulating quantum field theory with a quantum computer}",
    eprint = "1811.10085",
    archivePrefix = "arXiv",
    primaryClass = "hep-lat",
    doi = "10.22323/1.334.0024",
    journal = "PoS",
    volume = "LATTICE2018",
    pages = "024",
    year = "2018"
}

@article{PRXQuantum.5.037001,
  title = {Quantum Computing for High-Energy Physics: State of the Art and Challenges},
  author = {Di Meglio, Alberto and Jansen, Karl and Tavernelli, Ivano and Alexandrou, Constantia and Arunachalam, Srinivasan and Bauer, Christian W. and Borras, Kerstin and Carrazza, Stefano and Crippa, Arianna and Croft, Vincent and de Putter, Roland and Delgado, Andrea and Dunjko, Vedran and Egger, Daniel J. and Fern\'andez-Combarro, Elias and Fuchs, Elina and Funcke, Lena and Gonz\'alez-Cuadra, Daniel and Grossi, Michele and Halimeh, Jad C. and Holmes, Zo\"e and K\"uhn, Stefan and Lacroix, Denis and Lewis, Randy and Lucchesi, Donatella and Martinez, Miriam Lucio and Meloni, Federico and Mezzacapo, Antonio and Montangero, Simone and Nagano, Lento and Pascuzzi, Vincent R. and Radescu, Voica and Ortega, Enrique Rico and Roggero, Alessandro and Schuhmacher, Julian and Seixas, Joao and Silvi, Pietro and Spentzouris, Panagiotis and Tacchino, Francesco and Temme, Kristan and Terashi, Koji and Tura, Jordi and T\"uys\"uz, Cenk and Vallecorsa, Sofia and Wiese, Uwe-Jens and Yoo, Shinjae and Zhang, Jinglei},
  journal = {PRX Quantum},
  volume = {5},
  issue = {3},
  pages = {037001},
  numpages = {49},
  year = {2024},
  month = {Aug},
  publisher = {American Physical Society},
  doi = {10.1103/PRXQuantum.5.037001},
  url = {https://link.aps.org/doi/10.1103/PRXQuantum.5.037001}
}

@article{Pavesic:2026yiz,
    author = "Pave{\v{s}}i{\'c}, Luka and Moss, Ian G. and Montangero, Simone",
    title = "{False vacuum decay in a two-dimensional quantum spin system}",
    eprint = "2607.01994",
    archivePrefix = "arXiv",
    primaryClass = "quant-ph",
    month = "7",
    year = "2026"
}

@article{Bepari:2020xqi,
    author = "Bepari, Khadeejah and Malik, Sarah and Spannowsky, Michael and Williams, Simon",
    title = "{Towards a quantum computing algorithm for helicity amplitudes and parton showers}",
    eprint = "2010.00046",
    archivePrefix = "arXiv",
    primaryClass = "hep-ph",
    reportNumber = "IPPP/20/41",
    doi = "10.1103/PhysRevD.103.076020",
    journal = "Phys. Rev. D",
    volume = "103",
    number = "7",
    pages = "076020",
    year = "2021"
}

@article{Bepari:2021kwv,
    author = "Bepari, Khadeejah and Malik, Sarah and Spannowsky, Michael and Williams, Simon",
    title = "{Quantum walk approach to simulating parton showers}",
    eprint = "2109.13975",
    archivePrefix = "arXiv",
    primaryClass = "hep-ph",
    doi = "10.1103/PhysRevD.106.056002",
    journal = "Phys. Rev. D",
    volume = "106",
    number = "5",
    pages = "056002",
    year = "2022"
}

@article{Gustafson:2022dsq,
    author = {Gustafson, G{\"o}sta and Prestel, Stefan and Spannowsky, Michael and Williams, Simon},
    title = "{Collider events on a quantum computer}",
    eprint = "2207.10694",
    archivePrefix = "arXiv",
    primaryClass = "hep-ph",
    doi = "10.1007/JHEP11(2022)035",
    journal = "JHEP",
    volume = "11",
    pages = "035",
    year = "2022"
}

@article{deLejarza:2024pgk,
    author = "de Lejarza, Jorge J. Mart{\'\i}nez and Cieri, Leandro and Grossi, Michele and Vallecorsa, Sofia and Rodrigo, Germ{\'a}n",
    title = "{Loop Feynman integration on a quantum computer}",
    eprint = "2401.03023",
    archivePrefix = "arXiv",
    primaryClass = "hep-ph",
    doi = "10.1103/PhysRevD.110.074031",
    journal = "Phys. Rev. D",
    volume = "110",
    number = "7",
    pages = "074031",
    year = "2024"
}

@article{Ramirez-Uribe:2021ubp,
    author = "Ram{\'\i}rez-Uribe, Selomit and Renter{\'\i}a-Olivo, Andr{\'e}s E. and Rodrigo, Germ{\'a}n and Sborlini, German F. R. and Vale Silva, Luiz",
    title = "{Quantum algorithm for Feynman loop integrals}",
    eprint = "2105.08703",
    archivePrefix = "arXiv",
    primaryClass = "hep-ph",
    reportNumber = "IFIC/21-15, DESY 21-067",
    doi = "10.1007/JHEP05(2022)100",
    journal = "JHEP",
    volume = "05",
    pages = "100",
    year = "2022"
}

@article{Williams:2025hza,
    author = "Williams, Ifan and Pellen, Mathieu",
    title = "{A general approach to quantum integration of cross sections in high-energy physics}",
    eprint = "2502.14647",
    archivePrefix = "arXiv",
    primaryClass = "quant-ph",
    reportNumber = "FR-PHENO-2025-02",
    doi = "10.1088/2058-9565/adf771",
    journal = "Quantum Sci. Technol.",
    volume = "10",
    number = "4",
    pages = "045017",
    year = "2025"
}

@article{Bauer:2019qxa,
    author = "Bauer, Christian W. and de Jong, Wibe A. and Nachman, Benjamin and Provasoli, Davide",
    title = "{Quantum Algorithm for High Energy Physics Simulations}",
    eprint = "1904.03196",
    archivePrefix = "arXiv",
    primaryClass = "hep-ph",
    doi = "10.1103/PhysRevLett.126.062001",
    journal = "Phys. Rev. Lett.",
    volume = "126",
    number = "6",
    pages = "062001",
    year = "2021"
}

@article{Chawdhry:2023jks,
    author = "Chawdhry, Herschel A. and Pellen, Mathieu",
    title = "{Quantum simulation of colour in perturbative quantum chromodynamics}",
    eprint = "2303.04818",
    archivePrefix = "arXiv",
    primaryClass = "hep-ph",
    reportNumber = "FR-PHENO-2023-02, OUTP-23-02P",
    doi = "10.21468/SciPostPhys.15.5.205",
    journal = "SciPost Phys.",
    volume = "15",
    number = "5",
    pages = "205",
    year = "2023"
}

@article{Chawdhry2026,
  author   = {Chawdhry, Herschel A. and Pellen, Mathieu and Williams, Simon},
  title    = {Quantum simulation of scattering amplitudes and interferences in perturbative {QCD}},
  journal  = {The European Physical Journal C},
  year     = {2026},
  month    = aug,
  volume   = {86},
  number   = {8},
  pages    = {961},
  issn     = {1434-6052},
  doi      = {10.1140/epjc/s10052-026-16121-0},
  url      = {https://doi.org/10.1140/epjc/s10052-026-16121-0}
}

@article{adc2021,
	author = {Govorkova, Ekaterina and Puljak, Ema and Aarrestad, Thea and Pierini, Maurizio and Wo{\'z}niak, Kinga Anna and Ngadiuba, Jennifer},
	date = {2022/03/29},
	doi = {10.1038/s41597-022-01187-8},
	id = {Govorkova2022},
	isbn = {2052-4463},
	journal = {Scientific Data},
	number = {1},
	pages = {118},
	title = {LHC physics dataset for unsupervised New Physics detection at 40 MHz},
	url = {https://doi.org/10.1038/s41597-022-01187-8},
	volume = {9},
	year = {2022}}

@misc{susy_279,
  author       = {Whiteson, Daniel},
  title        = {{SUSY}},
  year         = {2014},
  howpublished = {UCI Machine Learning Repository},
  note         = {{DOI}: https://doi.org/10.24432/C54606}
}

@article{anomalypreservingneuralembeddings,
  title = {Anomaly-preserving contrastive neural embeddings for end-to-end model-independent searches at the LHC},
  author = {Metzger, Kyle and Xu, Lana and Sodini, Mia and \AA{}rrestad, Thea K. and Govorkova, Katya and Grosso, Gaia and Harris, Philip},
  journal = {Phys. Rev. D},
  volume = {112},
  issue = {7},
  pages = {072011},
  numpages = {20},
  year = {2025},
  month = {Oct},
  publisher = {American Physical Society},
  doi = {10.1103/5n77-ynsp},
  url = {https://link.aps.org/doi/10.1103/5n77-ynsp}
}

@article{CMS-DP-2025-061,
      author        = "{CMS Collaboration}",
      collaboration = "CMS",
      title         = "{Anomaly detection with AXOL1TL at the CMS Level-1 Trigger
                       in 2024 and 2025}",
      year          = "2025",
      url           = "https://cds.cern.ch/record/2942560",
}

@article{Govorkova:2021utb,
    author = "Govorkova, Ekaterina and others",
    title = "{Autoencoders on field-programmable gate arrays for real-time, unsupervised new physics detection at 40 MHz at the Large Hadron Collider}",
    eprint = "2108.03986",
    archivePrefix = "arXiv",
    primaryClass = "physics.ins-det",
    reportNumber = "FERMILAB-PUB-21-487-CMS, FERMILAB-PUB-21-487-CMS",
    doi = "10.1038/s42256-022-00441-3",
    journal = "Nature Mach. Intell.",
    volume = "4",
    pages = "154--161",
    year = "2022"
}

@article{Cerri:2018anq,
    author = "Cerri, Olmo and Nguyen, Thong Q. and Pierini, Maurizio and Spiropulu, Maria and Vlimant, Jean-Roch",
    title = "{Variational Autoencoders for New Physics Mining at the Large Hadron Collider}",
    eprint = "1811.10276",
    archivePrefix = "arXiv",
    primaryClass = "hep-ex",
    doi = "10.1007/JHEP05(2019)036",
    journal = "JHEP",
    volume = "05",
    pages = "036",
    year = "2019"
}

@article{Schuld:2020enb,
    author = "Schuld, Maria and Sweke, Ryan and Meyer, Johannes Jakob",
    title = "{Effect of data encoding on the expressive power of variational quantum-machine-learning models}",
    eprint = "2008.08605",
    archivePrefix = "arXiv",
    primaryClass = "quant-ph",
    doi = "10.1103/PhysRevA.103.032430",
    journal = "Phys. Rev. A",
    volume = "103",
    number = "3",
    pages = "032430",
    year = "2021"
}

@article{Louppe:2016ylz,
    author = "Louppe, Gilles and Kagan, Michael and Cranmer, Kyle",
    title = "{Learning to Pivot with Adversarial Networks}",
    eprint = "1611.01046",
    archivePrefix = "arXiv",
    primaryClass = "stat.ML",
    month = "11",
    year = "2016"
}

@article{Englert:2018cfo,
    author = "Englert, Christoph and Galler, Peter and Harris, Philip and Spannowsky, Michael",
    title = "{Machine Learning Uncertainties with Adversarial Neural Networks}",
    eprint = "1807.08763",
    archivePrefix = "arXiv",
    primaryClass = "hep-ph",
    reportNumber = "IPPP/18/61",
    doi = "10.1140/epjc/s10052-018-6511-8",
    journal = "Eur. Phys. J. C",
    volume = "79",
    number = "1",
    pages = "4",
    year = "2019"
}

@article{Ge:2026eya,
    author = "Ge, Ivan and Addepalli, Sagar and Dave, Abhilasha and Gonski, Julia",
    title = "{Classical Hardware Acceleration of Quantum Autoencoders for Real-Time Anomaly Detection in Collider Experiments}",
    eprint = "2607.20302",
    archivePrefix = "arXiv",
    primaryClass = "cs.LG",
    month = "7",
    year = "2026"
}

@article{Liu:2019drq,
    author = "Liu, Nana and Wittek, Peter",
    title = "{Vulnerability of quantum classification to adversarial perturbations}",
    eprint = "1905.04286",
    archivePrefix = "arXiv",
    primaryClass = "quant-ph",
    doi = "10.1103/PhysRevA.101.062331",
    journal = "Phys. Rev. A",
    volume = "101",
    number = "6",
    pages = "062331",
    year = "2020"
}

@article{Dowling:2024eja,
    author = "Dowling, Neil and West, Maxwell T. and Southwell, Angus and Nakhl, Azar C. and Sevior, Martin and Usman, Muhammad and Modi, Kavan",
    title = "{Adversarial robustness guarantees for quantum classifiers}",
    eprint = "2405.10360",
    archivePrefix = "arXiv",
    primaryClass = "quant-ph",
    doi = "10.1038/s41534-025-01129-3",
    journal = "npj Quantum Inf.",
    volume = "12",
    number = "1",
    pages = "16",
    year = "2026"
}

@article{mott2017solving,
  title={Solving a Higgs optimization problem with quantum annealing for machine learning},
  author={Mott, Alex and Job, Joshua and Vlimant, Jean-Roch and Lidar, Daniel and Spiropulu, Maria},
  journal={Nature},
  volume={550},
  number={7676},
  pages={375--379},
  year={2017},
  publisher={Nature Publishing Group UK London}
}
 
\end{document}